\documentclass[sigconf, nonacm]{acmart}
\newcommand\vldbdoi{10.14778/3749646.3749655}
\newcommand\vldbpages{3797 - 3811}
\newcommand\vldbvolume{18}
\newcommand\vldbissue{11}
\newcommand\vldbyear{2025}
\newcommand\vldbauthors{\authors}
\newcommand\vldbtitle{\shorttitle} 
\newcommand\vldbpagestyle{empty} 
\newcommand\vldbavailabilityurl{https://github.com/fpgasystems/Falcon-accelerate-graph-vector-search}

\usepackage[most]{tcolorbox}

\usepackage[T1]{fontenc}
\usepackage{aecompl}

\usepackage{hyperxmp}  % Load hyperxmp after hyperref

\usepackage{algorithm}
\usepackage{algpseudocode}
\algtext*{EndWhile}% Remove "end while" text
\algtext*{EndFor}% Remove "end for" text
\algtext*{EndIf}% Remove "end if" text
\usepackage{balance}
\usepackage{enumitem}
\usepackage{multicol}
\usepackage{multirow}
\usepackage{tablefootnote}

\usepackage[super]{nth}

\usepackage{tikz}
\usepackage{xcolor}

\usepackage{tabularx}
\newcolumntype{M}[1]{>{\centering\arraybackslash}m{#1}} % centered
\newcolumntype{R}[1]{>{\raggedleft\arraybackslash}m{#1}} % right centered
\newcolumntype{L}[1]{>{\raggedright\arraybackslash}m{#1}} % right centered
\usepackage{booktabs}

\usepackage{arydshln}

\usepackage{wrapfig}
\usepackage{graphicx}
\usepackage{caption}  % in your preamble

\usepackage{subcaption}  % Required for correct subfigure formatting
\usepackage{amsmath}
\setlist[itemize]{align=parleft,left=0pt..1.2em} % 

\usepackage{microtype}
\usepackage{booktabs} % for professional tables
\usepackage[export]{adjustbox}

\begin{document}
%-------------------------------------------------------------------------------

% \title{Fast Graph Vector Search via Hardware-Algorithm Co-design}
% \title{Fast Graph-based Vector Search via Hardware Acceleration and Delayed-Synchronization Traversal}
\title{Fast Graph Vector Search via Hardware Acceleration and Delayed-Synchronization Traversal}
% \title{
% Chameleon: a Disaggregated CPU, GPU, and FPGA System for Retrieval-Augmented Language Models}

\date{}

% \thispagestyle{empty}

% %for single author (just remove % characters)
% \author{
% {\rm Wenqi Jiang}\\
% ETH Zurich
% \and
% {\rm Marco Zeller}\\
% ETH Zurich
% % copy the following lines to add more authors
% \and
% {\rm Roger Waleffe}\\
% University of Wisconsin-Madison
% \and
% \and
% {\rm Torsten Hoefler}\\
% ETH Zurich
% \and
% {\rm Gustavo Alonso}\\
% ETH Zurich
% } % end author

% multiple authors of different affiliations: https://tex.stackexchange.com/questions/9594/adding-more-than-one-author-with-different-affiliation
% \author[1]{Wenqi Jiang}
% \author[1]{Marco Zeller}
% \author[2]{Roger Waleffe}
% \author[1]{Torsten Hoefler}
% \author[1]{Gustavo Alonso}
% \affil[1]{Department of Computer Science, ETH Zurich}
% \affil[2]{Department of Computer Science, University of Wisconsin-Madison}
% % \affil[1]{Systems Group, Department of Computer Science, ETH Zurich}
% % \affil[2]{Department of Computer Science, University of Wisconsin-Madison}
% % \affil[3]{Scalable Parallel Computing Laboratory, Department of Computer Science, ETH Zurich}

\author{Wenqi Jiang}
\affiliation{%
  % \institution{Systems Group}
  % \institution{Department of Computer Science\\ETH Zurich}
  % \city{Zurich}
  % \country{ETH Zurich, Switzerland}
  \country{Systems Group, ETH Zurich}
}
\email{wenqi.jiang@inf.ethz.ch}

\author{Hang Hu}
\affiliation{%
  % \institution{Systems Group}
  % \institution{ETH Zurich}
  % \city{Zurich}
  % \country{ETH Zurich, Switzerland}
  \country{Systems Group, ETH Zurich}
}
\email{hanghu@student.ethz.ch}

\author{Torsten Hoefler}
\affiliation{%
  % \institution{SPCL}
  % \institution{Scalable Parallel Computing Laboratory\\Department of Computer Science\\ETH Zurich}
  % \city{Zurich}
  % \country{ETH Zurich, Switzerland}
  \country{SPCL, ETH Zurich}
}
\email{torsten.hoefler@inf.ethz.ch}

\author{Gustavo Alonso}
\affiliation{%
  % \institution{Systems Group}
  % \city{Zurich}
  % \country{ETH Zurich, Switzerland}
  \country{Systems Group, ETH Zurich}
}
\email{alonso@inf.ethz.ch}

\sloppy
%-------------------------------------------------------------------------------
\begin{abstract}
%-------------------------------------------------------------------------------

Vector search systems are indispensable in large language model (LLM) serving, search engines, and recommender systems, where minimizing online search latency is essential.
Among various algorithms, graph-based vector search (GVS) is particularly popular due to its high search performance and quality. 
However, reducing GVS latency by intra-query parallelization remains challenging due to limitations imposed by both existing hardware architectures (CPUs and GPUs) and the inherent difficulty of parallelizing graph traversals.  
To efficiently serve low-latency GVS, we co-design hardware and algorithm by proposing Falcon and Delayed-Synchronization Traversal (DST). 
Falcon is a hardware GVS accelerator that implements efficient GVS operators, pipelines these operators, and reduces memory accesses by tracking search states with an on-chip Bloom filter. 
DST is an efficient graph traversal algorithm that simultaneously improves search performance and quality by relaxing traversal orders to maximize accelerator utilization.
Evaluation across various graphs and datasets shows that Falcon, prototyped on FPGAs, together with DST, achieves up to 4.3$\times$ and 19.5$\times$ lower latency and up to 8.0$\times$ and 26.9$\times$ improvements in energy efficiency over CPU- and GPU-based GVS systems. 
%
% The remarkable efficiency of Falcon and DST demonstrates their potential to become the standard solutions for future GVS acceleration.

% Compared to CPU and CPU-GPU vector search systems, Chameleon's accelerators achieve up to 23.72$\times$ speedup and 26.2$\times$ energy efficiency. 
% However, RALMs introduce unique system design challenges due to (a) the diverse workload characteristics between LLM inference and retrieval and (b) the various system requirements and bottlenecks for different RALM configurations including model sizes, database sizes, and retrieval frequencies. 

% requiring substantial memory capacity and rapid quantized vector decoding, with the CPU server managing the vector index and FPGA-based disaggregated memory nodes scanning database vectors using near-memory accelerators.
% The node-level pipeline mechanism in Chameleon further improves system throughput. The disaggregated design also allows flexible resource scaling. 

\end{abstract}

\maketitle % should come after the abstract

\pagestyle{\vldbpagestyle}
\begingroup\small\noindent\raggedright\textbf{PVLDB Reference Format:}\\
\vldbauthors. \vldbtitle. PVLDB, \vldbvolume(\vldbissue): \vldbpages, \vldbyear.\\
\href{https://doi.org/\vldbdoi}{doi:\vldbdoi}
\endgroup
\begingroup
\renewcommand\thefootnote{}\footnote{\noindent
This work is licensed under the Creative Commons BY-NC-ND 4.0 International License. Visit \url{https://creativecommons.org/licenses/by-nc-nd/4.0/} to view a copy of this license. For any use beyond those covered by this license, obtain permission by emailing \href{mailto:info@vldb.org}{info@vldb.org}. Copyright is held by the owner/author(s). Publication rights licensed to the VLDB Endowment. \\
\raggedright Proceedings of the VLDB Endowment, Vol. \vldbvolume, No. \vldbissue\ %
ISSN 2150-8097. \\
\href{https://doi.org/\vldbdoi}{doi:\vldbdoi} \\
}\addtocounter{footnote}{-1}\endgroup

\ifdefempty{\vldbavailabilityurl}{}{
\vspace{.3cm}
\begingroup\small\noindent\raggedright\textbf{PVLDB Artifact Availability:}\\
The source code, data, and/or other artifacts have been made available at \url{\vldbavailabilityurl}.
\endgroup
}

%%%% VLDB version

\setcounter{page}{1}
\section{Introduction}
\label{sec:intro}

Vector search is essential in large language model (LLM) serving systems~\cite{guu2020realm, lewis2020retrieval, borgeaud2022improving}, recommender systems~\cite{google_recommendation, suchal2010full}, and search engines~\cite{chen2021spann, xiong2020approximate, karpukhin2020dense}. 
Upon receiving a query vector, a vector search system retrieves the most similar vectors from a database approximately,  a process known as approximate nearest neighbor (ANN) search.  
For example, search engines represent web pages as database vectors, and user’s textual queries are encoded as query vectors~\cite{chen2021spann, facebook_EBR, xiong2020approximate, karpukhin2020dense, khattab2020colbert}. 
Similarly, recommender systems identify advertisements that are potentially appealing to users by searching through encoded advertisement vectors~\cite{google_recommendation, suchal2010full}. 
More recently, LLM systems have also adopted ANN search to improve content generation quality by retrieving reliable textual knowledge, an approach known as \textit{Retrieval-Augmented Generation (RAG)}~\cite{guu2020realm, lewis2020retrieval, borgeaud2022improving}. 
% The prompt is encoded as a query to retrieve relevant textual knowledge from the databases, which is then integrated into the generation process, thereby facilitating more reliable knowledge-intensive LLM serving~\cite{lewis2020retrieval, izacard2020leveraging, izacard2022few, borgeaud2022improving, khandelwal2019generalization}.

Among various ANN search algorithms, \textit{graph-based vector search (GVS)} algorithms are particularly popular due to their high search performance and quality~\cite{li2019approximate, malkov2018efficient, fu2017fast}, with the latter measured by recall, the percentage of true nearest neighbors correctly identified by the search. The key idea of GVS is to construct a proximity graph on database vectors: each vector is a node, and similar vectors are linked by edges. During a search, the query vector is compared to a subset of database vectors by iteratively traversing the graph using best-first-search (BFS), which greedily selects the best candidate node to evaluate for each search iteration.% (Figure~\ref{fig:graph_traversal}).

% \begin{figure}[t]
% 	% \vspace*{-5mm} % to shrink gap between figures
% 	% full width, can be adjusted
% 	\centering
%   \includegraphics[width=1.0\linewidth]{fig/graph_traversal.pdf}
%   % \vspace{-2em}
%   \caption{Vector search on a graph using best-first search.}
%   % \vspace{-1em}
%   \label{fig:graph_traversal}
% 	% \vspace*{-5mm} % to shrink gap between figures
% \end{figure}

% \begin{figure*}[t]
% \begin{tcolorbox}[colframe=black, boxrule=0.2mm, colback=white, arc=0mm]
\begin{table*}[t]

\centering
\caption{Comparison between Falcon + DST and vector search on existing hardware platforms, including CPUs, GPUs, and prior FPGA-based accelerators. $\checkmark$ and $\times$ indicate supported and unsupported features, respectively.}
\label{tab:overview_comparison}
\vspace{-1em} 
% \begin{tabular}{lcccc}
\begin{tabular}{>{\color{black}}l >{\color{black}}c >{\color{black}}c >{\color{black}}c >{\color{black}}c}

\toprule
\textbf{Feature} & \textbf{CPU}~\cite{malkov2018efficient, malkov2014approximate, fu2017fast} & \textbf{GPU}~\cite{groh2022ggnn, zhao2020song} & \textbf{Prior FPGA}~\cite{zeng2023df, peng2021optimizing} & \textbf{Falcon + DST (ours)} \\
\midrule
Minimize Memory Access (\S\ref{sec:accelerator_pe})                   & $\times$ & $\times$ & $\times$ & $\checkmark$ \\
Intra-query Parallelism (\S\ref{sec:accelerator_intra_inter})                 & $\times$ & $\times$ & $\times$ & $\checkmark$  \\
Support Various Graphs  (\S\ref{sec:accelerator_network})                   & $\checkmark$ & $\times$ & $\times$ & $\checkmark$ \\
Hardware-Efficient Traversal Algorithm (\S\ref{sec:dst})         & $\times$ & $\times$ & $\times$ & $\checkmark$ \\
\midrule
Latency                                  & Moderate & High    & Low to Moderate & \textbf{Low} \\
Throughput                               & High & \textbf{Very High} & High    & High \\
Throughput / Bandwidth                               & High & High & High    & \textbf{High} \\
Energy Efficiency                        & Low      & Low & Moderate    & \textbf{High} \\
\bottomrule

\end{tabular}
\end{table*}
% \end{tcolorbox}
% \end{figure*}

Given the rising adoption of ANN search in online systems, an ideal GVS system should \textit{achieve low search latency for real-time query batches}, while being cost- and energy-efficient. 
% For example, in a RAG system, the LLM serving engine may perform on-demand retrievals in the middle of the generation process~\cite{borgeaud2022improving, khandelwal2019generalization, jiang2023active, jeong2024adaptive}. These retrievals typically involve small query batches or even individual queries because (a) the sequence batch sizes are constrained by accelerator memory capacity~\cite{yu2022orca, kwon2023efficient}, and (b) these sequences can trigger retrievals asynchronously due to their different generation contexts~\cite{jiang2023active, jeong2024adaptive, trivedi2022interleaving}.
% Consequently, high search latency not only prolongs the overall generation time but also leads to idleness of the inference accelerators such as GPUs and TPUs, which have to wait for search results before proceeding~\cite{jiang2025rago, jiang2024piperag, zhang2024accelerating}.
%
However, reducing GVS latency remains challenging due to limitations imposed by common hardware architectures (CPUs and GPUs) and the inherent difficulty of parallelizing graph traversals.
First, CPUs and GPUs operate on a time-multiplexed basis, executing GVS operations sequentially, thus leading to accumulated latency across traversal iterations. 
Second, if intra-query parallelization is employed, the synchronization overhead among CPU cores or GPU streaming multi-processors~\cite{lagrone2011set, zhang2020study} is disproportionately high relative to a single iteration of graph traversal, which typically takes only microseconds and involves just dozens of distance computations.
% 
% CPUs and GPUs operate on a time-multiplexed basis, executing GVS operations --- such as database vector fetching, distance computation, and result insertion --- sequentially, with only limited overlap between them, even if data prefetching is applied.
% % 
% Thus, given the classic BFS traversal algorithm~\cite{malkov2018efficient, fu2017fast}, query latency accumulates over multiple iterations as the search progresses through each operator.  
% 
% While improving throughput of queries per second (QPS) is straightforward by parallelizing execution across a large batch of queries, reducing search latency for a single query is significantly more challenging.
%
% This is because, when implementing intra-query parallelization, the synchronization overhead among CPU cores or GPU streaming multi-processors~\cite{lagrone2011set, zhang2020study} is disproportionately high relative to a single iteration of graph traversal, which typically takes only microseconds and involves just dozens of distance computations.

While previous research has explored hardware accelerator designs for GVS based on FPGA prototyping~\cite{zeng2023df, peng2021optimizing}, these approaches have three main limitations.
% They implemented Hierarchical Navigable Small World (HNSW), a popular GVS algorithm, on FPGAs~\cite{zeng2023df, peng2021optimizing}. 
% While showing performance improvements compared to software,
Firstly, they only support the Hierarchical Navigable Small World (HNSW) graph.
% which possesses a unique multi-layer structure uncommon in GVS algorithms.
While HNSW is widely used today, more efficient graph construction algorithms are emerging that offer improved recall~\cite{malkov2014approximate, malkov2018efficient, fu2017fast, zhao2023towards, zuo2023arkgraph, lu2021hvs, peng2023efficient}. 
% For example, the Navigating Spreading-out Graph (NSG)~\cite{fu2017fast}, with additional time invested in index construction, can achieve better recall than HNSW.
Secondly, directly implementing the software-oriented BFS algorithm on these accelerators results in sub-optimal search latency, because it significantly under-utilizes the accelerators, as we will further explain in conjunction with the hardware designs.
% given that only one type of hardware unit is used at any given time.
% because the greedy and synchronous nature of BFS restricts the amount of parallelizable workload and thus leads to significant accelerator under-utilization.  
Thirdly, existing architectures are mainly throughput-oriented and either do not support~\cite{peng2021optimizing} or suboptimally support intra-query parallelism for low-latency search~\cite{zeng2023df}.
% such that different hardware processing elements cannot collaboratively process the same graph. 

% thus limiting their versatility in serving other types of graphs that offer better performance or recall. 
% Despite the popularity of HNSW, recent research has shown more effective graph construction methods that improve search performance without relying on multiple graph layers.

\textit{To achieve low-latency GVS while supporting various graphs, we argue that both algorithm-level and hardware-level optimizations are essential.}
To this end, we propose a hardware-algorithm co-design solution including \textit{Falcon}, a specialized GVS accelerator, and \textit{delayed-synchronization traversal (DST)}, an accelerator-optimized graph traversal algorithm designed to simultaneously improve accelerator search performance and recall.
We summarize the advantages of Falcon and DST over existing solutions in Table~\ref{tab:overview_comparison}, and elaborate on the key features of our approach below.

% % In this paper, we first introduce Falcon, a hardware accelerator designed for online graph-based vector search. 

\textit{Falcon is an in-memory GVS accelerator with four key features.}
Firstly, Falcon involves fast distance computations and sorting units, and minimizes off-chip memory accesses by using an on-chip Bloom filter to track visited nodes. 
Secondly, Falcon supports both intra-query parallelism, utilizing all compute and memory resources to process a single query, and across-query parallelism, handling multiple queries through separate processing pipelines.
Thirdly, Falcon supports general GVS, allowing it to leverage emerging algorithms offering better recall and performance.
Finally, Falcon functions as a networked service with an integrated TCP/IP stack, thus reducing end-to-end service latency by bypassing the accelerator's host server from the communication path. 
%

% \textit{The design of the hardware-efficient delayed-synchronization traversal (DST) algorithm is motivated by two key observations.} 

\textit{Delayed-synchronization traversal (DST) relaxes the greedy graph traversal order to improve accelerator utilization.}
The design of the algorithm is motivated by two key observations.
First, from a system performance perspective, the synchronous and greedy nature of the software-oriented best-first search (BFS) limits the amount of parallelism the accelerator can exploit and thus leads to significant accelerator under-utilization. 
Second, from a traversal-pattern perspective, we found that relaxing the order of candidate evaluations does not compromise recall. 
Building on these observations and drawing inspiration from label-correcting algorithms for parallel shortest path computation on graphs~\cite{bertsekas1993simple, meyer2003delta}, DST relaxes synchronizations that enforce the greedy traversal order, thereby increasing the amount of parallel workloads that Falcon can handle.
Consequently, DST both reduces search latency by improving accelerator utilization and improves recall by allowing the exploration of search paths that the greedy BFS would otherwise overlook.

We prototype Falcon on FPGAs and evaluate it on various vector search benchmarks across different types of graphs. 
In combination with DST, Falcon achieves up to 4.3$\times$ and 19.5$\times$ lower online search latency and up to 8.0$\times$ and 26.9$\times$ better energy efficiency compared to CPU and GPU-based GVS systems, respectively.
Besides, the proposed DST algorithm outperforms the classic BFS by 1.7$\sim$2.9$\times$ in terms of latency on Falcon and simultaneously improves recall.% by up to 4.93\%.
% The excellent performance and efficiency pave the way for their future adoptions in production GVS systems. 

The paper makes the following \textbf{contributions}:
\begin{itemize}%[leftmargin=*]
    \item We identify the hardware primitives essential for efficient GVS, design Falcon, a specialized GVS accelerator, prototype it on FPGAs, and expose it as a networked service.
    \item We analyze the graph traversal patterns of best-first search and propose DST, an accelerator-optimized graph traversal algorithm that reduces GVS latency by relaxing traversal order.
    \item We evaluate Falcon and DST across diverse graphs and datasets, demonstrating their high performance and energy efficiency.
\end{itemize}

    % \item We propose DST, an accelerator-optimized GVS algorithm inspired by label correcting to minimize search latency.
    % \item We demonstrate the generalizability of both Falcon and DST across various types of graphs and datasets.
\section{Background and Motivation}
\label{sec:background}

In this section, we define the vector search problem~(\S\ref{background:vector_search}), introduce GVS algorithms~(\S\ref{sec:background_gvs}), discuss the limitations of existing processors for online GVS~(\S\ref{sec:background_limitataions}), and motivate the need for an algorithm-hardware co-design solution for low-latency GVS~(\S\ref{sec:background_motivation}).

% (TODO: summarize)

\subsection{Vector Search: Problem Definition}
\label{background:vector_search}
% \subsection{Vector Search: Problem Definition and Use Cases}

% Vector search has been widely used in large language model (LLM) serving systems~\cite{guu2020realm, lewis2020retrieval, borgeaud2022improving}, recommender systems~\cite{google_recommendation, suchal2010full}, and search engines~\cite{chen2021spann, xiong2020approximate, karpukhin2020dense}. 

% \textit{Problem definition.} 
A $k$ nearest neighbor (\textit{kNN}) search takes a $d$-dimensional query vector $q$ as input and retrieves the $k$ most similar vectors from a database $Y$ containing $d$-dimensional vectors, based on metrics such as L2 distances, dot product, or cosine similarity. 

Real-world vector search systems typically adopt \textit{approximate nearest neighbor (ANN) search} instead of exact kNN search to boost search performance (latency and throughput) by avoiding exhaustive scans of all database vectors.
% Due to the curse of dimensionality~\cite{beyer1999nearest}, an exact nearest neighbor search on high-dimensional vector datasets requires a costly brute-force scan over all database vectors. On the contrary, ANN search trades search quality for much higher search performance (latency and throughput) by various algorithms that we will discuss in \S\ref{sec:background_gvs}.
%
The quality of an ANN search is measured by the recall at $k$ ($R@k$). 
Let \( \mathit{NN}_k(q) \) be the set of true \( k \) nearest neighbors to a query \( q \) and \( \mathit{ANN}_k(q) \) be the set of \( k \) results returned by the ANN search, recall at $k$ measures the proportion of the true \( k \) nearest neighbors that are successfully retrieved by the ANN search: \( R@k = \frac{| ANN_k(q) \cap NN_k(q) |}{| NN_k(q) |} \).

\subsection{Graph-based Vector Search}
\label{sec:background_gvs}

Graph-based vector search (GVS) is among the most popular ANN search methods, renowned for its high search performance and quality~\cite{malkov2014approximate, malkov2018efficient, fu2017fast, zhao2023towards, zuo2023arkgraph, lu2021hvs, gao2023high}. 
It involves constructing a proximity graph \( G(V, E) \), where \( V \) represents the set of nodes, each is a database vector, and \( E \) represents the set of edges between nodes, with each edge indicating high similarity between the two connected nodes. 
Some notable examples of graph construction algorithms include HNSW\cite{malkov2018efficient}, NSG\cite{fu2017fast}, and DiskANN~\cite{jayaram2019diskann}.

% A well-designed graph construction algorithm should ensure that vectors in the graph are well-connected such that typically more than one path between arbitrary pairs of similar nodes.

\subsubsection{Best-First Search (BFS) for Query Processing.}
Once the graph is constructed, query vectors can traverse the graph to find their nearest neighbors.
While various graph construction algorithms exist~\cite{malkov2014approximate, malkov2018efficient, fu2017fast, zhao2023towards, zuo2023arkgraph, lu2021hvs}, the traversal on those constructed graphs all converges to the classic best-first search (BFS) algorithm. 

% Although there are many ways to construct a graph for vector search~\cite{malkov2014approximate, malkov2018efficient, fu2017fast, zhao2023towards, zuo2023arkgraph, lu2021hvs}, they all process ANN queries by applying the classic best-first search (BFS) algorithm. 
% As visualized the Figure~\ref{fig:graph_traversal}, BFS walks over the graph starting from a specific (typically fixed) entry node and tries to reach the nearest neighbors of the query greedily. 

\begin{algorithm}[t]
\caption{Best-First Search (BFS)}
% \caption{Best-First-Search (G, p, q, l, k)}
\label{algo:bfs}
\begin{algorithmic}[1]
\Require graph $G$, entry node $p$, query vector $q$, maximum result queue size $l$, number of results to return $k$ ($k \leq l$)
\Ensure $k$ approximate nearest neighbors of query $q$

\State $C \gets \{p\}, R \gets \{p\}, Visited \gets \{p\}$ 
\While{$C \neq \emptyset$ \textbf{and} $\Call{Min}{C.dist} \leq \Call{Max}{R.dist}$}
    \State $c \gets \Call{Extract-Min}{C}$ \Comment{pop the nearest candidate}
    \For{all neighbors $n$ of $c$}
        \If{$n \notin Visited$}
            \State $dist \gets  \Call{Compute-Dist}{q, n} $
            \State $Visited.\text{add}(n), C.\text{add}(n,dist), R.\text{add}(n,dist)$ 
        \EndIf
    \EndFor
    \State $R.\text{resize}(l)$ \Comment{keep only the closest $l$ elements}
\EndWhile
\State \textbf{return} $\Call{Sort}{R}[:k]$ \Comment{return the first $k$ elements}
\end{algorithmic}
\end{algorithm}

\textit{BFS traverses a graph by greedily evaluating the best candidate node in each search iteration.}
As illustrated in Algorithm~\ref{algo:bfs}, BFS begins by adding the typically fixed entry node \(p\) to the candidate queue \(C\), which stores nodes for potential exploration; the result queue \(R\), which holds the nearest neighbors found so far; and the visited set \(Visited\), which tracks nodes that have already been visited. It then searches on the graph iteratively as long as there is at least one candidate that is reasonably close to the query \(q\). Here, reasonably close means that the minimum distance from the candidates in \(C\) to \(q\) is less than the maximum distance of the nodes currently in \(R\). The algorithm then pops and evaluates the best candidate \(c\) by visiting all of its neighbors. Each neighbor that has not been visited is added to the visited set, the candidate queue, and the result queue, ensuring that no node is processed more than once. Following the exploration of neighbors, \(R\) is adjusted to maintain only the closest $l$ elements. 

The maximum size of the result queue \(l\) (\(k \leq l\)) controls the trade-off between search performance and quality. A larger \(l\) increases the threshold distance for considering a candidate, thereby expanding the number of candidate nodes evaluated during the search. Although visiting more nodes increases the likelihood of finding the nearest neighbors, it also leads to higher search latency.

\subsection{Limitations of Existing Processors for GVS}
\label{sec:background_limitataions}

% \textcolor{red}{TODO: challenges in CPU and GPU based search (detailed analysis of bottlenecks including bottlenecks @Hang Hu), and also existing FPGA-based work.}

% \textcolor{red}{TODO: add more evidence that optimizing search queries with small batches is important.}

Existing GVS systems have been mostly CPU-based, and recent research has explored their deployments on GPUs and FPGAs. 
All these systems adopt the classic BFS algorithm.
However, current solutions remain sub-optimal for latency-sensitive online GVS.
% However, as we will explain below, current solutions remain sub-optimal for online vector search involving small batches of queries.
% Specifically, while modern multi-core CPUs and GPUs both have some degree of parallelism, they struggle to fully leverage those parallelism due to the following reasons. 

\subsubsection{Search on CPU} 
CPUs have several limitations in online GVS systems.
Firstly, CPUs operate on a time-multiplexing basis, executing GVS operators such as fetching, computing, and insertion sequentially, with only limited timeline overlaps due to data prefetching. 
This sequential processing leads to cumulative search latency for each operator.
% This sequential processing leads to cumulative search latency for each operator, in contrast to Falcon's design as we will introduce in this paper.
Secondly, software implementations typically employ a byte array to track visited nodes for each query~\cite{malkov2018efficient, fu2017fast}, resulting in additional read and write operations per visited node.
Thirdly, CPUs struggle with random memory accesses to fetch vectors, which are typically less than 1 KB, and to update the visited arrays (one byte per read or write).

% \begin{figure}[t]
% 	% \vspace*{-5mm} % to shrink gap between figures
% 	% full width, can be adjusted
% 	\centering
%   \includegraphics[width=1.0\linewidth]{fig/graph_traversal.pdf}
%   \vspace{-1em}
%   \caption{An example of best-first search (BFS) on graphs.}
%   \vspace{-1em}
%   \label{fig:bfs}
% 	% \vspace*{-5mm} % to shrink gap between figures
% \end{figure}

% In a single-core CPU search setting, the search is typically limited by the compute speed. Figure~\ref{fig:roofline_analysis} shows the performance of vector search and the distance comparison, as can be seen in the Figure, the compute limited the amount of compute that can be done.  

% A natural thought to improve the performance is to leveraging multi-core CPUs to parallelize the compute, and there are two ways of parallelizing it, as depicted by Figure~\ref{fig:cpu_parallel_approaches}. 

% The first way is to stick to algorithm~\ref{algo:backtrack_search}, in which the computation of all neighbors of the most relevant candidate node is parallelized (\textbf{CPU-P1}). A second way is to pop multiple candidates a time and parallelize across different neighbors~\cite{} (\textbf{CPU-P2}). 

\subsubsection{High-throughput GVS on GPUs}
GPUs are known for their massive parallelism, featuring thousands of cores~\cite{choquette20213}. 
% GPUs are known for their massive parallelism, featuring thousands of cores grouped into many streaming multi-processors~\cite{choquette20213}. 
Thus, GPUs are well-suited for high-throughput GVS applications, as evidenced by recent studies~\cite{groh2022ggnn, zhao2020song}. 
However, GPUs exhibit two shortcomings for online GVS.
Firstly, GPUs show much higher GVS latency than CPUs as shown in our evaluation, because the limited amount of workload per search iteration makes it infeasible to effectively parallelize one query across multiple streaming multi-processors.
Secondly, the scale of graphs that GPUs can efficiently serve is constrained by memory capacity.
GPUs typically use high-bandwidth memory (HBM), which offers high bandwidth but several times less capacity compared to DDR memory given the same cost~\cite{hbmcost}. Although utilizing CPU-side memory is a potential option, search performance remains a concern: the throughput of fast CPU-GPU interconnects like the NVLink in NVIDIA Grace Hopper~\cite{gracehopper} is still an order of magnitude lower than that of GPU memory.

% Given that each query is executed on one thread block,
% This latency issue arises because of the higher memory access and instruction latency of GPU~\cite{} contributes to high search latency of each individual query, 

\begin{figure*}[t]
	% full width, can be adjusted
  \centering
  \includegraphics[width=1.0\linewidth]{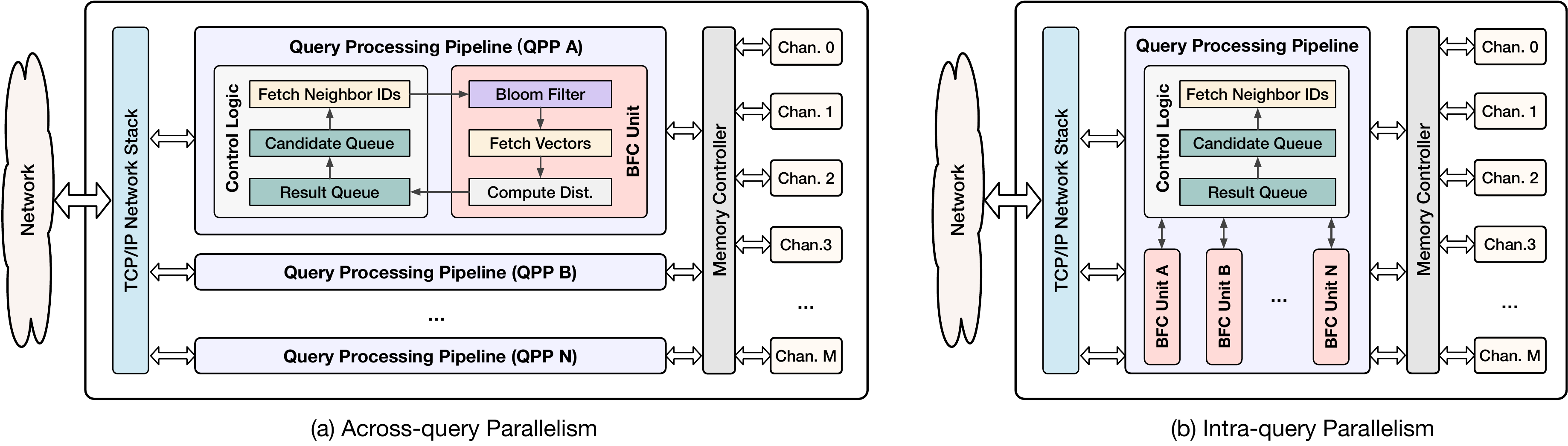}
  \vspace{-2em}
  \caption{Falcon overview. It has two architecture variants supporting across-query and intra-query parallelisms.}% given a single client.}
  \vspace{-1em}
  \label{fig:accelerator_overview}
\end{figure*}

\subsubsection{Specialized GVS Accelerators}
Two recent studies~\cite{zeng2023df, peng2021optimizing} implemented HNSW, a popular GVS algorithm, on FPGAs. Peng et al.\cite{peng2021optimizing} presented the first FPGA-based implementation, while Zeng et al.\cite{zeng2023df} further optimized the design by introducing data prefetching and enabling multi-FPGA search.

However, they are still not optimal for online GVS for the following reasons. 
Firstly, supporting only one type of graph (HNSW) may be inadequate given the rapid emergence of efficient GVS algorithms~\cite{malkov2014approximate, malkov2018efficient, fu2017fast, zhao2023towards, zuo2023arkgraph, lu2021hvs}. 
For example, NSG~\cite{fu2017fast}, given longer graph construction time, can achieve better performance-recall trade-offs than HNSW. 
Specializing the accelerator for HNSW~\cite{peng2021optimizing, zeng2023df} restricts the accelerator's flexibility in supporting various types of graphs: HNSW has a unique multi-level architecture, while the vast majority of graphs in GVS do not incorporate a leveled structure.
Secondly, applying the software-friendly BFS on the accelerators leads to sub-optimal search performance. This is because BFS can cause significant under-utilization of the accelerators, as we will specify in \S\ref{sec:dst}.
Thirdly, although Zeng et al.~\cite{zeng2023df} supports intra-query parallelism, an improvement over Peng et al.~\cite{peng2021optimizing}, the parallel strategy remains suboptimal. 
Specifically, the method of partitioning the graph into several sub-graphs and searching all sub-graphs in parallel~\cite{zeng2023df} leads to significantly more nodes being visited per query compared to traversing a single, larger graph, as we will explain further in \S\ref{sec:accelerator_intra_inter}.

% Finally, \citet{zeng2023df} implemented a simplified traversal algorithm, assuming a fixed number of iterations of traversal for each query, in contrast to the BFS algorithm described in Algorithm~\ref{algo:bfs}. Applying the same number of iterations per query contrasts the self-adaptive nature offered by BFS: according to our experiments on the SIFT1M dataset using HNSW, the number of iterations per using BFS query can range can range from AA to BB --- a CCx difference between queries. Using a fixed iteration however, the traversal assumes same number of iterations for harder queries and easier queries, and thus the recall drops from R@1=xx using BFS to R@1=yy using the fixed iteration. 
% \textcolor{red}{@Hang Hu add experiment here}.

\subsection{Motivation: Algorithm-Hardware Co-Design}
% \subsection{The Need of Algorithm-Hardware Co-Design for Efficient GVS}
\label{sec:background_motivation}

\textit{In this paper, we aim to achieve low-latency, energy-efficient, and general GVS.}
Given the insufficient hardware support (\S\ref{sec:background_limitataions}) and the inherent difficulty of parallelizing BFS (\S\ref{sec:background_gvs}), we argue that achieving this goal requires both algorithm-level and hardware-level optimizations.
In the following sections, we present a hardware-algorithm co-design solution, which includes \textit{Falcon} (\S\ref{sec:accelerator}), a hardware accelerator for GVS, and \textit{delayed-synchronization traversal (DST)} (\S\ref{sec:delayed_sync_traversal}), an accelerator-optimized graph traversal algorithm that simultaneously improves search performance and recall.
\section{Falcon for GVS Acceleration}
\label{sec:accelerator}

We present Falcon, a low-latency GVS accelerator that we prototype on FPGAs but also applicable to ASICs~(\S\ref{sec:accelerator_overview}).
% We present Falcon, a low-latency GVS accelerator~(\S\ref{sec:accelerator_overview}). Its FPGA prototype is ready-to-use for data centers~\cite{putnam2014reconfigurable, fowers2018configurable}, and its architecture is also applicable to ASICs.
Falcon consists of various high-performance hardware processing elements~(\S\ref{sec:accelerator_pe}).
It has two variants supporting across-query and intra-query parallelisms, optimized for processing batches of queries and individual queries, respectively~(\S\ref{sec:accelerator_intra_inter}). 
The accelerator is directly accessible as a networked service and supports various types of graphs~(\S\ref{sec:accelerator_network}).

\subsection{Design Overview}
\label{sec:accelerator_overview}

\textbf{Accelerator components.} Figure~\ref{fig:accelerator_overview} shows Falcon, a spatial dataflow accelerator for GVS. Each \textit{query processing pipeline (QPP)} handles one query at a time, containing both control logics and \textit{Bloom-fetch-compute (BFC) units}. Falcon is composed of various processing elements (PEs) interconnected via FIFOs, including systolic priority queues for storing candidate nodes and search results, Bloom filters to avoid revisiting nodes, and compute PEs for efficient distance calculations between query and database vectors.

% The interconnections and control flows of the PEs depends on which mode Falcon is operated with.
\textbf{Parallel modes.} Falcon has two variants that support \textit{across-query parallelism} and \textit{intra-query parallelism}, as shown in Figure~\ref{fig:accelerator_overview}(a) and (b), respectively. 
Across-query parallelism processes different queries across QPPs, while intra-query parallelism minimizes per-query latency by utilizing all compute and memory resources (multiple BFC units) to process one query at a time. 

\textbf{Differences compared to existing accelerators.} 
Falcon distinguishes itself from previous GVS accelerators~\cite{zeng2023df, peng2021optimizing} in four aspects, as summarized in Table~\ref{tab:overview_comparison}.
Firstly, Falcon utilizes on-chip Bloom filters to manage the list of visited nodes, thereby minimizing memory accesses~(\S\ref{sec:accelerator_pe}).
Secondly, Falcon's intra-query parallel design utilizes all compute and memory resources to traverse a single graph rather than partitioned sub-graphs~(\S\ref{sec:accelerator_intra_inter}). 
Thirdly, Falcon supports various GVS algorithms, rather than being limited to a specific one such as HNSW, allowing it to benefit from emerging algorithms that offer improved search quality and performance~(\S\ref{sec:accelerator_network}).
Finally, Falcon employs the proposed accelerator-optimized traversal algorithm that significantly reduces vector search latency~(\S\ref{sec:delayed_sync_traversal}).

\subsection{Hardware Processing Elements}
\label{sec:accelerator_pe}

We now introduce the main types of PEs in the order of their appearance in Algorithm ~\ref{algo:bfs}.

\subsubsection{Priority Queues}

% \begin{figure}[t]
% 	% \vspace*{-5mm} % to shrink gap between figures
% 	% full width, can be adjusted
% 	\centering
%   \includegraphics[width=1.0\linewidth]{fig/systolic-priority-queue.pdf}
%   \vspace{-1em}
%   \caption{A systolic priority queue with $s=8$ elements.}
%   \vspace{-1em}
%   \label{fig:priority-queue}
% 	% \vspace*{-5mm} % to shrink gap between figures
% \end{figure}

We adopt the systolic priority queue architecture~\cite{huang2014scalable, leiserson1979systolic} for the candidate and result queues in Algorithm~\ref{algo:bfs}. 
% As shown in Figure~\ref{fig:priority-queue}, a systolic priority queue is a register array of $s$ elements interconnected by $s-1$ compare-swap units. 
A systolic priority queue is a register array of $s$ elements interconnected by $s-1$ compare-swap units.
It enables high-throughput input ingestion of one insertion per two clock cycles by comparing and swapping neighboring elements in parallel in alternating odd and even cycles. The queue can be sorted in $s-1$ cycles. 
% \red{Potential Cut}

% Optionally add: run-time configurable sizes. 

\subsubsection{Bloom Filters}

Once the candidate queue pops a candidate to be explored, the next step is to check whether each of the candidate's neighbors is already visited. 

Previous software and specialized hardware implementations either maintain a visited array or a hash table, but neither is ideal for Falcon.
For example, software-based implementations~\cite{malkov2018efficient, fu2017fast} maintain an array with a length as the number of nodes in the graph. Node IDs are used as the array addresses to access the visited tags. However, this approach leads to extra memory accesses, requiring one read operation per check and one extra write operation to update the array for unvisited nodes.
% unless the graph is small enough that the array resides in the CPU cache completely.
Zeng et al.~\cite{zeng2023df} developed on-chip hash tables as part of the accelerators to track the visited nodes to avoid off-chip memory accesses. Each entry of the hash table stores up to four visited node IDs. However, given the limited on-chip SRAM, it is unlikely to instantiate large hash tables, and thus collisions would appear during the search. A collision would not only lead to redundant node visits, but those visited nodes will be inserted into the candidate and result queues repetitively, thus eventually degrading recall.  

Falcon, in contrast to existing solutions, adopts on-chip Bloom filters to track visited nodes. A Bloom filter is a space-efficient probabilistic data structure designed to test whether an element is a member of a set, e.g., determining whether a node has been visited based on its ID. 
A Bloom filter uses multiple ($h$) hash functions to map each input to several positions in a $b$-bit array. 
% As shown in Figure~\ref{fig:bloom}, a Bloom filter uses multiple ($h$) hash functions to map each input to several positions in a $b$-bit array. 
To check if a node has been visited, the same hash functions are used to check the status of these specific positions: if any of the bits are not set, the node is definitely not visited; if all are set, the node is highly likely visited (but not guaranteed, a scenario known as false positive). Given $m$ inserted elements, the false positive rates can be calculated by \( \left(1 - e^{-\frac{{hm}}{{b}}}\right)^h \)~\cite{bloom1970space}.

% \begin{figure}[t]
% 	% \vspace*{-5mm} % to shrink gap between figures
% 	% full width, can be adjusted
% 	\centering
%   \includegraphics[width=1.0\linewidth]{fig/Bloom_filter.pdf}
%   \vspace{-2em}
%   \caption{A Bloom filter for visited nodes filtering with $h=2$.}
%   \vspace{-1em}
%   \label{fig:bloom}
% 	% \vspace*{-5mm} % to shrink gap between figures
% \end{figure}

% numbers: 
% https://hur.st/bloomfilter/?n=1000&p=&m=1000&k=1
% https://hur.st/bloomfilter/?n=1000&p=&m=32000&k=3
\textit{Compared to hash tables, Bloom filters are significantly more space efficient for identifying visited nodes}. For example, instantiating a hash table with 1K slots for 4-byte node IDs requires 32Kbit SRAM. 
Using a chaining strategy to resolve hash collisions~\cite{mehta2004handbook}, where collided elements are moved to DRAM, the collision probability for a new incoming node ID is as high as 63.2\% when 1K nodes have already been visited.
In contrast, using the same amount of SRAM, a Bloom filter can provide 32K slots. With an equivalent number of nodes visited, the false positive rate for a new node ID is only 3.0\% and 0.07\% using a single hash function and three hash functions, respectively.
As we will show in our experiments, the very few false positives, meaning that an unvisited node is reported as visited, will not visibly degrade recall. 
This is because a well-constructed graph typically offers multiple paths from the query vector to the nearest neighbors, mitigating the effects of these very few false positives.

Falcon implements Bloom filters in the following manner. Both the number of hash functions and the size of the Bloom filters are configurable. Currently, Falcon uses three Murmur2 hashes~\cite{murmur} per filter. These hash functions are computed in parallel, and each hash function pipeline can yield a hash code every clock cycle. The size of the bitmap is set to 256Kbit, which translates to low false positive rates --- only one in 600K for 1K visited nodes.
% The size of the bitmap is calculated according to the maximal amount of visited node across example queries. We cap the false positive rate to less than 0.01\% after the evaluated maximal amount of insertion. 

% The bottleneck, though, is to lookup and update the bitmap, in which processing every node takes 2 cycles (read and write to the on-chip memory) multiplied by the number of hash functions. We empirically chose four hash functions, such that the cost of Bloom filter is less than half of a random memory access, otherwise Bloom filter cannot save much performance. 

\subsubsection{Fetching Vectors}
Upon identifying nodes to visit, the next step is reading the vectors for each node.

Falcon optimizes bandwidth utilization by pipelining vector fetches. Rather than waiting for the first vector to return before issuing a second read, each fetch unit pipelines up to 64 read requests (configurable), thus improving read throughput by hiding the latency associated with memory and the memory controller. The data width of the FIFO connecting a fetch unit to the memory controller is set to 64 bytes.

\subsubsection{Distance Computations} 
Each vector fetch unit is connected to a compute PE that calculates L2 distances, dot product, or cosine similarity between queries and database vectors. A compute PE instantiates multiple multipliers and adders and pipelines different compute stages, such that the compute throughput can match the maximum read throughput of a vector fetch unit.

\subsection{Intra-query and Across-query Parallelism}
\label{sec:accelerator_intra_inter}

While across-query parallelism for batched queries can be straightforwardly implemented by instantiating multiple query processing pipelines (QPP) on the accelerator, there are two design choices for intra-query parallelism, which aim to minimize latency for individual queries.
One option involves adopting the architecture of across-query parallelism by partitioning the dataset into multiple subsets, querying each subset with an individual QPP, and aggregating the results, as Zeng et al.~\cite{zeng2023df} described.

% \begin{figure}[t]
% 	% \vspace*{-5mm} % to shrink gap between figures
% 	% full width, can be adjusted
% 	\centering
%   \includegraphics[width=1.0\linewidth]{fig/subgraph_vs_full_graph_SIFT1M.pdf}
%   % \vspace{-2em}
%   \caption{Traversing one graph versus several sub-graphs.}
%   % \vspace{-1em}
%   \label{fig:subgraph}
% 	% \vspace*{-5mm} % to shrink gap between figures
% \end{figure}

\begin{figure}[t]
	% \vspace*{-5mm} % to shrink gap between figures
	% full width, can be adjusted
	\centering
  \includegraphics[width=0.7\linewidth]{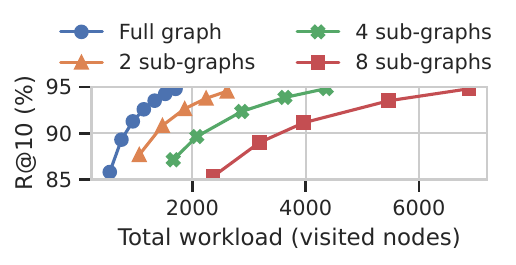}
  \vspace{-1.5em}
  \caption{Traversing one graph versus several sub-graphs.}
  \vspace{-1em}
  \label{fig:subgraph} 
	% \vspace*{-5mm} % to shrink gap between figures
\end{figure}

% python plot_subgraph_vs_full_graph.py --dataset SPACEV1M --min_recall 0.85 --max_recall 0.95 --compari
% son_recall 0.9
% Baseline: Total workload: 950.08
% Subgraph num: 2 Total workload: 1473.07 Total workload ratio: 1.55 (theoretical speedup: 1.29))
% Subgraph num: 4 Total workload: 2873.51 Total workload ratio: 3.02 (theoretical speedup: 1.32))
% Subgraph num: 8 Total workload: 3967.96 Total workload ratio: 4.18 (theoretical speedup: 1.92))
Alternatively, our choice is to \textit{speed up the traversal of a single graph} by instantiating multiple BFC units in a single QPP to utilize all the compute and memory resources for a single query (Figure~\ref{fig:accelerator_overview}(b)).
This decision stems from the observation that traversing several sub-graphs significantly increases the total amount of workload per query compared to traversing a single graph. 
% \begin{wrapfigure}{r}{0.55\linewidth}
% 	% full width, can be adjusted
%   \includegraphics[width=\linewidth]{fig/subgraph_vs_full_graph_SPACEV1M.pdf}
%   \caption{Traversing one graph versus several sub-graphs.}
%   \label{fig:subgraph} 
% \end{wrapfigure}
Figure~\ref{fig:subgraph} shows that, to achieve a recall of  $R@10=90\%$ on the SPACEV natural language embedding dataset~\cite{spacev}, the total number of visited nodes per query when using eight subgraphs is 4.2$\times$ of that for a single graph. Thus, the maximum speedup (assuming perfect load balancing) that eight partitions and eight QPPs can achieve is only 1.9$\times$ that of traversing a single graph with one QPP.

When traversing a single graph using intra-query parallelism, Falcon leverages its direct message-passing mechanism via FIFOs to enable low-overhead, fine-grained task dispatching among different BFC units.
This is a significant architectural advantage compared to CPUs and GPUs, where synchronization overhead among CPU cores or GPU streaming processors~\cite{lagrone2011set, zhang2020study} is too high compared to a single iteration of graph traversal, which only takes microseconds typically involving dozens of distance computations.

\begin{figure*}[t]
	% full width, can be adjusted
  \centering
  \includegraphics[width=1.0\linewidth]{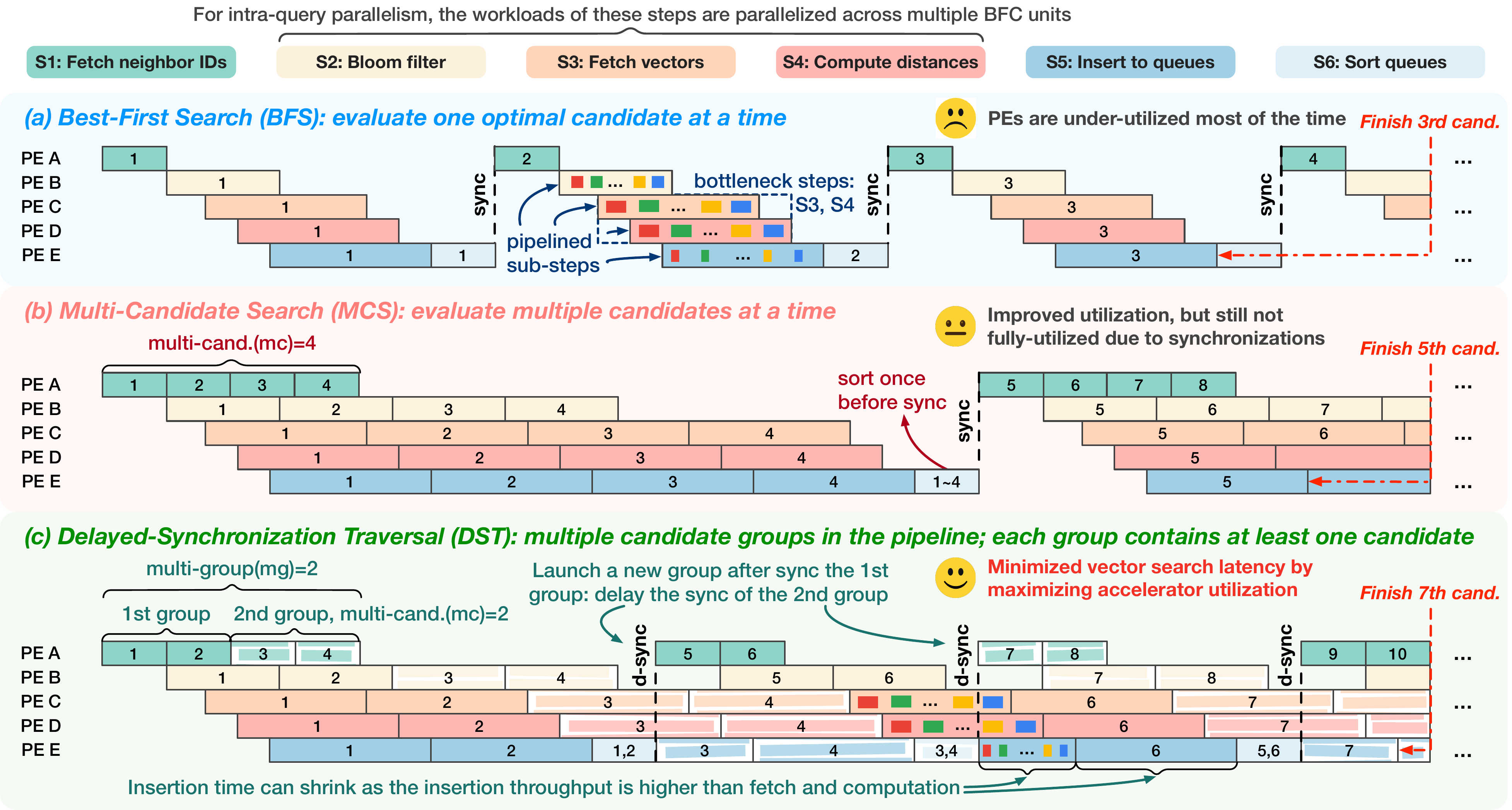}
  \vspace{-2em}
  \caption{The proposed Delayed-Synchronization Traversal (DST) reduces vector search latency by maximizing accelerator utilization. It delays synchronizations and allows multiple candidates to be evaluated simultaneously in the processing pipeline.}%
% Implementing the software-friendly BFS on Falcon results in significant under-utilization of the accelerator.
  \vspace{-1em}
  \label{fig:timeline}
\end{figure*}

\subsection{Accelerator-as-a-Service Implementation}
\label{sec:accelerator_network}

\hypertarget{fpga_backend}{}
Falcon is implemented with a total of 6.6K lines of code, including 3.6K lines of High-Level Synthesis (HLS) code for the accelerator kernel, developed using Vitis 2022.1, and 3K lines of C++ code for the CPU host and client programs.
We instantiate and evaluate Falcon on AMD Alveo U250 FPGAs, but the architecture is portable to arbitrary FPGA platforms.
Falcon operates as a networked service through a TCP/IP stack and supports various types of graphs.

\subsubsection{Network Stack Integration} 
Vector search systems are typically wrapped as services for real-time LLM serving or recommender systems. To minimize service latency, we integrate a TCP/IP network stack~\cite{100gbps} into Falcon, as shown in Figure~\ref{fig:accelerator_overview}. This integration allows Falcon to function as a networked accelerator service in data centers~\cite{putnam2014reconfigurable, fowers2018configurable}, facilitating direct communication with clients. This approach differs from common setups where the accelerator operates as a PCIe-based operator offloading engine, which involves additional latency including CPU handling requests from the network, accelerator kernel invocation, and data copying between the CPU and the accelerator. 
Compared to CPU and GPU-based services, Falcon can partially overlap communication and query latency: for a batch of queries, it begins processing the first query upon its arrival rather than waiting for the entire batch to be received.

\subsubsection{Supporting Various Graphs} 
Falcon supports arbitrary graphs by representing them with a unified graph format, accommodating common graph elements including nodes, edges, entry nodes, and degrees.
% regardless of which graph construction algorithm is used. 
This approach is naturally compatible with the vast majority of graphs~\cite{malkov2014approximate, fu2017fast, zhao2023towards, zuo2023arkgraph}, except for HNSW~\cite{malkov2018efficient} that has a unique multiple-layer structure. The upper layers of HNSW are designed to identify a high-quality entry point into the base layer, which contains all the database vectors --- thus the base layer is comparable to the entire graph in other GVS algorithms~\cite{malkov2014approximate, fu2017fast}. Instead of customizing the accelerator for this case, we prioritize the Falcon's versatility by initiating searches from a fixed entry point on the base layer of HNSW, which is the same (and only) entry point used at the top level. We found that this approach, without starting from the optimal entry node at the base layer for each query, does not lead to recall degrades (Figure~\ref{fig:dst}), although more hops might be necessary to reach the nearest neighbors, a finding also supported by existing research~\cite{wang2021comprehensive, lin2019graph}.

% A graph constructed by arbitrary GVS algorithms can be adapted to a unified data format for Falcon using a format conversion script. 
% The format conversion script requires several inputs, including the graph's metadata (which specifies the entry node, the maximum degree of the nodes, and the vector dimensionality), the raw vector data, and the edges connecting the nodes. 

\hypertarget{memory_management}{}
\subsubsection{Memory Management}
To leverage multiple memory channels, we partition the data --- including links and vectors --- evenly across all channels in a round-robin fashion based on node IDs. 
This contrasts with the approach of splitting a single vector across multiple channels, which provides limited bandwidth benefits when fetching a vector due to the random access latency incurred by each channel.
The number of BFC units is set equal to the number of memory channels, as each compute unit in a BFC is designed to fully utilize the maximum bandwidth of a single memory channel.

% \subsection{Implementation}

% We develop the Falcon accelerator using Vitis HLS 2022.1 in C/C++ and integrate an open-source FPGA TCP/IP network stack~\cite{100gbps} that connects to the accelerator kernel. 

\section{Delayed-Synchronization Traversal}
\label{sec:delayed_sync_traversal}

Realizing the inefficiencies of BFS on Falcon~(\S\ref{sec:bfs_inefficiency}), we investigate its graph traversal patterns~(\S\ref{sec:goal_improve_util}) and propose DST, an accelerator-optimized traversal algorithm applicable for both intra- and across-query parallelisms~(\S\ref{sec:dst}).

\subsection{Inefficiency of BFS on Accelerators}
\label{sec:bfs_inefficiency}

% \textbf{Inefficiency of BFS on accelerators.}

Figure~\ref{fig:timeline}(a) visualizes the timeline of BFS on Falcon, where each unique color represents one of the six search steps (S1$\sim$S6), and each PE handles a specific step, except for the priority queues that manage two steps, including distance insertions and sorting (S5 and S6).
Some steps must wait for the previous step to complete: sorting only begins after all distances are inserted into the queues. 
Other steps like filtering, fetching vectors, computing distances, and insertions can partially overlap because these PEs pipeline the execution of sub-steps, where each sub-step involves one of the neighbors of the candidate being evaluated. 
Between search iterations, an \textit{implicit synchronization} between all of the PEs ensures that the queues are sorted, such that the best candidate can be popped for evaluation in the next iteration.

Unfortunately, directly implementing the software-oriented BFS algorithm on a parallel GVS accelerator like Falcon can lead to sub-optimal search performance due to under-utilization of the accelerator.
As shown in Figure~\ref{fig:timeline}(a), only a fraction of the PEs are utilized simultaneously because of the inherently greedy nature of BFS, which processes only one candidate at a time, offering little opportunity for parallelization.

% An iteration of BFS begins by ejecting the top candidate, followed by fetching the candidate’s neighbors' IDs from memory, which involves a short sequential read as all neighbor IDs are stored consecutively. Subsequent steps include checking whether each node has been visited using a Bloom filter, fetching vectors of unvisited nodes, and computing distances. 
% The most time-consuming processes so far are vector fetching and distance computation, as each fetch operation requires a random memory access. Following these, the results are inserted into and sorted by the candidate and result queues. A synchronization occurs after the sorting to ensure that the next iteration can use the most optimal candidate so far.

\subsection{Goal: Improving Accelerator Performance through Traversal Algorithm Redesign}
\label{sec:goal_improve_util}

% mc: 1, mg: 1, total steps: 1472
% mc: 4, mg: 1, total steps: 1997
% mc: 1, mg: 4, total steps: 2047
% mc: 2, mg: 2, total steps: 1991
% \begin{figure}[h!]
% 	% \vspace*{-5mm} % to shrink gap between figures
%   \centering
%   \begin{subfigure}[b]{1.0\linewidth}
%     \includegraphics[width=1.0\linewidth]{fig/distances_over_steps_mc1_mg1.pdf}
%   \end{subfigure}
%   % To add some spacing between the horizontal aligned figures we'll use the \hfill command
%   % \hfill
%   \begin{subfigure}[b]{1.0\linewidth}
%     \includegraphics[width=1.0\linewidth]{fig/distances_over_steps_mc4_mg1.pdf}
%   \end{subfigure}
%   \begin{subfigure}[b]{1.0\linewidth}
%     \includegraphics[width=1.0\linewidth]{fig/distances_over_steps_mc2_mg2.pdf}
%   \end{subfigure}
%   \caption{The convergence of BFS, MCS, and DST.}
%   \label{fig:dist_trend}
%   % \vspace*{-5mm} % to shrink gap between figures
% \end{figure}

% 2684
% 3172
% 3202

A natural idea to optimize accelerator performance is to \textit{maximize accelerator utilization by minimizing PE  idleness}. 
Given the imbalanced workloads across different search steps, this approach does not necessitate all PEs to be always active but rather focuses on keeping those PEs involved in bottleneck steps consistently busy. 
In the context of GVS, the bottleneck steps usually include fetching neighbor vectors (S3) and calculating their distances relative to the query vectors (S4). 
% Unfortunately, even these critical PEs are significantly under-utilized as shown in Figure~\ref{fig:timeline}(a).

\begin{figure*}[t] 
	% full width, can be adjusted
  \centering
  \begin{subfigure}[b]{0.49\linewidth}
    \includegraphics[width=\linewidth]{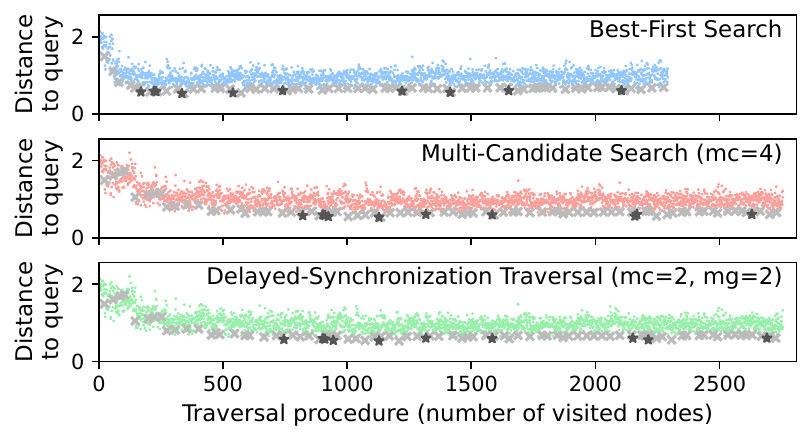}
  \vspace{-2em}    
    \caption{HNSW graph on Deep1M dataset}
  \label{fig:dist_trend_hnsw_deep}
  \end{subfigure}
  \begin{subfigure}[b]{0.49\linewidth}
    \includegraphics[width=\linewidth]{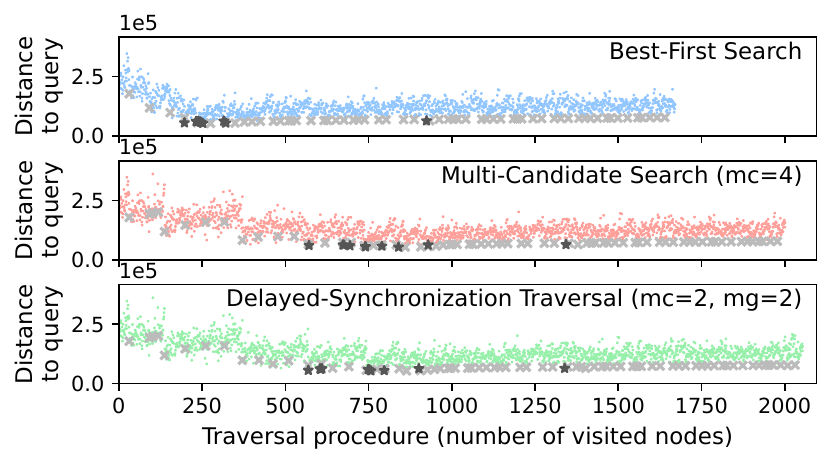}
  \vspace{-2em}
    \caption{NSG graph on SIFT1M dataset}
  \label{fig:dist_trend_nsg_sift}
  \end{subfigure}
  \vspace{-1em}
  \caption{Traversal convergence of BFS, MCS, and DST across graphs (HNSW and NSG) and datasets (Deep and SIFT). Each grey cross is an evaluated candidate, each colored dot a visited neighbor node, and each dark star one of the ten nearest neighbors.}
  \vspace{-1em}
  \label{fig:dist_trend}
\end{figure*}

\subsubsection{Algorithm-specific Observations.} Given the critical role of accelerator utilization in search performance, we ask: \textit{Is it necessary to strictly follow the BFS traversal order and synchronization pattern to achieve high search quality?}

To answer this question, we examine the traversal patterns of GVS. 
To address this question, we analyze the traversal patterns of GVS.
The upper section of Figure~\ref{fig:dist_trend} illustrates the BFS traversal process for a sample query on the Deep1M dataset~\cite{babenko2016efficient} using HNSW (Figure~\ref{fig:dist_trend_hnsw_deep}) and on the SIFT1M dataset~\cite{SIFT} using NSG (Figure~\ref{fig:dist_trend_nsg_sift}). 
Each grey cross represents an evaluated candidate node, colored dots denote its neighbor nodes, and black stars mark the ten nearest neighbors.
Notably, while the node distances to the query decrease at the beginning of the traversal, most subsequent candidates maintain similar distances rather than showing a monotonically decreasing trend --- an observation consistent across queries, datasets (SIFT, Deep, and SPACEV), and graphs (HNSW and NSG), although not all data are visualized due to space limits.

This observation suggests that \textbf{traversals in GVS do not have to adhere to a strictly greedy approach --- relaxing the traversal order of different candidate nodes should result in comparable search quality}, assuming the same or a similar set of candidates is evaluated.

\subsubsection{Naive Solution: MCS}
Leveraging the intuition above, one straightforward way to improve accelerator utilization is increasing the number of candidates evaluated per iteration, a strategy we term \textit{multi-candidate search (MCS)}. As illustrated in Figure~\ref{fig:timeline}(b), each iteration evaluates $\mathit{mc}=4$ candidates instead of just the closest one, because the second to the fourth best candidates per iteration may also be close to the query and could be on the search path of BFS.
% This approach results in higher utilization rates of the fetch and compute PEs compared to BFS (Figure~\ref{fig:timeline}(a)), due to the increased workload per iteration.

However, the PE utilization is not yet optimal due to the implicit synchronization operations required between iterations, where the candidate queue must be sorted before evaluating the next $\mathit{mc}$ nearest candidates.
While increasing$\mathit{mc}$ could push PE utilization rates towards 100\%, this approach can potentially degrade end-to-end search performance as we will show in the evaluation, because evaluating many candidates per iteration means potentially processing irrelevant candidates. 

\subsection{Low-latency GVS via DST}
\label{sec:dst}

To maximize accelerator utilization with minimal overhead (the number of extra nodes visited), we propose \textit{Delayed-Synchronization Traversal (DST)}, a parallel, low latency graph traversal algorithm for GVS. 
\textbf{The key idea of DST is to allow on-the-fly processing of multiple groups of candidates within the query processing pipeline by delaying synchronizations between search iterations, thereby improving the utilization of compute units and memory bandwidth. }
Here, each candidate group can contain one or multiple candidate nodes.
% as in BFS and MCS, respectively.

\subsubsection{DST Procedure.} Figure~\ref{fig:timeline}(c) demonstrates how DST enhances accelerator utilization. In this example, there are two candidate groups (\(mg=2\)), each with two candidates (\(mc=2\)), thus allowing four candidates to be processed simultaneously in the pipeline, mirroring the MCS setup (\(mc=4\)) in Figure~\ref{fig:timeline}(b). 
Unlike MCS, DST introduces \textit{delayed synchronization}: as the evaluation of the candidate group containing the 5th and 6th candidates begins, only the first group, containing the 1st and 2nd candidates, has been fully evaluated --- the delayed synchronization sorts the existing results, while the synchronization of the second group (with 3rd and 4th candidates) is deferred. This strategy ensures that the processing pipeline remains filled and that the bottleneck-step PEs for fetching vectors and computing distances are fully utilized, thereby avoiding the periods of idleness around synchronizations as shown in Figure~\ref{fig:timeline}(a) and (b). When applying DST to intra-query parallelism, steps S2$\sim$S4 can be parallelized across multiple BFC units, unlike across-query parallelism, which utilizes one BFC unit per QPP.

Algorithm~\ref{algo:dst} details the procedure of DST from the accelerator controller's perspective. DST starts by evaluating the entry node as the first candidate group. As soon as a candidate group is evaluated, DST tries to fill the accelerator pipeline by launching the evaluation of additional candidate groups, where both the number of groups in the pipeline (\(mg\)) and the number of candidates per group (\(mc\)) can be set by the user. DST terminates when there are no active groups in the pipeline and there are no more valid candidates.

\begin{algorithm}[t]
\caption{Delayed-Synchronization Traversal (DST)}
\label{algo:dst}
\begin{algorithmic}[1]
\Require graph $G$, entry node $p$, query vector $q$, result queue size $l$,  number of candidate groups $mg$, number of candidates per group $mc$, number of results $k$ ($k \leq l$)
\Ensure $k$ approximate nearest neighbors of query $q$

% \State \Comment{Initialize candidate, result, and visited queues}
\State $C \gets \{p\}, R \gets \{p\}, Visited \gets \{p\}$ 
\State \Call{Launch-Eval-Non-Block}{$\{p\}$}, $GroupCnt \gets 1$

\While{$GroupCnt > 0$ \textbf{or} $\Call{Min}{C.dist} \leq \Call{Max}{R.dist}$} \Comment{stop if no active groups and qualified candidates}

    \If{\Call{Earliest-Eval-Done}{}} \Comment{check task status}
        \State $GroupCnt \gets GroupCnt - 1$ % \Comment{finish one group}
        \While{$GroupCnt < mg$} \Comment{fill the pipeline}
            % \State \Comment{Get up to $mc$ candidates}
            \State $threshold \gets \Call{Max}{R.dist}$
            \State $Group \gets \Call{Extract-Min}{C, mc, threshold}$ 
            \If{$\Call{Size}{Group} > 0$}
                \State \Call{Launch-Eval-Non-Block}{$Group$}
                \State $GroupCnt \gets GroupCnt + 1$
            \EndIf
        \EndWhile
    \EndIf
\EndWhile
\State \textbf{return} $\Call{Sort}{R}[:k]$ \Comment{return the first $k$ elements}
\end{algorithmic}
\end{algorithm}

\subsubsection{Performance Benefits.} 
DST achieves significantly higher throughput than BFS and MCS in terms of the number of candidates processed per unit of time.
Figure~\ref{fig:timeline} marks the count of processed candidates by the end of the timeline on the right side.
In this example, BFS completes only three candidates, meaning that the results for the 3rd candidate have been inserted into the candidate queue. 
MCS shows improved throughput, managing to finish processing five candidates in the same time frame.
DST, given an equivalent number of candidates in the pipeline as MCS (four), achieves the highest throughput by completing seven candidates by the end of the timeline. 
Notably, DST fully utilizes the critical PEs for vector fetching and distance computations, thanks to the delayed-synchronization mechanism.

\subsubsection{Search Quality.} Given the algorithmic relaxations in DST compared to BFS, one might immediately question: \textit{Will the reordered traversal in DST degrade recall?} Contrary to this concern, DST can actually improve recall while lowering search latency as our experiments will demonstrate (Figure~\ref{fig:dst}) for the following reasons. 
On one hand, BFS traverses the graph in a greedy manner, striving to avoid visiting nodes that are not sufficiently close to the query. On the other, DST, by delaying synchronizations and allowing multiple candidates to be processed in the pipeline, relaxes the threshold for node evaluation. 
Considering that the termination condition remains consistent with BFS (when there is no qualified candidate left), DST likely evaluates the high-quality candidates on the search path of BFS and additionally explores other potentially relevant candidates. Thus, the evaluation of these extra sub-optimal candidates (a) does not prevent the evaluation of better candidates close to the queries and (b) may uncover extra paths leading to the nearest neighbors, thereby potentially improving recall.

Figure~\ref{fig:dist_trend} compares the search convergence of BFS, MCS, and DST. All of them find the nearest neighbors in this example, with DST and MCS visiting more nodes than BFS. This trend remains consistent across various datasets (SIFT, Deep, and SPACEV) and graph structures (HNSW and NSG), though Figure~\ref{fig:dist_trend} only visualized a subset of experiments due to space constraints.

\subsubsection{Parameter Configuration.} DST introduces two additional runtime configurable parameters compared to BFS: the number of candidate groups in the pipeline (\(mg\)) and candidates per group (\(mc\)). The optimal configuration depends on several factors, including vector dimensionalities, data distributions, and degrees (number of neighbors per node). 
We found it challenging to determine the optimal parameters by performance modeling due to (a) the significant variance in node degrees and (b) the unpredictable proportion of visited nodes as traversal progresses.
Thus, to ensure optimal search performance, it is advisable to perform an empirical parameter search using a set of sample queries before system deployment. Typically, this process only takes minutes, as the search space is relatively small, with both \(mg\) and \(mc\) usually not exceeding ten according to our experiments.

\section{Evaluation}
\label{sec:eval}

% ===== Speedup across all experiments =====
% Across-query latency speedup over CPU (Graph): median: 1.17~2.52x P95: 1.13~2.69x
% Intra-query latency speedup over CPU (Graph): median: 0.70~4.31x P95: 0.73~4.80x
% Across-query latency speedup over GPU (Graph): median: 6.12~17.44x P95: 9.26~38.39x
% Intra-query latency speedup over GPU (Graph): median: 3.66~19.49x P95: 5.85~62.20x
% Falcon (best of inter/intra-query) latency speedup over CPU (Graph): median: 1.17~4.31x P95: 1.13~4.80x
% Falcon (best of inter/intra-query) latency speedup over GPU (Graph): median: 6.12~19.49x P95: 9.26~62.20x
% Falcon (best of inter/intra-query) latency speedup over CPU (Faiss): median: 10.86~102.11x P95: 11.23~116.21x
% Falcon (best of inter/intra-query) latency speedup over GPU (Faiss): median: 1.99~6.48x P95: 1.82~8.94x
\begin{figure*}[t]
	% full width, can be adjusted
  \centering
  \begin{subfigure}[b]{1.0\linewidth}
    \includegraphics[width=\linewidth]{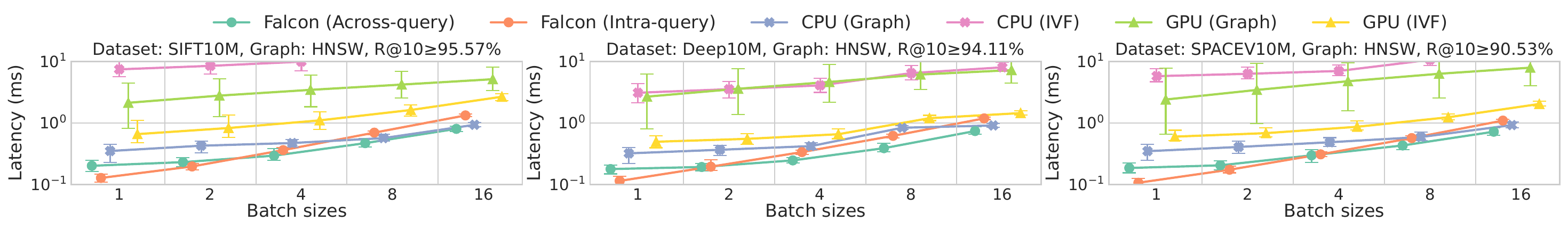}
    % optional to add distance before caption
    % \vspace{0.1cm}
    % \caption{Coffee.}
    % if its two vertical subfigures, can add distance between them
    % \vspace{0.5cm}
  \end{subfigure}
  \hfill
  \begin{subfigure}[b]{1.0\linewidth}
    \includegraphics[width=\linewidth]{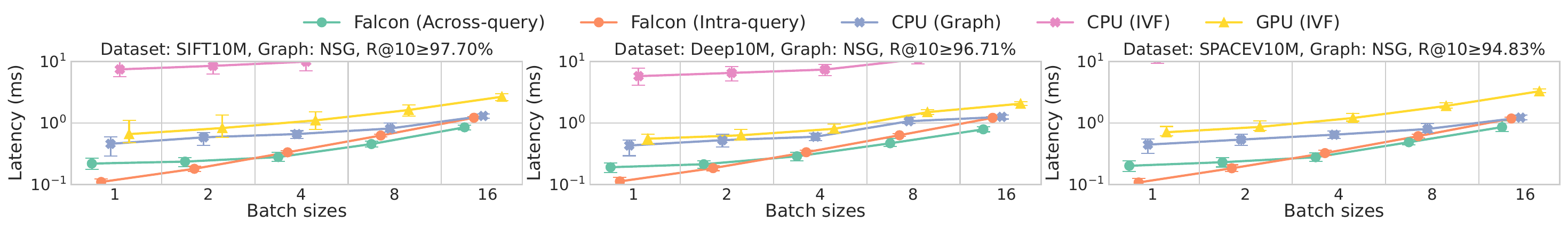}
    % optional to add distance before caption
    % \vspace{0.1cm}
    % \caption{Coffee.}
    % if its two vertical subfigures, can add distance between them
    % \vspace{0.5cm}
  \end{subfigure}
  
  \vspace{-1em}
  \caption{End-to-end GVS latency distribution of CPU, GPU, and Falcon across various graphs (rows) and datasets (columns). 
  The error bar shows the range within which 95\% of query latencies fall; CPU latency with IVF may surpass the $y$-axis limit. }
  \vspace{-1em}
  \label{fig:latency}
\end{figure*}

Our evaluation aims to answer the following questions:

\begin{itemize}[leftmargin=*]
    \item How does Falcon's search performance and energy efficiency compare to that of CPUs and GPUs? \S~\ref{sec:eval_e2e}
    % (esp. latency) / energy efficiency compare to CPU and GPU? 
    \item How much speedup and recall improvement can DST achieve on Falcon over BFS? \S~\ref{sec:eval_dst}
    \item Where is the performance cross-over point between intra-query and across-query parallelism? \S~\ref{sec:eval_inter_intra}
\end{itemize}

% \item How scalable is the proposed arch & algorithm? 
% How does the accelerator compared to other specialized accelerators without latency-oriented design? 

% FPGA throughput speedup over CPU: 1.09~1.46
% FPGA throughput speedup over GPU: 0.05~0.08
% FPGA throughput speedup over CPU Faiss: 32.32~101.99
% FPGA throughput speedup over GPU Faiss: 1.19~3.86
% Normalized FPGA throughput speedup over CPU: 3.39~4.54
% Normalized FPGA throughput speedup over GPU: 0.60~0.94
% Normalized FPGA throughput speedup over CPU Faiss: 100.74~317.90
% Normalized FPGA throughput speedup over GPU Faiss: 14.48~46.92
\begin{figure}[t]
	% full width, can be adjusted
  \centering
  \includegraphics[width=1.0\linewidth]{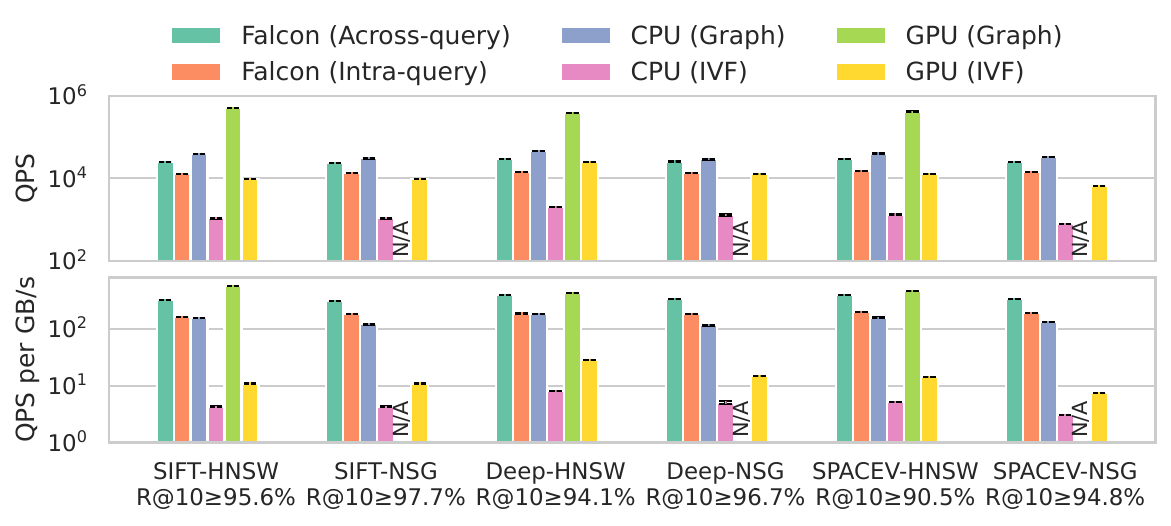}
  \vspace{-2em}
  \caption{GVS throughput across processors, without latency constraints, is strongly related to memory bandwidth.}% given a single client.}  
  % \caption{Throughput in queries-per-second (QPS) of different processors and indexes without latency constraints.}% given a single client.}
  \vspace{-1em}
  \label{fig:throughput}
\end{figure}

\subsection{Experimental Setup}
\label{sec:experiment_setup}

\textbf{Baseline systems.}
For CPUs, we evaluate two popular graphs, namely HNSW~\cite{malkov2014approximate} and NSG~\cite{fu2017fast}, using their official implementations.
For GPUs, we evaluate GGNN~\cite{groh2022ggnn}, an approximate version of HNSW optimized for GPU architectures.
Additionally, we evaluate the inverted-file (IVF) index~\cite{IVF}, a clustering-based index, using the Faiss library~\cite{faiss} for both CPUs and GPUs.
As the previous FPGA GVS implementations~\cite{zeng2023df, peng2021optimizing} are not open-sourced, we mainly compare their traversal strategies with DST based on Falcon in \S\ref{sec:eval_dst}.

\hypertarget{nic}{}
\textbf{Hardware.} 
We use server-class hardware manufactured in similar generations of technology (12$\sim$16 nm), where the CPU and GPU hold advantages over the FPGA in terms of bandwidth. 
We develop Falcon using Vitis HLS 2022.1, instantiate it on the AMD Alveo U250 FPGA (16 nm) with 64 GB of DDR4 memory (four channels x 16 GB, 77 GB/s in total), and set the accelerator frequency to 200 MHz.
We use a CPU server with 48 cores of Intel Xeon Platinum 8259CL operating at 2.5 GHz and 384 GB DDR4 memory (12 channels, 256 GB/s).
GPU evaluations are performed on NVIDIA V100 with 16 GB HBM2 memory (900 GB/s). 
Both the CPU and GPU servers are equipped with Mellanox ConnectX-4 NICs, providing network bandwidth equivalent to the FPGA at 100 Gbps.

% CPU experiments are conducted on a server with 16-core Intel Xeon E5-2630 v3 operating at 2.4 GHz, paired with 256 GB DDR4 memory (16 channels, 240 GB/s).
% GPU evaluations are performed on NVIDIA RTX 3090 with 24 GB GDDR6X memory (935 GB/s). 

\textbf{Datasets.} 
We use the SIFT~\cite{SIFT}, Deep~\cite{babenko2016efficient}, and SPACEV~\cite{spacev} datasets, containing 128, 96, and 100-dimensional vectors, respectively, thus covering both vision features (SIFT and Deep) and text embeddings (SPACEV).
We evaluate their subsets of the first ten million vectors, such that the constructed graphs can fit within the GPU and FPGA memory. 
% Each dataset contains 10K query vectors along with their exact nearest neighbors. 

\textbf{Algorithm settings.}
Unless specified otherwise, we set the maximum degree of the graphs to 64, balancing between graph size and search quality.
We set the candidate queue size as 64, which ensures at least 90\% recall for ten nearest neighbors across datasets. 
Falcon uses the best-performing DST parameters unless otherwise specified.
For IVF indexes, we set the number of IVF lists as 4096, approximately the square root of the number of vectors as a common practice.

% \textbf{Hardware.} 
% We instantiate Falcon on the AMD Alveo U250 FPGA (16 nm) that is equipped with 64 GB of DDR4 memory (four channels x 16 GB) and set the accelerator frequency to 200 MHz.
% For CPU-based solutions, we use a server with 16-core AMD EPYC 7313 processor (7 nm) that operates at a base frequency of 3.0 GHz and can turbo up to 3.7 GHz, accompanied by 128 GB memory (eight channels).
% For GPU experiments, we evaluate NVIDIA RTX 3090 (8nm) with 24 GB GDDR6X memory. 
% As we will show later, Falcon achieves superior performance compared to both CPUs and GPUs, despite being implemented on an FPGA with older technology.

\subsection{End-to-end Performance and Efficiency}
\label{sec:eval_e2e}

We compare Falcon with baseline systems on the six combinations between datasets and graphs.  
The software recall of these experiments is noted in Figure~\ref{fig:latency}: NSG consistently achieves better recall than HNSW. Falcon always achieves better recall than software because DST explores more search paths per query than BFS, as we will analyze in \S\ref{sec:eval_dst}.

\subsubsection{End-to-end Online Search Latency.} 
For online search, we treat all systems as a service where both the client and the server are connected to the same network switch. The network transmission time between CPU servers and between CPUs and FPGAs are similar --- around 50$\mu$s given a batch size of one, only a tiny fraction of the end-to-end query latency. 
Figure~\ref{fig:latency} shows the distributions of vector search latency for various batch sizes across six graph-dataset combinations.
We set the IVF-based index parameters for each scenario to achieve at least the same recall as GVS.

% ===== Speedup across all experiments =====
% Across-query latency speedup over CPU (Graph): median: 1.17~2.52x P95: 1.13~2.69x
% Intra-query latency speedup over CPU (Graph): median: 0.70~4.31x P95: 0.73~4.80x
% Across-query latency speedup over GPU (Graph): median: 6.12~17.44x P95: 9.26~38.39x
% Intra-query latency speedup over GPU (Graph): median: 3.66~19.49x P95: 5.85~62.20x
% Falcon (best of inter/intra-query) latency speedup over CPU (Graph): median: 1.17~4.31x P95: 1.13~4.80x
% Falcon (best of inter/intra-query) latency speedup over GPU (Graph): median: 6.12~19.49x P95: 9.26~62.20x
% Falcon (best of inter/intra-query) latency speedup over CPU (Faiss): median: 10.86~102.11x P95: 11.23~116.21x
% Falcon (best of inter/intra-query) latency speedup over GPU (Faiss): median: 1.99~6.48x P95: 1.82~8.94x
\textit{Falcon consistently outperforms all baselines in median latency, achieving speedups of up to 4.3$\times$ over CPU with graphs, 19.5$\times$ over GPU with graphs, 102.1$\times$ over CPU with IVF, and 6.5$\times$ over GPU with IVF.} 
Falcon achieves the lowest search latency among the compared systems, with its intra-query and across-query parallel modes preferable for different batch sizes as we will discuss in \S\ref{sec:eval_inter_intra}. 
% Falcon, with its specialized architecture strengthened by the latency-oriented DST algorithm, enables the lowest search latency among the compared systems, with its intra-query and across-query parallel modes preferable for different batch sizes as we will discuss in \S\ref{sec:eval_inter_intra}.
For CPUs, GVS outperforms the IVF index as the latter requires more database vectors to scan to achieve comparable recall~\cite{gao2023high, li2019approximate}. 
As batch sizes increase, CPU GVS latency becomes closer to that of Falcon, mainly benefiting from the CPU server's 3.3$\times$ higher bandwidth than the FPGA, whose bandwidth is saturated at a batch size of four. 
On GPUs, the embarrassingly parallel scan pattern of IVF results in better latency than GVS.
Despite their high bandwidth and numerous cores, GPUs struggle to efficiently handle queries with small batch sizes due to the GPU’s throughput-oriented architecture, which prioritizes parallel processing of many queries but results in high latency for individual queries. 

% Falcon's intra-query parallel mode delivers superior performance for smaller batch sizes (e.g., one), while its across-query parallel mode is better for larger batches (e.g., 16). 

\subsubsection{Throughput without Latency Constraints.} 
Figure~\ref{fig:throughput} presents search throughput in queries-per-second (QPS) without latency constraints by setting the batch size as 10K.%, where all 10K queries are sent at once. 
% Falcon adopts across-query parallelism here.

% FPGA throughput (inter):
%  {'SIFT-HNSW': 25062.028520588457, 'SIFT-NSG': 23604.947597016337, 'Deep-HNSW': 29807.414296232044, 'Deep-NSG': 25787.219338351526, 'SPACEV-HNSW': 30193.966037827, 'SPACEV-NSG': 25619.019560121436}
% FPGA throughput (intra):
%  {'SIFT-HNSW': 12547.964594663099, 'SIFT-NSG': 13780.900223526201, 'Deep-HNSW': 14281.042344718655, 'Deep-NSG': 13772.4542651225, 'SPACEV-HNSW': 15471.1347303768, 'SPACEV-NSG': 14328.76198063611}
% FPGA throughput speedup over CPU: 0.64~0.88
% FPGA throughput speedup over GPU: 0.05~0.08
% FPGA throughput speedup over CPU Faiss: 14.54~32.88
% FPGA throughput speedup over GPU Faiss: 1.19~3.86
% Normalized FPGA throughput speedup over CPU: 2.13~2.92
% Normalized FPGA throughput speedup over GPU: 0.58~0.90
% Normalized FPGA throughput speedup over CPU Faiss: 48.33~109.29
% Normalized FPGA throughput speedup over GPU Faiss: 13.92~45.13
\textit{Unsurprisingly, when latency constraints are removed, GVS throughput primarily becomes a competition of memory bandwidth.} 
For both CPUs and GPUs, graph-based indexes outperform IVF, which necessitates scanning more database vectors to reach the same recall~\cite{gao2023high, li2019approximate}.
For GVS, the GPU exhibits superior throughput thanks to its 12$\times$ memory bandwidth over the FPGA, as shown in the upper half of Figure~\ref{fig:throughput}.
Upon normalization by bandwidth (Figure~\ref{fig:throughput} lower), the performance of Falcon and GPUs becomes comparable, with GPUs showing a slight edge for SIFT. This is because the GPU adopts the greedy BFS algorithm, whereas Falcon uses DST that trades off additional nodes to visit for reduced latency, as we will analyze in \S\ref{sec:eval_dst}. 
The CPU performs the worst in QPS per unit bandwidth due to additional memory accesses required to check and update the visit status array.
% as well as less optimized memory accesses compared to the other platforms.
% The CPU performs the worst due to the additional memory accesses needed to check and update the visit status array, along with suboptimal memory access patterns compared to other platforms. 

% \begin{table}[t]
% \begin{small}
%   \begin{center}
%     \caption{\textcolor{red}{TODO: replace content}Average energy consumption per query (in mJ) on CPUs, GPUs, and Falcon using various batch sizes (1 and 16).}
%     % \vspace{-1em}
%     \label{tab:energy}
%     \begin{small}
%     \scalebox{0.8} {
%     \begin{tabular}{L{8em}     M{3em} M{0em} M{3em} M{0em}      M{2em} M{0em} M{2em} M{0em} M{2em} M{0em} M{2em}    }\\% M{2.8em}} % <-- Alignments: 1st column left, 2nd middle and 3rd right, with vertical lines in between
%       \toprule
%       & \multicolumn{3}{c}{CPU} & \phantom{}& \multicolumn{3}{c}{GPU} & \phantom{}& \multicolumn{3}{c}{Falcon} \\
%         \cmidrule{2-4} \cmidrule{6-8} \cmidrule{10-12}
%       & b=1 & \phantom{}& b=16 & \phantom{}&   b=1 & \phantom{}& b=16 & \phantom{}&   b=1 & \phantom{}& b=16  \\
%       \midrule
% SIFT10M-HNSW &  - && - && - && - && - && - \\
% SIFT10M-NSG &   - && - && - && - && - && - \\
% Deep10M-HNSW &   - && - && - && - && - && - \\
% Deep10M-NSG &  - && - && - && - && - && - \\
%       \bottomrule
%     \end{tabular}
%     } % scalebox
%     \end{small}
%   \end{center}
% \end{small}
% \end{table}

\begin{figure*}[t]
  \centering
  \begin{subfigure}[t]{0.32\linewidth}
    \includegraphics[width=\linewidth]{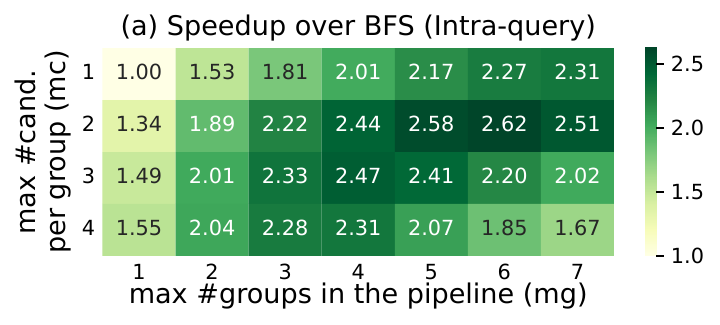}
    % \caption{DST speedup over BFS given intra-query parallelism}
    \label{fig:dst:intra}
  \end{subfigure}
  % \hfill
  \begin{subfigure}[t]{0.32\linewidth}
    \includegraphics[width=\linewidth]{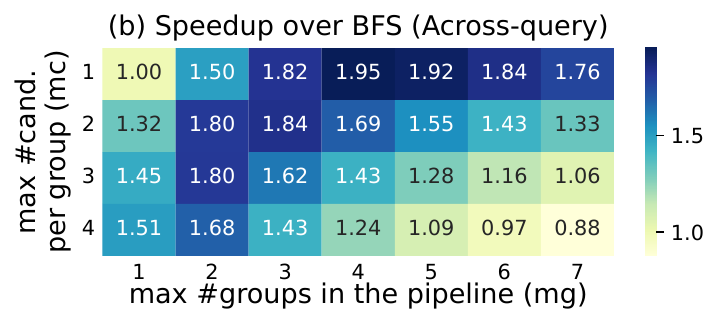}
    % \caption{DST speedup over BFS given across-query parallelism}
    \label{fig:dst:across}
  \end{subfigure}
  % \hfill
  \begin{subfigure}[t]{0.32\linewidth}
    \includegraphics[width=\linewidth]{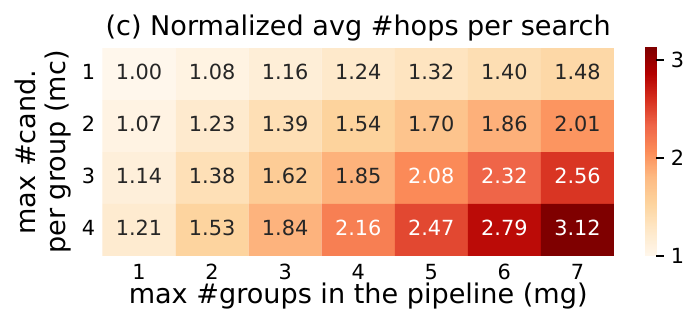}
    % \caption{Normalized number of evaluated nodes per query}
    \label{fig:dst:nodes}
  \end{subfigure}
    \vspace{-1em}
    
  \begin{subfigure}[t]{0.32\linewidth}
    \includegraphics[width=\linewidth]{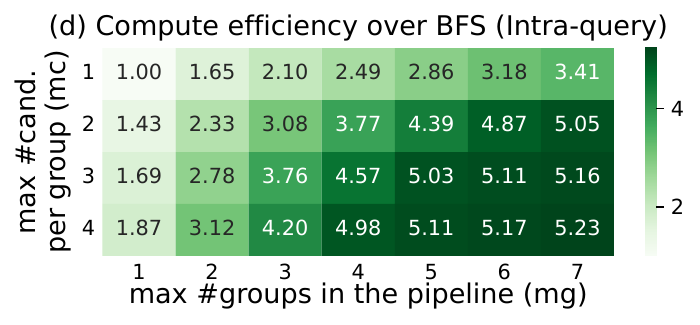}
    % \caption{DST speedup over BFS given intra-query parallelism}
    \label{fig:dst:intra}
  \end{subfigure}
  % \hfill
  \begin{subfigure}[t]{0.32\linewidth}
    \includegraphics[width=\linewidth]{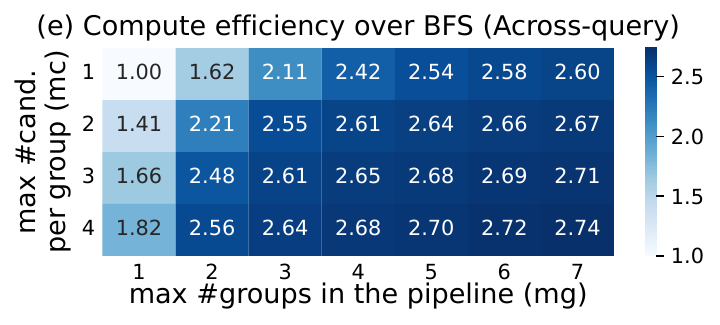}
    % \caption{DST speedup over BFS given across-query parallelism}
    \label{fig:dst:across}
  \end{subfigure}
  % \hfill
  \begin{subfigure}[t]{0.32\linewidth}
    \includegraphics[width=\linewidth]{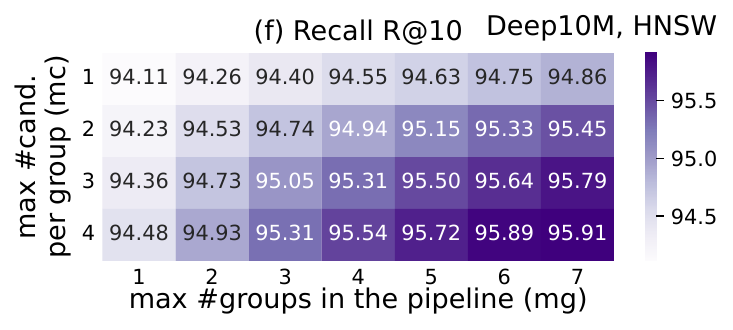}
    % \caption{Recall R@10}
    \label{fig:dst:recall}
  \end{subfigure}

  \vspace{-1.5em}
  \caption{The speedup, number of evaluated nodes, compute efficiency, and recall given various traversal configurations (HNSW on the Deep10M dataset). The x- and y-axes represent DST parameters~\(mg\) and \(mc\), where BFS corresponds to \(mg=1\) and \(mc=1\). }
  \vspace{-1em}

  % DST significantly outperforms BFS in terms of both search performance and quality. DST outperforms BFS in both search performance and quality.
  % 
  \label{fig:dst}
\end{figure*}

% ===== Energy efficiency across datasets =====
% Batch size: 1
% Efficiency over CPU: 4.98 ~ 8.03 x
% Efficiency over GPU: 17.52 ~ 26.87 x
% Efficiency over CPU Faiss: 59.87 ~ 231.13 x
% Efficiency over GPU Faiss: 3.09 ~ 5.46 x
% Batch size: 16
% Efficiency over CPU: 2.23 ~ 3.66 x
% Efficiency over GPU: 11.90 ~ 24.29 x
% Efficiency over CPU Faiss: 38.29 ~ 93.65 x
% Efficiency over GPU Faiss: 1.22 ~ 3.31 x
% Batch size: 10000
% Efficiency over CPU: 1.92 ~ 3.86 x
% Efficiency over GPU: 0.08 ~ 0.19 x
% Efficiency over CPU Faiss: 47.34 ~ 134.89 x
% Efficiency over GPU Faiss: 2.44 ~ 7.92 x
\subsubsection{Energy Efficiency.} 
We measure the power consumption (in Watt) of CPU, GPU, and Falcon using \textit{Intel RAPL}, \textit{NVIDIA System Management Interface}, and \textit{AMD's Vitis Analyzer}. 
The energy consumption per query batch (in Joule) is calculated by multiplying power with batch latency. 

\textit{Falcon is energy efficient, achieving up to 8.0$\times$, 26.9$\times$, 231.1$\times$, and 5.5$\times$ better energy efficiency than CPU graph, GPU graph, CPU IVF, and GPU IVF, respectively.} 
For online GVS with batch sizes up to 16, the power consumption of CPU, GPU, and Falcon ranges from 136.9$\sim$209.2W, 183.4$\sim$324.2W, and 55.2$\sim$62.3W, respectively. Considering energy consumption per batch, Falcon achieves 2.2$\sim$8.0$\times$ and 11.9$\sim$26.9$\times$ better energy efficiency than CPUs and GPUs. 
For offline GVS without latency constraints (using batch size of 10K), Falcon still achieves 1.9$\sim$3.9$\times$ energy efficiency over CPUs, but is outperformed by GPUs by 5.3$\sim$11.1$\times$, indicating that GPUs remain the preferred option for scenarios requiring high-throughput thanks to their superior memory bandwidth.

\begin{figure}[t]
	% full width, can be adjusted
  \centering
  \includegraphics[width=1.0\linewidth]{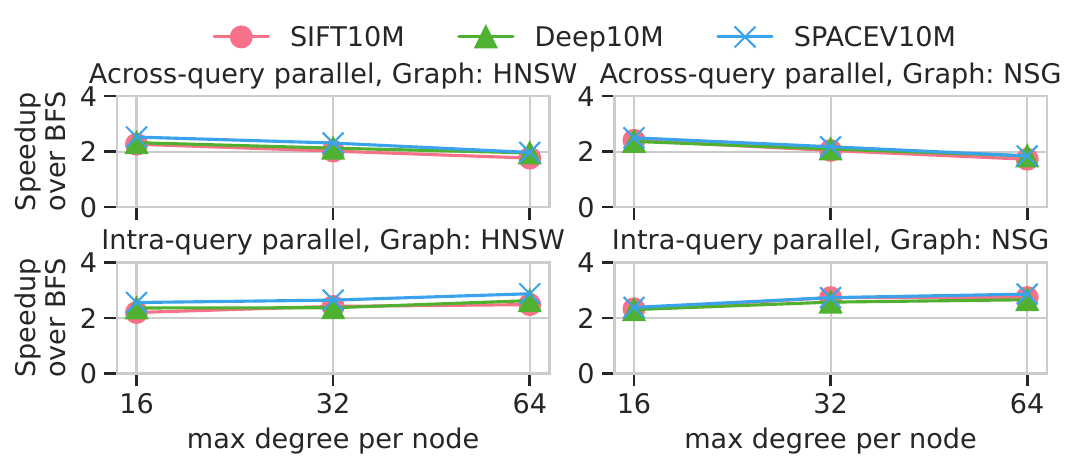}
  \vspace{-2em}
  \caption{DST consistently outperforms BFS across various datasets, graph configurations, and parallel modes.}% given a single client.}
  \vspace{-1em}
  \label{fig:dst_speedup_across_settings}
\end{figure}

% \begin{figure}[t]
% 	% full width, can be adjusted
%   \centering
%   \begin{subfigure}[t]{0.49\linewidth}
%     \includegraphics[width=\linewidth]
%     {fig/dst_speedup_across_settings_HNSW_inter_query.pdf}
%     % \caption{Intra-query speedup.}
%   \end{subfigure}
%   \hfill
%   \begin{subfigure}[t]{0.49\linewidth}
%     \includegraphics[width=\linewidth]
%     {fig/dst_speedup_across_settings_NSG_inter_query.pdf}
%     % \caption{Intra-query speedup.}
%   \end{subfigure}
%   \hfill
%   \begin{subfigure}[t]{0.49\linewidth}
%     \includegraphics[width=\linewidth]
%     {fig/dst_speedup_across_settings_HNSW_intra_query.pdf}
%     % \caption{Intra-query speedup.}
%   \end{subfigure}
%   \hfill
%   \begin{subfigure}[t]{0.49\linewidth}
%     \includegraphics[width=\linewidth]
%     {fig/dst_speedup_across_settings_NSG_intra_query.pdf}
%     % \caption{Intra-query speedup.}
%   \end{subfigure}
%   \hfill
  
%   \caption{DST consistently outperforms BFS across various datasets, graph configurations, and parallel modes.}% given a single client.}
%   % \vspace{-1.5em}
%   \label{fig:dst_speedup_across_settings}
% \end{figure}

\subsection{DST Efficiency on Accelerators}
\label{sec:eval_dst}

\subsubsection{Performance Benefits}
% DST achieves 1.7$\sim$2.9$\times$ speedup over BFS on Falcon. 
We now discuss the speedup achieved with different DST parameters and the maximum speedup across various experimental setups.

% purely intra-query parallel
\textbf{The impact of DST configurations on performance.} 
We evaluate the impact of the numbers of candidate groups in the pipeline (\(mg\)) and candidates per group (\(mc\)) on DST performance. 
Figure~\ref{fig:dst} (a) and (b) shows the throughput speedup achieved by DST compared to BFS on the Deep10M dataset with HNSW, across both the intra-query and across-query parallel versions of Falcon.
BFS is equivalent to \(mg=1, mc=1\) (upper-left corner), whereas MCS, evaluating multiple candidates per iteration without delayed synchronization, is shown in the first column \((mg=1, mc\geq 1)\). All the other setups are considered as DST.
Note that previous FPGA designs~\cite{peng2021optimizing, zeng2023df} adopts BFS, with Zeng et al.~\cite{zeng2023df} implementing a prefetching strategy on BFS that, at best (zero miss rate), matches the performance of MCS with \(mc=2\).

% \textit{DST significantly outperforms BFS and MCS on Falcon.}
% Benefiting from improved accelerator utilization through delayed synchronization, DST attains speedups of 2.62$\times$ and 1.84$\times$ BFS in intra-query and inter-query parallel modes, and 1.69$\times$ and 1.27$\times$ speedup over MCS.

\textit{The optimal configuration for DST varies across use cases, with intra-query parallelism typically requiring higher parameter values than across-query parallelism.} As shown Figure~\ref{fig:dst}, the optimal parameters are \(mg=6, mc=2\) for intra-query parallelism (Figure~\ref{fig:dst} (a)) and \(mg=4, mc=1\) for across-query parallelism (Figure~\ref{fig:dst} (b)). This is because the intra-query version parallelizes the distance computations, thus achieving a higher throughput of workload processing per query, leading to a higher throughput of processing nodes and thus necessitating a greater workload intensity to fully utilize the accelerator. However, higher \(mg=6\) and \(mc=2\) also lead to a greater amount of query-wise workloads as more hops are needed before the search terminates, as shown in Figure~\ref{fig:dst} (c). Thus, the maximum speedup is determined by the balance between accelerator utilization and the number of extra hops per query.
Figures~\ref{fig:dst}~(d) and~(e) show the compute efficiency—measured as the number of nodes evaluated per unit time—under different DST configurations. As \(mg\) and \(mc\) increase, compute efficiency improves but eventually plateaus. For instance, in Figure~\ref{fig:dst}~(d), setting \(mg = 5\) and \(mc = 3\) yields a 5.03\(\times\) improvement over BFS. However, further increasing parallelism to \(mg = 7\) and \(mc = 4\) results in only a marginal gain, reaching 5.23\(\times\) over BFS.

% NSG/HNSW x across/intra query, each with 3 different degress
\textbf{Maximum speedup in various experimental setups.} Figure~\ref{fig:dst_speedup_across_settings} shows the speedup of DST over BFS across various settings, including parallel modes, datasets, and graph types, and the maximum degrees of each graph. 
% Figure~\ref{fig:dst_speedup_across_settings} shows the speedup of DST over BFS across various settings, including parallel modes, datasets, graph types, and the maximum degrees of each graph. 

% The benefits are more significant when graph degree is lower and the vector dimensionality is smaller (SIFT). This reason is that lower node degrees and vector dimensionalities reduce the time required to fetch neighbors and compute distances when applying BFS, leading to significant accelerator under-utilization, as we explained in Figure~\ref{fig:timeline}.

\hypertarget{graph-degree}{}
\textit{DST consistently outperforms BFS across all setups, achieving speedups from 1.7$\sim$2.9$\times$.} 
DST is particularly advantageous in intra-query parallelism: with a maximum degree size of 64, it achieves speedups of 2.5$\sim$2.9$\times$ over BFS for intra-query parallelism, compared to 1.7$\sim$2.5$\times$ for across-query parallelism. This is because intra-query parallelism utilizes more BFC units for a single query and thus benefits more from increased workloads in the pipeline when adopting DST. 
% For across-query parallelism, DST shows better performance given lower graph degrees.
This reason is that lower node degrees reduce the time required to fetch neighbors and compute distances when applying BFS, leading to significant accelerator under-utilization, as we explained in Figure~\ref{fig:timeline}.
% For intra-query parallelism, this does not necessarily applies, because 
% \blue{Although not shown in Figure~\ref{fig:dst_speedup_across_settings}, we also evaluate the speedup of DST over BFS on graphs with lower maximum degrees of 32 and 16 (instead of the default high degree of 64).   
% For a maximum degree of 32, DST achieves a 2.37$\times$ speedup with intra-query parallelism and 2.12$\times$ with across-query parallelism over BFS. For a degree of 16, the speedups are 2.36$\times$ and 2.26$\times$, respectively.
% }

\subsubsection{Recall Benefits.} The rightmost heatmap in Figure~\ref{fig:dst} shows the improvements in search quality achieved by DST.

\textit{In general, larger numbers of candidates in the processing pipeline (higher \(mg\) and \(mc\)) lead to increased recall.} This is due to the evaluation of a broader range of candidates. Although some candidates may not be on the optimal search path, they could still lead to paths that reach the nearest neighbors.

\textit{DST consistently achieves better recall than BFS across all experiments.}
In Figure~\ref{fig:dst}(f), employing the performance-optimal DST configurations enhances R@10 from 94.11\% to 94.55\% and 95.33\% for across-query and intra-query parallelism, respectively. 
Given various experimental setups as in Figure~\ref{fig:dst_speedup_across_settings}, the R@10 improvements range from 0.14\% to 4.93\%.

\subsection{Across-query and Intra-query Parallelism}
\label{sec:eval_inter_intra}

\begin{figure}[t]
	% \vspace*{-5mm} % to shrink gap between figures
  \centering
  \begin{subfigure}[b]{0.49\linewidth}
    \includegraphics[width=\linewidth]{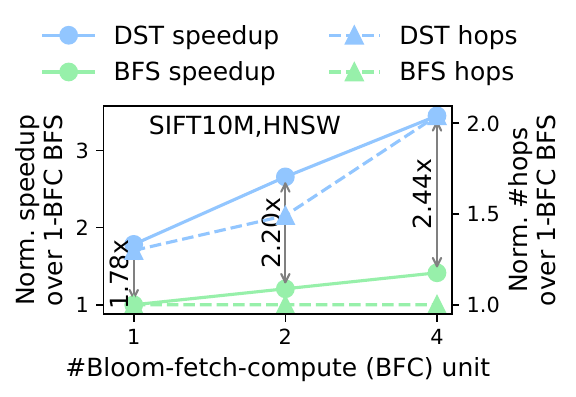}
  \vspace{-2em}
    \caption{HNSW on SIFT}
\label{fig:intra_query_scalability:hnsw_sift}
  \end{subfigure}
  \begin{subfigure}[b]{0.49\linewidth}
\includegraphics[width=\linewidth]{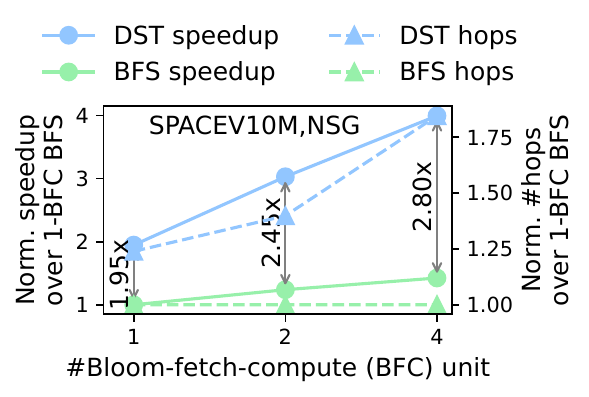}
  \vspace{-2em}
    \caption{NSG on SPACEV}
\label{fig:intra_query_scalability:nsg_spacev}
  \end{subfigure}
  \vspace{-1em}
  \caption{DST achieves significantly better performance scalability than BFS given intra-query parallelism.}
  \vspace{-1em}
  % \caption{The scalability of DST and BFS for intra-query parallelism across various numbers of BFC units.}
  \label{fig:intra_query_scalability}
  % \vspace*{-5mm} % to shrink gap between figures
\end{figure}

\subsubsection{Scalability of Intra-query Parallelism}
Figure~\ref{fig:intra_query_scalability} compares the scalability of DST and BFS given various numbers of Bloom-fetch-compute (BFC) units across datasets, with all units sharing a common control unit to form a query processing pipeline (QPP). For DST, we use \(mc\) and \(mg\) that achieve the highest performance.

\textit{DST demonstrates better performance scalability than BFS.}
For example, for HNSW on the SIFT dataset (Figure~\ref{fig:intra_query_scalability:hnsw_sift}), the speedup of DST over BFS increases from 1.78$\times$ to 2.44$\times$ as the number of BFC units grows from one to four. 
BFS, with four BFC units, achieves only a speedup of 1.41$\times$ over the single BFC version. This limited scalability of BFS stems from its greedy traversal pattern, which processes only one candidate at a time, resulting in minimal parallelizable workloads per iteration while the control overhead associated with the queues remains constant. 
In contrast, DST expands the workloads in the pipeline, ensuring that each BFC unit has sufficient workload to work with.
% The similar speedup trend of DST over BFS is consistent across datesets and graphs (e.g., Figure~\ref{fig:intra_query_scalability:nsg_spacev}), although not all results are elaborated here.
A similar speedup trend of DST over BFS is observed across various datasets and graphs (e.g., Figure~\ref{fig:intra_query_scalability:nsg_spacev}), although not all results are shown here.
% The similar speedup trend of DST over BFS can be found in Figure~\ref{fig:intra_query_scalability:hnsw_spacev}, \ref{fig:intra_query_scalability:nsg_sift}, and \ref{fig:intra_query_scalability:nsg_spacev}.

\subsubsection{Performance Trade-offs between Intra-query and Across-query Parallelism.}
Figure~\ref{fig:latency} compares the performance of the two types of parallelism, where each accelerator contains four BFC units, configured as either a single QPP for intra-query parallelism or four QPPs for across-query parallelism.

\textit{The optimal choice of parallel mode is related to batch sizes.} As shown in as shown in Figure~\ref{fig:latency}, intra-query parallelism is always advantageous for a query size of one. However, since the latency speedup from intra-query parallelism does not scale linearly with the number of BFC units (Figure~\ref{fig:intra_query_scalability}), across-query parallelism performs better for queries with batch sizes at least equal to the number of QPPs (four in our case). For batch sizes that fall between these two scenarios, the preferred parallel mode depends on the dataset, vector dimensionality, and graph construction parameters.

\subsection{Speedup in Recommender and RAG Systems}
\label{sec:dlrm_rag}

We also evaluate the end-to-end speedup achieved by deploying Falcon in recommender and RAG systems.  
\textit{Due to the wide range of possible model configurations, the proportion of time spent on retrieval can vary significantly (and consequently, the achievable speedup), as we demonstrate below.}

\begin{figure}[t]
	% \vspace*{-5mm} % to shrink gap between figures
  \centering
  \begin{subfigure}[b]{0.49\linewidth}
    \includegraphics[width=\linewidth]{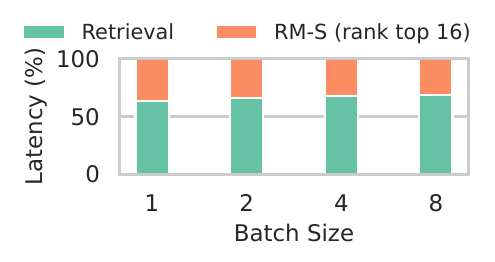}
    % \caption{HNSW on SIFT}
% \label{fig:intra_query_scalability:hnsw_sift}
  \end{subfigure}
  \begin{subfigure}[b]{0.49\linewidth}
    \includegraphics[width=\linewidth]{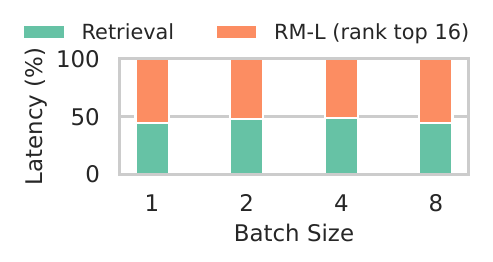}
    % \caption{HNSW on SIFT}
% \label{fig:intra_query_scalability:hnsw_spacev}
  \end{subfigure}  
  \vspace{-1.5em}
  
  \begin{subfigure}[b]{0.49\linewidth}  
    \includegraphics[width=\linewidth]{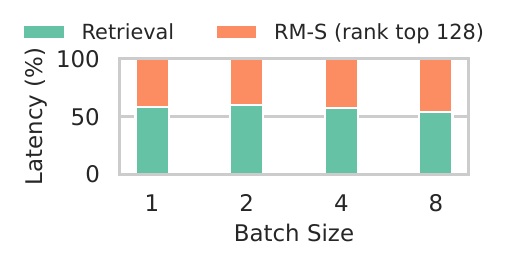}
    % \caption{HNSW on SIFT}
% \label{fig:intra_query_scalability:nsg_sift}
  \end{subfigure}
  \begin{subfigure}[b]{0.49\linewidth}
    \includegraphics[width=\linewidth]{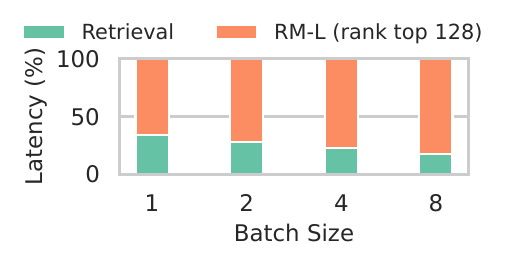}
    % \caption{HNSW on SIFT}
% \label{fig:dlrm:l_128}
  \end{subfigure}
  \vspace{-1em}
  \caption{Latency breakdown of end-to-end recommendation using a CPU for retrieval and a GPU for inference, across different model sizes and numbers of candidates to rank.}
  \vspace{-1em}
  % \caption{The scalability of DST and BFS for intra-query parallelism across various numbers of BFC units.}
  \label{fig:dlrm}
  % \vspace*{-5mm} % to shrink gap between figures
\end{figure}

\textit{Recommender Systems.}  
We instantiate two DLRM models of different sizes, as summarized in Table~\ref{tab:dlrm}. The smaller model (RM-S) is based on the Criteo TB dataset and includes 26 embedding tables, while the larger model (RM-L) comprises over 100 tables, aligning with recent industry-scale recommender systems~\cite{gupta2020deeprecsys, jiang2021fleetrec, jiang2021microrec}.
We assume that each recommendation request first performs an ANN search to identify candidate items, followed by ranking the top 16 or 128 candidates via model inference.
We adopt NVIDIA’s inference implementation~\cite{nvidia_dlrm} and evaluate it on a V100 GPU (same as the vector search baseline), assuming that the embedding tables fit in GPU memory.
For the ANN search, we use the latency on the SIFT dataset with HNSW as a reference. 
Figure~\ref{fig:dlrm} shows the latency breakdown when using a typical CPU-GPU architecture, where the CPU is responsible for ANN search. Depending on the model sizes and the number of candidates to rank per request, the percentage of time spent on 
retrieval can range from 17.56$\sim$68.92\%. Thus, by replacing the CPU with Falcon, the end-to-end speedup for recommender system ranges from 1.03$\sim$1.60$\times$.

\begin{table}[t]
\centering
% \small % or \footnotesize
\caption{Recommendation model configurations.}
\vspace{-1em}
\label{tab:dlrm}
% \begin{tcolorbox}[colframe=blue, colback=white, boxrule=0.5mm]
\scalebox{0.8}{
\begin{tabular}{l p{1.6cm} p{1.6cm} p{2cm} p{2.5cm}}
% \begin{tabular}{
%   >{\color{blue}}l
%   >{\color{blue}}p{1.6cm}
%   >{\color{blue}}p{1.6cm}
%   >{\color{blue}}p{2.cm}
%   >{\color{blue}}p{2.5cm}
% }
\toprule
% \textbf{Model} & \textbf{Embed.\\Tables} & \textbf{Embed.\\Dim} & \textbf{Bot. Layers} & \textbf{Top Layers} \\
\textbf{Model} & \textbf{Embedding Table Num} & \textbf{Embedding Dimension} & \textbf{Bot. Layers} & \textbf{Top Layers} \\
\midrule
\textbf{RM-S} & 26 & 64 & 256,128,64 & 512,512,256,128,1 \\
\textbf{RM-L} & 104 & 64 & 512,256,128,64 & 1024,1024,512,256,1 \\
\bottomrule
\end{tabular}
}
% \end{tcolorbox}
\end{table}

% ==== Model RM-S (rerank top 16) ====
% retrieval : inference (ms) = 0.36 : 0.21
% retrieval : inference (ms) = 0.43 : 0.22
% retrieval : inference (ms) = 0.47 : 0.23
% retrieval : inference (ms) = 0.56 : 0.25
% Retrieval (%): 63.30 - 68.92
% ==== Model RM-S (rerank top 16) ====
% E2E Speedup: 1.13 - 1.6
% ==== Model RM-S (rerank top 128) ====
% retrieval : inference (ms) = 0.36 : 0.25
% retrieval : inference (ms) = 0.43 : 0.29
% retrieval : inference (ms) = 0.47 : 0.35
% retrieval : inference (ms) = 0.56 : 0.48
% Retrieval (%): 53.72 - 60.03
% ==== Model RM-S (rerank top 128) ====
% E2E Speedup: 1.10 - 1.60 x
% ==== Model RM-L (rerank top 16) ====
% retrieval : inference (ms) = 0.36 : 0.45
% retrieval : inference (ms) = 0.43 : 0.46
% retrieval : inference (ms) = 0.47 : 0.50
% retrieval : inference (ms) = 0.56 : 0.70
% Retrieval (%): 44.38 - 48.66
% ==== Model RM-L (rerank top 16) ====
% E2E Speedup: 1.08 - 1.40 x
% ==== Model RM-L (rerank top 128) ====
% retrieval : inference (ms) = 0.36 : 0.70
% retrieval : inference (ms) = 0.43 : 1.13
% retrieval : inference (ms) = 0.47 : 1.59
% retrieval : inference (ms) = 0.56 : 2.63
% Retrieval (%): 17.56 - 33.67
% ==== Model RM-L (rerank top 128) ====
% E2E Speedup: 1.03 - 1.28 x

\begin{figure}[t]
	% \vspace*{-5mm} % to shrink gap between figures
  \centering
  \begin{subfigure}[b]{0.49\linewidth}
    \includegraphics[width=\linewidth]{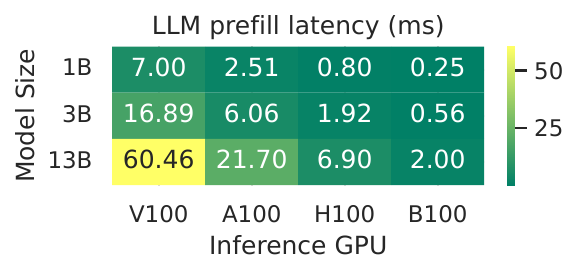}
    % \caption{HNSW on SIFT}
% \label{fig:intra_query_scalability:hnsw_spacev}
  \end{subfigure}  
  \begin{subfigure}[b]{0.49\linewidth}
    \includegraphics[width=\linewidth]{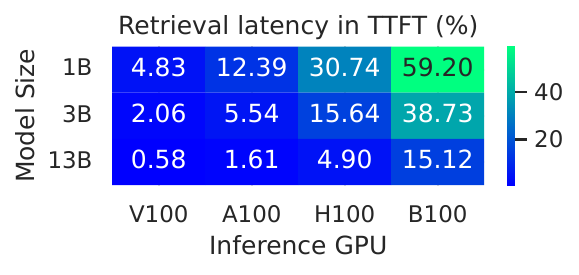}
    % \caption{HNSW on SIFT}
% \label{fig:intra_query_scalability:hnsw_sift}
  \end{subfigure}

  \vspace{-1em}
  \caption{RAG latency across inference and retrieval stages.
}
  \vspace{-2em}
  % \caption{The scalability of DST and BFS for intra-query parallelism across various numbers of BFC units.}
  \label{fig:rag}
  % \vspace*{-5mm} % to shrink gap between figures
\end{figure}

\textit{RAG Systems.} Similar to recommender systems, RAG pipelines can be built using various LLMs~\cite{jiang2025rago, jiang2024piperag}. We evaluate the performance of LLaMA models of varying sizes (ranging from 1B to 13B parameters) across different GPUs (from NVIDIA V100 to B100), using the Generative LLM Analyzer~\cite{bambhaniya2024demystifying}. As in the recommender setting, we use the latency of vector search on the SIFT dataset with HNSW as a reference for retrieval time.  
We assume the prompt, including both the query and retrieved documents, has a total length of 512 tokens~\cite{jiang2025rago}. Prompt computation is done with a batch size of one, as it already performs token-level batching inherently and can fully utilize the GPU without request-level batching~\cite{patel2023splitwise, zhong2024distserve}. 
The left side of Figure~\ref{fig:rag} shows the inference latency of prompt computation given a single GPU, while the right side illustrates the percentage of time-to-first-token (TTFT) latency spent on CPU-based retrieval. As GPU capability improves, inference latency decreases significantly. For example, for the 1B model, TTFT latency drops from 7\,ms on a V100 to just 0.25\,ms on a B100. As a result, the proportion of retrieval time in the overall TTFT latency increases—from 4.83\% to 59.20\%. Consequently, the end-to-end TTFT speedup by deploying Falcon depends on both model size and GPU backend, and can reach up to 1.60$\times$ given advanced GPUs.

% E2E Speedup (CPU vs FPGA):
%  [[1.03205153 1.08653411 1.24624165 1.61428819]
%  [1.01343001 1.03692014 1.11178101 1.33151551]
%  [1.00377217 1.01046901 1.03253969 1.1076376 ]]
% E2E Speedup: 1.00 - 1.61 x

\section{Discussion}
\label{sec:discussion}

% So far, we have introduced the design and implementation of Falcon and DST. We now discuss their potential future extensions. 

We have shown the performance advantages of Falcon and DST over CPUs and GPUs through FPGA-prototyping. 
We now discuss potential future extensions of the prototype to enable broader deployments, including adding more functionalities, supporting larger-scale searches, and achieving even higher efficiency.

\textbf{Handling insertions and updates.}
To support data insertions, deletions, or updates in Falcon, one could refer to the designs of software vector search systems. They typically manage a primary index for a dataset snapshot, an incremental (smaller) index for newly added vectors since the last snapshot, and a bitmap marking deleted vectors~\cite{adb-v}. These two indexes are merged periodically, e.g., daily, into a new primary index. 
Falcon can adopt this approach by focusing on serving the primary index, while the incremental index remains small enough to be efficiently managed by CPUs.

\textbf{Scale-out the system.}
We have not yet scaled out Falcon due to the limited number of FPGAs available.
However, we expect the scale-out design to be similar to software-based GVS systems~\cite{doshi2020lanns}. Specifically, the dataset is partitioned into subsets, each associated with a graph managed by a separate Falcon node. Queries are then directed to one or several of these partitions, with the results subsequently aggregated.

\textbf{Effectiveness on various memory and storage backends.}
Although Falcon is evaluated using an FPGA prototype with DDR memory, DST is applicable to a wide range of memory and storage backends. This is because its key idea --- improving the utilization of compute units and memory bandwidth --- is not tied to any specific hardware.
For instance, SSDs exhibit access latencies that are over an order of magnitude higher (e.g., 20µs~\cite{koh2018exploring}) than DRAM (e.g., <100ns~\cite{chang2016understanding}). 
Under the high-latency condition and given the BFS traversal strategy, compute units often remain idle while waiting for data, as illustrated in stages S1 and S3 of Figure~\ref{fig:timeline}.
By contrast, DST’s aggressive node exploration strategy enables on-the-fly processing of many nodes, thereby improving compute unit utilization and reducing overall search latency.

\textbf{Extensions for alternative hardware.}
The Falcon architecture is not specific to the FPGA platform used in our evaluation and can be applied to other hardware backends, including ASICs and various systems containing FPGAs.  
For example, Falcon can be implemented on FPGA-based data processing units (DPUs) acting as SmartNICs~\cite{intel_dpu, amd_dpu}. In such deployments, the FPGA may access not only its local device memory, but also host server memory and potentially remote memory, enabling larger-scale vector search.  
ASIC-based instantiation will not only offer higher energy efficiency, but also provide flexibility in memory technology choices. For instance, integrating HBM into an ASIC-based Falcon design can deliver both low latency and high throughput that is comparable to GPU-based systems.
% For FPGAs, Falcon can be instantiated not only on accelerator-oriented FPGAs~\cite{amd_v80}. 
% The remaining decision involves choosing between prioritizing memory capacity or bandwidth --- opting for DDR to serve larger graphs or HBM to process smaller datasets more rapidly.
% Based on the data ingestion speed measured for each BFC units and the total memory bandwidth, the number of PEs to be instantiated on the ASIC accelerator can then be calculated.

% \textbf{Extensions for ASICs.}
% Both Falcon's architecture and DST are applicable to ASICs. 
% The remaining decision involves choosing between prioritizing memory capacity or bandwidth --- opting for DDR to serve larger graphs or HBM to process smaller datasets more rapidly.
% Based on the data ingestion speed measured for each BFC units and the total memory bandwidth, the number of PEs to be instantiated on the ASIC accelerator can then be calculated.
% % There has been studies shown that for FPGAs with HBM, barely use all the channels due to the limited resources and cross bar limitation. 

% Can you merge intra-query and inter-query parallelism? Potentially yes. 

% Searching parameters for async search expensive? No, just a few minutes for each dataset. Considering the service should be deployed for a while, e.g., at least days to months, this is negligible initialization overhead. 
\section{Related Work}
\label{sec:related_work}

\textbf{ANN Search Algorithms.}  
Researchers have developed various ANN search algorithms~\cite{huang2021point, ouyang2020progressive, zhu2016range, zheng2016lazylsh, wang2015optimal, jiang2015exact, sun2014srs, dallachiesa2014top, yang2015reverse, lu2012efficient, gao2024rabitq} and vector data management systems~\cite{wang2021milvus, guo2022manu, adb-v, mohoney2023high, yang2020pase, pan2023survey, liu2025tigervector}.
Many variants of graph construction algorithms for GVS have also been proposed~\cite{malkov2014approximate, malkov2018efficient, fu2017fast, wu2014fast, zhao2023towards, zuo2023arkgraph, lu2021hvs, peng2023efficient, gao2023high, azizi2023elpis}, as GVS can achieve high recall with low latency.
Apart from GVS, other ANN search indexes offer different trade-offs between indexing cost and search performance. 
For example, locality-sensitive hashing (LSH) and inverted-file (IVF) indexes are indexing techniques that partition the vector space. 
Locality-sensitive hashing (LSH)~\cite{gionis1999similarity, datar2004locality} offers theoretical guarantees for ANN search, but empirically does not perform as well as graph-based algorithms. 
IVF indexes empirically outperform LSH, but still require scanning more database vectors than GVS to achieve the same recall~\cite{gao2023high, li2019approximate}. %, particularly in high-dimensional spaces.  
Beyond indexing, product quantization (PQ)~\cite{PQ, OPQ} is a widely adopted approach to compress high-dimensional vectors into compact byte codes. 
Often combined with IVF~\cite{PQ, OPQ} or graph indexes~\cite{gou2025symphonyqg, jayaram2019diskann}, PQ is particularly prevalent in large-scale ANN search, where reducing memory footprint is crucial, although this lossy compression technique can degrade recall beyond that caused by the indexes. 
% Thus, GVS is favored for high-recall ANN search on smaller datasets (e.g., millions of vectors)~\cite{wang2021comprehensive, li2019approximate}.

\textbf{Vector search on modern hardware.}
Beyond software optimizations~\cite{andre2016cache, peng2023iqan}, researchers have proposed various hardware-based solutions for vector search.
Exact kNN search can be accelerated by TPUs~\cite{chern2022tpu} or FPGAs~\cite{zeng2022faery, lu2020chip}.
For ANN search, Faiss is a popular GPU-accelerated library~\cite{johnson2019billion}, and there are several other implementations for PQ-based vector search~\cite{wieschollek2016efficient, chen2019robustiq, chen2019vector, liu2023juno} and GVS~\cite{groh2022ggnn, zhao2020song}. 
Lee et al.~\cite{lee2022anna} study ASIC designs for IVF-PQ, and several works~\cite{jiang2023chameleon, jiang2023co, zhang2018efficient} implement IVF-PQ on an FPGA, although their designs are constrained by either the limited HBM capacity or the speed of the CPU-FPGA interconnect. 
Several works propose to push down vector search to storage to improve performance by reducing data movements~\cite{wang2023storage, hu2022ice, xu2023proxima, kim2022accelerating, liang2022vstore, liang2019cognitive}.
Besides, the database vectors can be stored in non-volatile memory~\cite{ren2020hm} or CXL~\cite{jang2023cxl} to scale up GVS, while on-disk GVS must carefully manage I/O costs~\cite{chen2021spann, jayaram2019diskann, lejsek2008nv, wang2024starling}, similar to other graph processing workloads~\cite{vora2016load, roy2015chaos}.

% \textbf{Vector search on modern memory and storage.}
% % Apart from accelerator-based solutions, researchers also study modern storage for vector search. 
% Several works propose to push down vector search to storage to improve performance by reducing data movements~\cite{hu2022ice, xu2023proxima, wang2023storage, kim2022accelerating, liang2022vstore, liang2019cognitive}.
% Besides, Ren et al.~\cite{ren2020hm} suggest storing vectors in non-volatile memory to scale up GVS, while on-disk GVS must carefully manage I/O costs~\cite{chen2021spann, jayaram2019diskann, lejsek2008nv}, similar to other graph processing workloads~\cite{vora2016load, roy2015chaos}.
% Additionally, the emerging CXL technology has introduced another level of memory hierarchy as an option for ANN search~\cite{jang2023cxl}.

% \textbf{Hardware-software co-design.} 
% Maximizing system or accelerator performance often involves the co-design of hardware, software, and algorithms~\cite{tan2021aurora, liu2022overgen, li2023duet}. This design principle applies to various workloads, such as sparse matrix multiplications~\cite{hojabr2021spaghetti}, clustering~\cite{lee2017streaming, wang2021tiacc}, database engines~\cite{wu2014q100}, and deep learning training~\cite{li2018network}. 

\section{Conclusion}
\label{sec:conclusion}

% Very short version for submission only; camera-ready version is below
% To meet the surging demands of online GVS, we co-design algorithm and hardware by proposing Falcon and DST.
% They show up to 4.3$\times$ and 19.5$\times$ speedup over CPU and GPUs, showcasing great potential in future GVS deployments.

To meet the surging demands of online GVS, we propose Falcon, a high-performance GVS accelerator, and DST, an accelerator-optimized traversal algorithm.
Evaluated across various graphs and datasets, they shows up to 4.3$\times$ and 19.5$\times$ speedup in online search latency compared to CPUs and GPUs, while being up to 8.0$\times$ and 26.9$\times$ more energy efficient.
These compelling results show the potential for Falcon and DST to become the standard solutions for GVS acceleration.

% % camera-ready version conclusion
% We present Falcon, a high-performance GVS accelerator, as well as DST, an accelerator-optimized GVS algorithm that improves search latency and quality simultaneously by relaxing graph traversal orders. 
% Evaluated across various graphs and datasets, Falcon shows up to 4.3$\times$ and 19.5$\times$ speedup in online search latency compared to CPUs and GPUs, while being up to 8.0$\times$ and 26.9$\times$ more energy efficient.
% These compelling results show the potential for Falcon and DST to become the standard solutions for GVS acceleration.

% %%% MLSys version (10 pages)
% \input{mlsys_version/introduction}
% \input{mlsys_version/background_motivation}
% \input{mlsys_version/solution_accelerator}
% \input{mlsys_version/solution_algorithm}
% % \input{implementation}
% \input{mlsys_version/evaluation}
% % \input{discussion}
% % \input{related_work}
% \input{mlsys_version/conclusion}

% %%%% ASPLOS version (10.5 pages)
% \input{asplos_version/introduction}
% \input{asplos_version/background_motivation}
% \input{asplos_version/solution_accelerator}
% \input{asplos_version/solution_algorithm}
% % \input{asplos_version/implementation}
% \input{asplos_version/evaluation}
% \input{asplos_version/discussion}
% \input{asplos_version/related_work}
% \input{asplos_version/conclusion}

% \section*{Acknowledgements}

\begin{acks} 
We thank AMD for their generous donation of the Heterogeneous Accelerated Compute Clusters (HACC) at ETH Zurich (\url{https://systems.ethz.ch/research/data-processing-on-modern-hardware/hacc.html}), on which the experiments were conducted. 
% The HACC cluster is publically available to academic researchers: . 
\end{acks}

%-------------------------------------------------------------------------------

\balance
\bibliographystyle{ACM-Reference-Format}
\bibliography{ref}


\begin{thebibliography}{111}

%%% the Web address.

\ifx \showCODEN    \undefined \def \showCODEN     #1{\unskip}     \fi
\ifx \showDOI      \undefined \def \showDOI       #1{#1}\fi
\ifx \showISBNx    \undefined \def \showISBNx     #1{\unskip}     \fi
\ifx \showISBNxiii \undefined \def \showISBNxiii  #1{\unskip}     \fi
\ifx \showISSN     \undefined \def \showISSN      #1{\unskip}     \fi
\ifx \showLCCN     \undefined \def \showLCCN      #1{\unskip}     \fi
\ifx \shownote     \undefined \def \shownote      #1{#1}          \fi
\ifx \showarticletitle \undefined \def \showarticletitle #1{#1}   \fi
\ifx \showURL      \undefined \def \showURL       {\relax}        \fi
% invisible to TeX
\providecommand\bibfield[2]{#2}
\providecommand\bibinfo[2]{#2}
\providecommand\natexlab[1]{#1}
\providecommand\showeprint[2][]{arXiv:#2}

\bibitem[\protect\citeauthoryear{??}{amd}{[n.d.]}]%
        {amd_dpu}
 \bibinfo{year}{[n.d.]}\natexlab{}.
\newblock \bibinfo{title}{AMD Alveo SN1000 SmartNIC Accelerator Card}.
\newblock \bibinfo{howpublished}{\url{ https://www.amd.com/en/products/accelerators/alveo/sn1000/a-sn1022-p4.html}}.
\newblock


\bibitem[\protect\citeauthoryear{??}{fai}{[n.d.]}]%
        {faiss}
 \bibinfo{year}{[n.d.]}\natexlab{}.
\newblock \bibinfo{title}{Faiss}.
\newblock \bibinfo{howpublished}{\url{https://github.com/facebookresearch/faiss /}}.
\newblock


\bibitem[\protect\citeauthoryear{??}{int}{[n.d.]}]%
        {intel_dpu}
 \bibinfo{year}{[n.d.]}\natexlab{}.
\newblock \bibinfo{title}{Intel FPGA SmartNIC N6000-PL Platform}.
\newblock \bibinfo{howpublished}{\url{ https://www.intel.com/content/www/us/en/products/details/fpga/platforms/smartnic/n6000-pl-platform.html}}.
\newblock


\bibitem[\protect\citeauthoryear{??}{hbm}{[n.d.]}]%
        {hbmcost}
 \bibinfo{year}{[n.d.]}\natexlab{}.
\newblock \bibinfo{title}{The Memory Wall: Past, Present, and Future of DRAM}.
\newblock \bibinfo{howpublished}{\url{ https://semianalysis.com/2024/09/03/the-memory-wall/ }}.
\newblock


\bibitem[\protect\citeauthoryear{??}{mur}{[n.d.]}]%
        {murmur}
 \bibinfo{year}{[n.d.]}\natexlab{}.
\newblock \bibinfo{title}{The MurmurHash family}.
\newblock \bibinfo{howpublished}{\url{ https://github.com/aappleby/smhasher }}.
\newblock


\bibitem[\protect\citeauthoryear{??}{nvi}{[n.d.]}]%
        {nvidia_dlrm}
 \bibinfo{year}{[n.d.]}\natexlab{}.
\newblock \bibinfo{title}{NVIDIA Deep Learning Recommender Model Implementation}.
\newblock \bibinfo{howpublished}{\url{ https://github.com/NVIDIA/DeepLearningExamples/tree/master/PyTorch/Recommendation/DLRM}}.
\newblock


\bibitem[\protect\citeauthoryear{??}{gra}{[n.d.]}]%
        {gracehopper}
 \bibinfo{year}{[n.d.]}\natexlab{}.
\newblock \bibinfo{title}{The NVIDIA GH200 Grace Hopper Superchip}.
\newblock \bibinfo{howpublished}{\url{ https://www.nvidia.com/en-us/data-center/grace-hopper-superchip }}.
\newblock


\bibitem[\protect\citeauthoryear{??}{SIF}{[n.d.]}]%
        {SIFT}
 \bibinfo{year}{[n.d.]}\natexlab{}.
\newblock \bibinfo{title}{SIFT ANNS dataset}.
\newblock
\newblock
\urldef\tempurl%
\url{http://corpus-texmex.irisa.fr/}
\showURL{%
\tempurl}


\bibitem[\protect\citeauthoryear{??}{spa}{[n.d.]}]%
        {spacev}
 \bibinfo{year}{[n.d.]}\natexlab{}.
\newblock \bibinfo{title}{The SPACEV Web Embedding Dataset}.
\newblock \bibinfo{howpublished}{\url{ https://github.com/microsoft/SPTAG/tree/main/datasets/SPACEV1B}}.
\newblock


\bibitem[\protect\citeauthoryear{Andr{\'e}, Kermarrec, and Le~Scouarnec}{Andr{\'e} et~al\mbox{.}}{2016}]%
        {andre2016cache}
\bibfield{author}{\bibinfo{person}{Fabien Andr{\'e}}, \bibinfo{person}{Anne-Marie Kermarrec}, {and} \bibinfo{person}{Nicolas Le~Scouarnec}.} \bibinfo{year}{2016}\natexlab{}.
\newblock \showarticletitle{Cache locality is not enough: High-performance nearest neighbor search with product quantization fast scan}. In \bibinfo{booktitle}{\emph{42nd International Conference on Very Large Data Bases}}, Vol.~\bibinfo{volume}{9}. \bibinfo{pages}{12}.
\newblock


\bibitem[\protect\citeauthoryear{Azizi, Echihabi, and Palpanas}{Azizi et~al\mbox{.}}{2023}]%
        {azizi2023elpis}
\bibfield{author}{\bibinfo{person}{Ilias Azizi}, \bibinfo{person}{Karima Echihabi}, {and} \bibinfo{person}{Themis Palpanas}.} \bibinfo{year}{2023}\natexlab{}.
\newblock \showarticletitle{Elpis: Graph-based similarity search for scalable data science}.
\newblock \bibinfo{journal}{\emph{Proceedings of the VLDB Endowment}} \bibinfo{volume}{16}, \bibinfo{number}{6} (\bibinfo{year}{2023}), \bibinfo{pages}{1548--1559}.
\newblock


\bibitem[\protect\citeauthoryear{Babenko and Lempitsky}{Babenko and Lempitsky}{2016}]%
        {babenko2016efficient}
\bibfield{author}{\bibinfo{person}{Artem Babenko} {and} \bibinfo{person}{Victor Lempitsky}.} \bibinfo{year}{2016}\natexlab{}.
\newblock \showarticletitle{Efficient indexing of billion-scale datasets of deep descriptors}. In \bibinfo{booktitle}{\emph{Proceedings of the IEEE Conference on Computer Vision and Pattern Recognition}}. \bibinfo{pages}{2055--2063}.
\newblock


\bibitem[\protect\citeauthoryear{Bambhaniya, Raj, Jeong, Kundu, Srinivasan, Elavazhagan, Kumar, and Krishna}{Bambhaniya et~al\mbox{.}}{2024}]%
        {bambhaniya2024demystifying}
\bibfield{author}{\bibinfo{person}{Abhimanyu Bambhaniya}, \bibinfo{person}{Ritik Raj}, \bibinfo{person}{Geonhwa Jeong}, \bibinfo{person}{Souvik Kundu}, \bibinfo{person}{Sudarshan Srinivasan}, \bibinfo{person}{Midhilesh Elavazhagan}, \bibinfo{person}{Madhu Kumar}, {and} \bibinfo{person}{Tushar Krishna}.} \bibinfo{year}{2024}\natexlab{}.
\newblock \bibinfo{title}{Demystifying Platform Requirements for Diverse LLM Inference Use Cases}.
\newblock
\newblock
\showeprint[arxiv]{2406.01698}~[cs.AR]


\bibitem[\protect\citeauthoryear{Bertsekas}{Bertsekas}{1993}]%
        {bertsekas1993simple}
\bibfield{author}{\bibinfo{person}{Dimitri~P Bertsekas}.} \bibinfo{year}{1993}\natexlab{}.
\newblock \showarticletitle{A simple and fast label correcting algorithm for shortest paths}.
\newblock \bibinfo{journal}{\emph{Networks}} \bibinfo{volume}{23}, \bibinfo{number}{8} (\bibinfo{year}{1993}), \bibinfo{pages}{703--709}.
\newblock


\bibitem[\protect\citeauthoryear{Bloom}{Bloom}{1970}]%
        {bloom1970space}
\bibfield{author}{\bibinfo{person}{Burton~H Bloom}.} \bibinfo{year}{1970}\natexlab{}.
\newblock \showarticletitle{Space/time trade-offs in hash coding with allowable errors}.
\newblock \bibinfo{journal}{\emph{Commun. ACM}} \bibinfo{volume}{13}, \bibinfo{number}{7} (\bibinfo{year}{1970}), \bibinfo{pages}{422--426}.
\newblock


\bibitem[\protect\citeauthoryear{Borgeaud, Mensch, Hoffmann, Cai, Rutherford, Millican, Van Den~Driessche, Lespiau, Damoc, Clark, et~al\mbox{.}}{Borgeaud et~al\mbox{.}}{2022}]%
        {borgeaud2022improving}
\bibfield{author}{\bibinfo{person}{Sebastian Borgeaud}, \bibinfo{person}{Arthur Mensch}, \bibinfo{person}{Jordan Hoffmann}, \bibinfo{person}{Trevor Cai}, \bibinfo{person}{Eliza Rutherford}, \bibinfo{person}{Katie Millican}, \bibinfo{person}{George~Bm Van Den~Driessche}, \bibinfo{person}{Jean-Baptiste Lespiau}, \bibinfo{person}{Bogdan Damoc}, \bibinfo{person}{Aidan Clark}, {et~al\mbox{.}}} \bibinfo{year}{2022}\natexlab{}.
\newblock \showarticletitle{Improving language models by retrieving from trillions of tokens}. In \bibinfo{booktitle}{\emph{International conference on machine learning}}. PMLR, \bibinfo{pages}{2206--2240}.
\newblock


\bibitem[\protect\citeauthoryear{Chang, Kashyap, Hassan, Ghose, Hsieh, Lee, Li, Pekhimenko, Khan, and Mutlu}{Chang et~al\mbox{.}}{2016}]%
        {chang2016understanding}
\bibfield{author}{\bibinfo{person}{Kevin~K Chang}, \bibinfo{person}{Abhijith Kashyap}, \bibinfo{person}{Hasan Hassan}, \bibinfo{person}{Saugata Ghose}, \bibinfo{person}{Kevin Hsieh}, \bibinfo{person}{Donghyuk Lee}, \bibinfo{person}{Tianshi Li}, \bibinfo{person}{Gennady Pekhimenko}, \bibinfo{person}{Samira Khan}, {and} \bibinfo{person}{Onur Mutlu}.} \bibinfo{year}{2016}\natexlab{}.
\newblock \showarticletitle{Understanding latency variation in modern DRAM chips: Experimental characterization, analysis, and optimization}. In \bibinfo{booktitle}{\emph{Proceedings of the 2016 ACM SIGMETRICS International Conference on Measurement and Modeling of Computer Science}}. \bibinfo{pages}{323--336}.
\newblock


\bibitem[\protect\citeauthoryear{Chen, Zhao, Wang, Li, Liu, Li, Yang, and Wang}{Chen et~al\mbox{.}}{2021}]%
        {chen2021spann}
\bibfield{author}{\bibinfo{person}{Qi Chen}, \bibinfo{person}{Bing Zhao}, \bibinfo{person}{Haidong Wang}, \bibinfo{person}{Mingqin Li}, \bibinfo{person}{Chuanjie Liu}, \bibinfo{person}{Zengzhong Li}, \bibinfo{person}{Mao Yang}, {and} \bibinfo{person}{Jingdong Wang}.} \bibinfo{year}{2021}\natexlab{}.
\newblock \showarticletitle{SPANN: Highly-efficient Billion-scale Approximate Nearest Neighbor Search}.
\newblock \bibinfo{journal}{\emph{arXiv preprint arXiv:2111.08566}} (\bibinfo{year}{2021}).
\newblock


\bibitem[\protect\citeauthoryear{Chen, Chen, Zou, Li, Lu, Wang, and Zhao}{Chen et~al\mbox{.}}{2019b}]%
        {chen2019vector}
\bibfield{author}{\bibinfo{person}{Wei Chen}, \bibinfo{person}{Jincai Chen}, \bibinfo{person}{Fuhao Zou}, \bibinfo{person}{Yuan-Fang Li}, \bibinfo{person}{Ping Lu}, \bibinfo{person}{Qiang Wang}, {and} \bibinfo{person}{Wei Zhao}.} \bibinfo{year}{2019}\natexlab{b}.
\newblock \showarticletitle{Vector and line quantization for billion-scale similarity search on GPUs}.
\newblock \bibinfo{journal}{\emph{Future Generation Computer Systems}}  \bibinfo{volume}{99} (\bibinfo{year}{2019}), \bibinfo{pages}{295--307}.
\newblock


\bibitem[\protect\citeauthoryear{Chen, Chen, Zou, Li, Lu, and Zhao}{Chen et~al\mbox{.}}{2019a}]%
        {chen2019robustiq}
\bibfield{author}{\bibinfo{person}{Wei Chen}, \bibinfo{person}{Jincai Chen}, \bibinfo{person}{Fuhao Zou}, \bibinfo{person}{Yuan-Fang Li}, \bibinfo{person}{Ping Lu}, {and} \bibinfo{person}{Wei Zhao}.} \bibinfo{year}{2019}\natexlab{a}.
\newblock \showarticletitle{Robustiq: A robust ann search method for billion-scale similarity search on gpus}. In \bibinfo{booktitle}{\emph{Proceedings of the 2019 on International Conference on Multimedia Retrieval}}. \bibinfo{pages}{132--140}.
\newblock


\bibitem[\protect\citeauthoryear{Chern, Hechtman, Davis, Guo, Majnemer, and Kumar}{Chern et~al\mbox{.}}{2022}]%
        {chern2022tpu}
\bibfield{author}{\bibinfo{person}{Felix Chern}, \bibinfo{person}{Blake Hechtman}, \bibinfo{person}{Andy Davis}, \bibinfo{person}{Ruiqi Guo}, \bibinfo{person}{David Majnemer}, {and} \bibinfo{person}{Sanjiv Kumar}.} \bibinfo{year}{2022}\natexlab{}.
\newblock \showarticletitle{TPU-KNN: K Nearest Neighbor Search at Peak FLOP/s}.
\newblock \bibinfo{journal}{\emph{arXiv preprint arXiv:2206.14286}} (\bibinfo{year}{2022}).
\newblock


\bibitem[\protect\citeauthoryear{Choquette, Lee, Krashinsky, Balan, and Khailany}{Choquette et~al\mbox{.}}{2021}]%
        {choquette20213}
\bibfield{author}{\bibinfo{person}{Jack Choquette}, \bibinfo{person}{Edward Lee}, \bibinfo{person}{Ronny Krashinsky}, \bibinfo{person}{Vishnu Balan}, {and} \bibinfo{person}{Brucek Khailany}.} \bibinfo{year}{2021}\natexlab{}.
\newblock \showarticletitle{3.2 the a100 datacenter gpu and ampere architecture}. In \bibinfo{booktitle}{\emph{2021 IEEE International Solid-State Circuits Conference (ISSCC)}}, Vol.~\bibinfo{volume}{64}. IEEE, \bibinfo{pages}{48--50}.
\newblock


\bibitem[\protect\citeauthoryear{Covington, Adams, and Sargin}{Covington et~al\mbox{.}}{2016}]%
        {google_recommendation}
\bibfield{author}{\bibinfo{person}{Paul Covington}, \bibinfo{person}{Jay Adams}, {and} \bibinfo{person}{Emre Sargin}.} \bibinfo{year}{2016}\natexlab{}.
\newblock \showarticletitle{Deep neural networks for youtube recommendations}. In \bibinfo{booktitle}{\emph{Proceedings of the 10th ACM conference on recommender systems}}. \bibinfo{pages}{191--198}.
\newblock


\bibitem[\protect\citeauthoryear{Dallachiesa, Palpanas, and Ilyas}{Dallachiesa et~al\mbox{.}}{2014}]%
        {dallachiesa2014top}
\bibfield{author}{\bibinfo{person}{Michele Dallachiesa}, \bibinfo{person}{Themis Palpanas}, {and} \bibinfo{person}{Ihab~F Ilyas}.} \bibinfo{year}{2014}\natexlab{}.
\newblock \showarticletitle{Top-k nearest neighbor search in uncertain data series}.
\newblock \bibinfo{journal}{\emph{Proceedings of the VLDB Endowment}} \bibinfo{volume}{8}, \bibinfo{number}{1} (\bibinfo{year}{2014}), \bibinfo{pages}{13--24}.
\newblock


\bibitem[\protect\citeauthoryear{Datar, Immorlica, Indyk, and Mirrokni}{Datar et~al\mbox{.}}{2004}]%
        {datar2004locality}
\bibfield{author}{\bibinfo{person}{Mayur Datar}, \bibinfo{person}{Nicole Immorlica}, \bibinfo{person}{Piotr Indyk}, {and} \bibinfo{person}{Vahab~S Mirrokni}.} \bibinfo{year}{2004}\natexlab{}.
\newblock \showarticletitle{Locality-sensitive hashing scheme based on p-stable distributions}. In \bibinfo{booktitle}{\emph{Proceedings of the twentieth annual symposium on Computational geometry}}. \bibinfo{pages}{253--262}.
\newblock


\bibitem[\protect\citeauthoryear{Doshi, Das, Bhutani, Kumar, Bhatt, and Balasubramanian}{Doshi et~al\mbox{.}}{2020}]%
        {doshi2020lanns}
\bibfield{author}{\bibinfo{person}{Ishita Doshi}, \bibinfo{person}{Dhritiman Das}, \bibinfo{person}{Ashish Bhutani}, \bibinfo{person}{Rajeev Kumar}, \bibinfo{person}{Rushi Bhatt}, {and} \bibinfo{person}{Niranjan Balasubramanian}.} \bibinfo{year}{2020}\natexlab{}.
\newblock \showarticletitle{LANNS: a web-scale approximate nearest neighbor lookup system}.
\newblock \bibinfo{journal}{\emph{Proceedings of the VLDB Endowment}} (\bibinfo{year}{2020}).
\newblock


\bibitem[\protect\citeauthoryear{Fowers, Ovtcharov, Papamichael, Massengill, Liu, Lo, Alkalay, Haselman, Adams, Ghandi, Heil, Patel, Sapek, Weisz, Woods, Lanka, Reinhardt, Caulfield, Chung, and Burger}{Fowers et~al\mbox{.}}{2018}]%
        {fowers2018configurable}
\bibfield{author}{\bibinfo{person}{Jeremy Fowers}, \bibinfo{person}{Kalin Ovtcharov}, \bibinfo{person}{Michael Papamichael}, \bibinfo{person}{Todd Massengill}, \bibinfo{person}{Ming Liu}, \bibinfo{person}{Daniel Lo}, \bibinfo{person}{Shlomi Alkalay}, \bibinfo{person}{Michael Haselman}, \bibinfo{person}{Logan Adams}, \bibinfo{person}{Mahdi Ghandi}, \bibinfo{person}{Stephen Heil}, \bibinfo{person}{Prerak Patel}, \bibinfo{person}{Adam Sapek}, \bibinfo{person}{Gabriel Weisz}, \bibinfo{person}{Lisa Woods}, \bibinfo{person}{Sitaram Lanka}, \bibinfo{person}{Steven~K. Reinhardt}, \bibinfo{person}{Adrian~M. Caulfield}, \bibinfo{person}{Eric~S. Chung}, {and} \bibinfo{person}{Doug Burger}.} \bibinfo{year}{2018}\natexlab{}.
\newblock \showarticletitle{A configurable cloud-scale DNN processor for real-time AI}. In \bibinfo{booktitle}{\emph{2018 ACM/IEEE 45th Annual International Symposium on Computer Architecture (ISCA)}}. IEEE, \bibinfo{pages}{1--14}.
\newblock


\bibitem[\protect\citeauthoryear{Fu, Xiang, Wang, and Cai}{Fu et~al\mbox{.}}{2017}]%
        {fu2017fast}
\bibfield{author}{\bibinfo{person}{Cong Fu}, \bibinfo{person}{Chao Xiang}, \bibinfo{person}{Changxu Wang}, {and} \bibinfo{person}{Deng Cai}.} \bibinfo{year}{2017}\natexlab{}.
\newblock \showarticletitle{Fast approximate nearest neighbor search with the navigating spreading-out graph}.
\newblock \bibinfo{journal}{\emph{arXiv preprint arXiv:1707.00143}} (\bibinfo{year}{2017}).
\newblock


\bibitem[\protect\citeauthoryear{Gao and Long}{Gao and Long}{2023}]%
        {gao2023high}
\bibfield{author}{\bibinfo{person}{Jianyang Gao} {and} \bibinfo{person}{Cheng Long}.} \bibinfo{year}{2023}\natexlab{}.
\newblock \showarticletitle{High-dimensional approximate nearest neighbor search: with reliable and efficient distance comparison operations}.
\newblock \bibinfo{journal}{\emph{Proceedings of the ACM on Management of Data}} \bibinfo{volume}{1}, \bibinfo{number}{2} (\bibinfo{year}{2023}), \bibinfo{pages}{1--27}.
\newblock


\bibitem[\protect\citeauthoryear{Gao and Long}{Gao and Long}{2024}]%
        {gao2024rabitq}
\bibfield{author}{\bibinfo{person}{Jianyang Gao} {and} \bibinfo{person}{Cheng Long}.} \bibinfo{year}{2024}\natexlab{}.
\newblock \showarticletitle{RaBitQ: Quantizing High-Dimensional Vectors with a Theoretical Error Bound for Approximate Nearest Neighbor Search}.
\newblock \bibinfo{journal}{\emph{Proceedings of the ACM on Management of Data}} \bibinfo{volume}{2}, \bibinfo{number}{3} (\bibinfo{year}{2024}), \bibinfo{pages}{1--27}.
\newblock


\bibitem[\protect\citeauthoryear{Ge, He, Ke, and Sun}{Ge et~al\mbox{.}}{2013}]%
        {OPQ}
\bibfield{author}{\bibinfo{person}{Tiezheng Ge}, \bibinfo{person}{Kaiming He}, \bibinfo{person}{Qifa Ke}, {and} \bibinfo{person}{Jian Sun}.} \bibinfo{year}{2013}\natexlab{}.
\newblock \showarticletitle{Optimized product quantization}.
\newblock \bibinfo{journal}{\emph{IEEE transactions on pattern analysis and machine intelligence}} \bibinfo{volume}{36}, \bibinfo{number}{4} (\bibinfo{year}{2013}), \bibinfo{pages}{744--755}.
\newblock


\bibitem[\protect\citeauthoryear{Gionis, Indyk, Motwani, et~al\mbox{.}}{Gionis et~al\mbox{.}}{1999}]%
        {gionis1999similarity}
\bibfield{author}{\bibinfo{person}{Aristides Gionis}, \bibinfo{person}{Piotr Indyk}, \bibinfo{person}{Rajeev Motwani}, {et~al\mbox{.}}} \bibinfo{year}{1999}\natexlab{}.
\newblock \showarticletitle{Similarity search in high dimensions via hashing}. In \bibinfo{booktitle}{\emph{Vldb}}, Vol.~\bibinfo{volume}{99}. \bibinfo{pages}{518--529}.
\newblock


\bibitem[\protect\citeauthoryear{Gou, Gao, Xu, and Long}{Gou et~al\mbox{.}}{2025}]%
        {gou2025symphonyqg}
\bibfield{author}{\bibinfo{person}{Yutong Gou}, \bibinfo{person}{Jianyang Gao}, \bibinfo{person}{Yuexuan Xu}, {and} \bibinfo{person}{Cheng Long}.} \bibinfo{year}{2025}\natexlab{}.
\newblock \showarticletitle{SymphonyQG: Towards Symphonious Integration of Quantization and Graph for Approximate Nearest Neighbor Search}.
\newblock \bibinfo{journal}{\emph{Proceedings of the ACM on Management of Data}} \bibinfo{volume}{3}, \bibinfo{number}{1} (\bibinfo{year}{2025}), \bibinfo{pages}{1--26}.
\newblock


\bibitem[\protect\citeauthoryear{Groh, Ruppert, Wieschollek, and Lensch}{Groh et~al\mbox{.}}{2022}]%
        {groh2022ggnn}
\bibfield{author}{\bibinfo{person}{Fabian Groh}, \bibinfo{person}{Lukas Ruppert}, \bibinfo{person}{Patrick Wieschollek}, {and} \bibinfo{person}{Hendrik~PA Lensch}.} \bibinfo{year}{2022}\natexlab{}.
\newblock \showarticletitle{Ggnn: Graph-based gpu nearest neighbor search}.
\newblock \bibinfo{journal}{\emph{IEEE Transactions on Big Data}} \bibinfo{volume}{9}, \bibinfo{number}{1} (\bibinfo{year}{2022}), \bibinfo{pages}{267--279}.
\newblock


\bibitem[\protect\citeauthoryear{Guo, Luan, Xiang, Yan, Yi, Luo, Cheng, Xu, Luo, Liu, et~al\mbox{.}}{Guo et~al\mbox{.}}{2022}]%
        {guo2022manu}
\bibfield{author}{\bibinfo{person}{Rentong Guo}, \bibinfo{person}{Xiaofan Luan}, \bibinfo{person}{Long Xiang}, \bibinfo{person}{Xiao Yan}, \bibinfo{person}{Xiaomeng Yi}, \bibinfo{person}{Jigao Luo}, \bibinfo{person}{Qianya Cheng}, \bibinfo{person}{Weizhi Xu}, \bibinfo{person}{Jiarui Luo}, \bibinfo{person}{Frank Liu}, {et~al\mbox{.}}} \bibinfo{year}{2022}\natexlab{}.
\newblock \showarticletitle{Manu: A Cloud Native Vector Database Management System}.
\newblock \bibinfo{journal}{\emph{arXiv preprint arXiv:2206.13843}} (\bibinfo{year}{2022}).
\newblock


\bibitem[\protect\citeauthoryear{Gupta, Hsia, Saraph, Wang, Reagen, Wei, Lee, Brooks, and Wu}{Gupta et~al\mbox{.}}{2020}]%
        {gupta2020deeprecsys}
\bibfield{author}{\bibinfo{person}{Udit Gupta}, \bibinfo{person}{Samuel Hsia}, \bibinfo{person}{Vikram Saraph}, \bibinfo{person}{Xiaodong Wang}, \bibinfo{person}{Brandon Reagen}, \bibinfo{person}{Gu-Yeon Wei}, \bibinfo{person}{Hsien-Hsin~S Lee}, \bibinfo{person}{David Brooks}, {and} \bibinfo{person}{Carole-Jean Wu}.} \bibinfo{year}{2020}\natexlab{}.
\newblock \showarticletitle{Deeprecsys: A system for optimizing end-to-end at-scale neural recommendation inference}. In \bibinfo{booktitle}{\emph{2020 ACM/IEEE 47th Annual International Symposium on Computer Architecture (ISCA)}}. IEEE, \bibinfo{pages}{982--995}.
\newblock


\bibitem[\protect\citeauthoryear{Guu, Lee, Tung, Pasupat, and Chang}{Guu et~al\mbox{.}}{2020}]%
        {guu2020realm}
\bibfield{author}{\bibinfo{person}{Kelvin Guu}, \bibinfo{person}{Kenton Lee}, \bibinfo{person}{Zora Tung}, \bibinfo{person}{Panupong Pasupat}, {and} \bibinfo{person}{Ming-Wei Chang}.} \bibinfo{year}{2020}\natexlab{}.
\newblock \showarticletitle{Realm: Retrieval-augmented language model pre-training}.
\newblock \bibinfo{journal}{\emph{arXiv preprint arXiv:2002.08909}} (\bibinfo{year}{2020}).
\newblock


\bibitem[\protect\citeauthoryear{He, Korolija, and Alonso}{He et~al\mbox{.}}{2021}]%
        {100gbps}
\bibfield{author}{\bibinfo{person}{Zhenhao He}, \bibinfo{person}{Dario Korolija}, {and} \bibinfo{person}{Gustavo Alonso}.} \bibinfo{year}{2021}\natexlab{}.
\newblock \showarticletitle{EasyNet: 100 Gbps Network for HLS}. In \bibinfo{booktitle}{\emph{2021 31th International Conference on Field Programmable Logic and Applications (FPL)}}.
\newblock


\bibitem[\protect\citeauthoryear{Hu, Wang, Chang, Lee, Lin, Wang, Lin, Huang, Lee, Su, et~al\mbox{.}}{Hu et~al\mbox{.}}{2022}]%
        {hu2022ice}
\bibfield{author}{\bibinfo{person}{Han-Wen Hu}, \bibinfo{person}{Wei-Chen Wang}, \bibinfo{person}{Yuan-Hao Chang}, \bibinfo{person}{Yung-Chun Lee}, \bibinfo{person}{Bo-Rong Lin}, \bibinfo{person}{Huai-Mu Wang}, \bibinfo{person}{Yen-Po Lin}, \bibinfo{person}{Yu-Ming Huang}, \bibinfo{person}{Chong-Ying Lee}, \bibinfo{person}{Tzu-Hsiang Su}, {et~al\mbox{.}}} \bibinfo{year}{2022}\natexlab{}.
\newblock \showarticletitle{ICE: An Intelligent Cognition Engine with 3D NAND-based In-Memory Computing for Vector Similarity Search Acceleration}. In \bibinfo{booktitle}{\emph{2022 55th IEEE/ACM International Symposium on Microarchitecture (MICRO)}}. IEEE, \bibinfo{pages}{763--783}.
\newblock


\bibitem[\protect\citeauthoryear{Huang, Sharma, Sun, Xia, Zhang, Pronin, Padmanabhan, Ottaviano, and Yang}{Huang et~al\mbox{.}}{2020}]%
        {facebook_EBR}
\bibfield{author}{\bibinfo{person}{Jui-Ting Huang}, \bibinfo{person}{Ashish Sharma}, \bibinfo{person}{Shuying Sun}, \bibinfo{person}{Li Xia}, \bibinfo{person}{David Zhang}, \bibinfo{person}{Philip Pronin}, \bibinfo{person}{Janani Padmanabhan}, \bibinfo{person}{Giuseppe Ottaviano}, {and} \bibinfo{person}{Linjun Yang}.} \bibinfo{year}{2020}\natexlab{}.
\newblock \showarticletitle{Embedding-based retrieval in facebook search}. In \bibinfo{booktitle}{\emph{Proceedings of the 26th ACM SIGKDD International Conference on Knowledge Discovery \& Data Mining}}. \bibinfo{pages}{2553--2561}.
\newblock


\bibitem[\protect\citeauthoryear{Huang, Lim, and Cong}{Huang et~al\mbox{.}}{2014}]%
        {huang2014scalable}
\bibfield{author}{\bibinfo{person}{Muhuan Huang}, \bibinfo{person}{Kevin Lim}, {and} \bibinfo{person}{Jason Cong}.} \bibinfo{year}{2014}\natexlab{}.
\newblock \showarticletitle{A scalable, high-performance customized priority queue}. In \bibinfo{booktitle}{\emph{2014 24th International Conference on Field Programmable Logic and Applications (FPL)}}. IEEE, \bibinfo{pages}{1--4}.
\newblock


\bibitem[\protect\citeauthoryear{Huang, Lei, and Tung}{Huang et~al\mbox{.}}{2021}]%
        {huang2021point}
\bibfield{author}{\bibinfo{person}{Qiang Huang}, \bibinfo{person}{Yifan Lei}, {and} \bibinfo{person}{Anthony~KH Tung}.} \bibinfo{year}{2021}\natexlab{}.
\newblock \showarticletitle{Point-to-Hyperplane Nearest Neighbor Search Beyond the Unit Hypersphere}. In \bibinfo{booktitle}{\emph{Proceedings of the 2021 International Conference on Management of Data}}. \bibinfo{pages}{777--789}.
\newblock


\bibitem[\protect\citeauthoryear{Jang, Choi, Bae, Lee, Kwon, and Jung}{Jang et~al\mbox{.}}{2023}]%
        {jang2023cxl}
\bibfield{author}{\bibinfo{person}{Junhyeok Jang}, \bibinfo{person}{Hanjin Choi}, \bibinfo{person}{Hanyeoreum Bae}, \bibinfo{person}{Seungjun Lee}, \bibinfo{person}{Miryeong Kwon}, {and} \bibinfo{person}{Myoungsoo Jung}.} \bibinfo{year}{2023}\natexlab{}.
\newblock \showarticletitle{$\{$CXL-ANNS$\}$:$\{$Software-Hardware$\}$ Collaborative Memory Disaggregation and Computation for $\{$Billion-Scale$\}$ Approximate Nearest Neighbor Search}. In \bibinfo{booktitle}{\emph{2023 USENIX Annual Technical Conference (USENIX ATC 23)}}. \bibinfo{pages}{585--600}.
\newblock


\bibitem[\protect\citeauthoryear{Jayaram~Subramanya, Devvrit, Simhadri, Krishnawamy, and Kadekodi}{Jayaram~Subramanya et~al\mbox{.}}{2019}]%
        {jayaram2019diskann}
\bibfield{author}{\bibinfo{person}{Suhas Jayaram~Subramanya}, \bibinfo{person}{Fnu Devvrit}, \bibinfo{person}{Harsha~Vardhan Simhadri}, \bibinfo{person}{Ravishankar Krishnawamy}, {and} \bibinfo{person}{Rohan Kadekodi}.} \bibinfo{year}{2019}\natexlab{}.
\newblock \showarticletitle{Diskann: Fast accurate billion-point nearest neighbor search on a single node}.
\newblock \bibinfo{journal}{\emph{Advances in Neural Information Processing Systems}}  \bibinfo{volume}{32} (\bibinfo{year}{2019}).
\newblock


\bibitem[\protect\citeauthoryear{Jegou, Douze, and Schmid}{Jegou et~al\mbox{.}}{2010}]%
        {PQ}
\bibfield{author}{\bibinfo{person}{Herve Jegou}, \bibinfo{person}{Matthijs Douze}, {and} \bibinfo{person}{Cordelia Schmid}.} \bibinfo{year}{2010}\natexlab{}.
\newblock \showarticletitle{Product quantization for nearest neighbor search}.
\newblock \bibinfo{journal}{\emph{IEEE transactions on pattern analysis and machine intelligence}} \bibinfo{volume}{33}, \bibinfo{number}{1} (\bibinfo{year}{2010}), \bibinfo{pages}{117--128}.
\newblock


\bibitem[\protect\citeauthoryear{Jiang, Fu, and Wong}{Jiang et~al\mbox{.}}{2015}]%
        {jiang2015exact}
\bibfield{author}{\bibinfo{person}{Minhao Jiang}, \bibinfo{person}{Ada Wai-Chee Fu}, {and} \bibinfo{person}{Raymond Chi-Wing Wong}.} \bibinfo{year}{2015}\natexlab{}.
\newblock \showarticletitle{Exact top-k nearest keyword search in large networks}. In \bibinfo{booktitle}{\emph{Proceedings of the 2015 ACM SIGMOD international conference on management of data}}. \bibinfo{pages}{393--404}.
\newblock


\bibitem[\protect\citeauthoryear{Jiang, He, Zhang, Preu{\ss}er, Zeng, Feng, Zhang, Liu, Li, Zhou, et~al\mbox{.}}{Jiang et~al\mbox{.}}{2021a}]%
        {jiang2021microrec}
\bibfield{author}{\bibinfo{person}{Wenqi Jiang}, \bibinfo{person}{Zhenhao He}, \bibinfo{person}{Shuai Zhang}, \bibinfo{person}{Thomas~B Preu{\ss}er}, \bibinfo{person}{Kai Zeng}, \bibinfo{person}{Liang Feng}, \bibinfo{person}{Jiansong Zhang}, \bibinfo{person}{Tongxuan Liu}, \bibinfo{person}{Yong Li}, \bibinfo{person}{Jingren Zhou}, {et~al\mbox{.}}} \bibinfo{year}{2021}\natexlab{a}.
\newblock \showarticletitle{MicroRec: efficient recommendation inference by hardware and data structure solutions}.
\newblock \bibinfo{journal}{\emph{Proceedings of Machine Learning and Systems}}  \bibinfo{volume}{3} (\bibinfo{year}{2021}), \bibinfo{pages}{845--859}.
\newblock


\bibitem[\protect\citeauthoryear{Jiang, He, Zhang, Zeng, Feng, Zhang, Liu, Li, Zhou, Zhang, et~al\mbox{.}}{Jiang et~al\mbox{.}}{2021b}]%
        {jiang2021fleetrec}
\bibfield{author}{\bibinfo{person}{Wenqi Jiang}, \bibinfo{person}{Zhenhao He}, \bibinfo{person}{Shuai Zhang}, \bibinfo{person}{Kai Zeng}, \bibinfo{person}{Liang Feng}, \bibinfo{person}{Jiansong Zhang}, \bibinfo{person}{Tongxuan Liu}, \bibinfo{person}{Yong Li}, \bibinfo{person}{Jingren Zhou}, \bibinfo{person}{Ce Zhang}, {et~al\mbox{.}}} \bibinfo{year}{2021}\natexlab{b}.
\newblock \showarticletitle{Fleetrec: Large-scale recommendation inference on hybrid gpu-fpga clusters}. In \bibinfo{booktitle}{\emph{Proceedings of the 27th ACM SIGKDD Conference on Knowledge Discovery \& Data Mining}}. \bibinfo{pages}{3097--3105}.
\newblock


\bibitem[\protect\citeauthoryear{Jiang, Li, Zhu, de~Fine~Licht, He, Shi, Renggli, Zhang, Rekatsinas, Hoefler, et~al\mbox{.}}{Jiang et~al\mbox{.}}{2023}]%
        {jiang2023co}
\bibfield{author}{\bibinfo{person}{Wenqi Jiang}, \bibinfo{person}{Shigang Li}, \bibinfo{person}{Yu Zhu}, \bibinfo{person}{Johannes de Fine~Licht}, \bibinfo{person}{Zhenhao He}, \bibinfo{person}{Runbin Shi}, \bibinfo{person}{Cedric Renggli}, \bibinfo{person}{Shuai Zhang}, \bibinfo{person}{Theodoros Rekatsinas}, \bibinfo{person}{Torsten Hoefler}, {et~al\mbox{.}}} \bibinfo{year}{2023}\natexlab{}.
\newblock \showarticletitle{Co-design hardware and algorithm for vector search}. In \bibinfo{booktitle}{\emph{Proceedings of the International Conference for High Performance Computing, Networking, Storage and Analysis}}. \bibinfo{pages}{1--15}.
\newblock


\bibitem[\protect\citeauthoryear{Jiang, Subramanian, Graves, Alonso, Yazdanbakhsh, and Dadu}{Jiang et~al\mbox{.}}{2025a}]%
        {jiang2025rago}
\bibfield{author}{\bibinfo{person}{Wenqi Jiang}, \bibinfo{person}{Suvinay Subramanian}, \bibinfo{person}{Cat Graves}, \bibinfo{person}{Gustavo Alonso}, \bibinfo{person}{Amir Yazdanbakhsh}, {and} \bibinfo{person}{Vidushi Dadu}.} \bibinfo{year}{2025}\natexlab{a}.
\newblock \showarticletitle{RAGO: Systematic Performance Optimization for Retrieval-Augmented Generation Serving}. In \bibinfo{booktitle}{\emph{Proceedings of the 52th Annual International Symposium on Computer Architecture}}.
\newblock


\bibitem[\protect\citeauthoryear{Jiang, Zeller, Waleffe, Hoefler, and Alonso}{Jiang et~al\mbox{.}}{2025b}]%
        {jiang2023chameleon}
\bibfield{author}{\bibinfo{person}{Wenqi Jiang}, \bibinfo{person}{Marco Zeller}, \bibinfo{person}{Roger Waleffe}, \bibinfo{person}{Torsten Hoefler}, {and} \bibinfo{person}{Gustavo Alonso}.} \bibinfo{year}{2025}\natexlab{b}.
\newblock \showarticletitle{Chameleon: a heterogeneous and disaggregated accelerator system for retrieval-augmented language models}.
\newblock \bibinfo{journal}{\emph{Proceedings of the VLDB Endowment}}  \bibinfo{volume}{18} (\bibinfo{year}{2025}).
\newblock


\bibitem[\protect\citeauthoryear{Jiang, Zhang, Han, Wang, Wang, and Kraska}{Jiang et~al\mbox{.}}{2025c}]%
        {jiang2024piperag}
\bibfield{author}{\bibinfo{person}{Wenqi Jiang}, \bibinfo{person}{Shuai Zhang}, \bibinfo{person}{Boran Han}, \bibinfo{person}{Jie Wang}, \bibinfo{person}{Bernie Wang}, {and} \bibinfo{person}{Tim Kraska}.} \bibinfo{year}{2025}\natexlab{c}.
\newblock \showarticletitle{Piperag: Fast retrieval-augmented generation via adaptive pipeline parallelism}.
\newblock \bibinfo{journal}{\emph{Proceedings of the ACM SIGKDD International Conference on Knowledge Discovery \& Data Mining}} (\bibinfo{year}{2025}).
\newblock


\bibitem[\protect\citeauthoryear{Johnson, Douze, and J{\'e}gou}{Johnson et~al\mbox{.}}{2019}]%
        {johnson2019billion}
\bibfield{author}{\bibinfo{person}{Jeff Johnson}, \bibinfo{person}{Matthijs Douze}, {and} \bibinfo{person}{Herv{\'e} J{\'e}gou}.} \bibinfo{year}{2019}\natexlab{}.
\newblock \showarticletitle{Billion-scale similarity search with gpus}.
\newblock \bibinfo{journal}{\emph{IEEE Transactions on Big Data}} (\bibinfo{year}{2019}).
\newblock


\bibitem[\protect\citeauthoryear{Karpukhin, O{\u{g}}uz, Min, Lewis, Wu, Edunov, Chen, and Yih}{Karpukhin et~al\mbox{.}}{2020}]%
        {karpukhin2020dense}
\bibfield{author}{\bibinfo{person}{Vladimir Karpukhin}, \bibinfo{person}{Barlas O{\u{g}}uz}, \bibinfo{person}{Sewon Min}, \bibinfo{person}{Patrick Lewis}, \bibinfo{person}{Ledell Wu}, \bibinfo{person}{Sergey Edunov}, \bibinfo{person}{Danqi Chen}, {and} \bibinfo{person}{Wen-tau Yih}.} \bibinfo{year}{2020}\natexlab{}.
\newblock \showarticletitle{Dense passage retrieval for open-domain question answering}.
\newblock \bibinfo{journal}{\emph{arXiv preprint arXiv:2004.04906}} (\bibinfo{year}{2020}).
\newblock


\bibitem[\protect\citeauthoryear{Khattab and Zaharia}{Khattab and Zaharia}{2020}]%
        {khattab2020colbert}
\bibfield{author}{\bibinfo{person}{Omar Khattab} {and} \bibinfo{person}{Matei Zaharia}.} \bibinfo{year}{2020}\natexlab{}.
\newblock \showarticletitle{Colbert: Efficient and effective passage search via contextualized late interaction over bert}. In \bibinfo{booktitle}{\emph{Proceedings of the 43rd International ACM SIGIR conference on research and development in Information Retrieval}}. \bibinfo{pages}{39--48}.
\newblock


\bibitem[\protect\citeauthoryear{Kim, Park, Do, Ji, and Kim}{Kim et~al\mbox{.}}{2022}]%
        {kim2022accelerating}
\bibfield{author}{\bibinfo{person}{Ji-Hoon Kim}, \bibinfo{person}{Yeo-Reum Park}, \bibinfo{person}{Jaeyoung Do}, \bibinfo{person}{Soo-Young Ji}, {and} \bibinfo{person}{Joo-Young Kim}.} \bibinfo{year}{2022}\natexlab{}.
\newblock \showarticletitle{Accelerating large-scale graph-based nearest neighbor search on a computational storage platform}.
\newblock \bibinfo{journal}{\emph{IEEE Trans. Comput.}} \bibinfo{volume}{72}, \bibinfo{number}{1} (\bibinfo{year}{2022}), \bibinfo{pages}{278--290}.
\newblock


\bibitem[\protect\citeauthoryear{Koh, Lee, Kwon, and Jung}{Koh et~al\mbox{.}}{2018}]%
        {koh2018exploring}
\bibfield{author}{\bibinfo{person}{Sungjoon Koh}, \bibinfo{person}{Changrim Lee}, \bibinfo{person}{Miryeong Kwon}, {and} \bibinfo{person}{Myoungsoo Jung}.} \bibinfo{year}{2018}\natexlab{}.
\newblock \showarticletitle{Exploring system challenges of $\{$ultra-low$\}$ latency solid state drives}. In \bibinfo{booktitle}{\emph{10th USENIX Workshop on Hot Topics in Storage and File Systems (HotStorage 18)}}.
\newblock


\bibitem[\protect\citeauthoryear{LaGrone, Aribuki, and Chapman}{LaGrone et~al\mbox{.}}{2011}]%
        {lagrone2011set}
\bibfield{author}{\bibinfo{person}{James LaGrone}, \bibinfo{person}{Ayodunni Aribuki}, {and} \bibinfo{person}{Barbara Chapman}.} \bibinfo{year}{2011}\natexlab{}.
\newblock \showarticletitle{A set of microbenchmarks for measuring OpenMP task overheads}. In \bibinfo{booktitle}{\emph{Proceedings of the International Conference on Parallel and Distributed Processing Techniques and Applications (PDPTA)}}. Citeseer, \bibinfo{pages}{1}.
\newblock


\bibitem[\protect\citeauthoryear{Lee, Choi, Min, Lee, Beak, Jeong, Lee, and Ham}{Lee et~al\mbox{.}}{2022}]%
        {lee2022anna}
\bibfield{author}{\bibinfo{person}{Yejin Lee}, \bibinfo{person}{Hyunji Choi}, \bibinfo{person}{Sunhong Min}, \bibinfo{person}{Hyunseung Lee}, \bibinfo{person}{Sangwon Beak}, \bibinfo{person}{Dawoon Jeong}, \bibinfo{person}{Jae~W Lee}, {and} \bibinfo{person}{Tae~Jun Ham}.} \bibinfo{year}{2022}\natexlab{}.
\newblock \showarticletitle{ANNA: Specialized Architecture for Approximate Nearest Neighbor Search}. In \bibinfo{booktitle}{\emph{2022 IEEE International Symposium on High-Performance Computer Architecture (HPCA)}}. IEEE, \bibinfo{pages}{169--183}.
\newblock


\bibitem[\protect\citeauthoryear{Leiserson}{Leiserson}{1979}]%
        {leiserson1979systolic}
\bibfield{author}{\bibinfo{person}{Charles~E Leiserson}.} \bibinfo{year}{1979}\natexlab{}.
\newblock \bibinfo{booktitle}{\emph{Systolic Priority Queues.}}
\newblock \bibinfo{type}{{T}echnical {R}eport}. \bibinfo{institution}{CARNEGIE-MELLON UNIV PITTSBURGH PA DEPT OF COMPUTER SCIENCE}.
\newblock


\bibitem[\protect\citeauthoryear{Lejsek, {\'A}smundsson, J{\'o}nsson, and Amsaleg}{Lejsek et~al\mbox{.}}{2008}]%
        {lejsek2008nv}
\bibfield{author}{\bibinfo{person}{Herwig Lejsek}, \bibinfo{person}{Fri{\dh}rik~Hei{\dh}ar {\'A}smundsson}, \bibinfo{person}{Bj{\"o}rn~{\TH}{\'o}r J{\'o}nsson}, {and} \bibinfo{person}{Laurent Amsaleg}.} \bibinfo{year}{2008}\natexlab{}.
\newblock \showarticletitle{NV-Tree: An efficient disk-based index for approximate search in very large high-dimensional collections}.
\newblock \bibinfo{journal}{\emph{IEEE Transactions on Pattern Analysis and Machine Intelligence}} \bibinfo{volume}{31}, \bibinfo{number}{5} (\bibinfo{year}{2008}), \bibinfo{pages}{869--883}.
\newblock


\bibitem[\protect\citeauthoryear{Lewis, Perez, Piktus, Petroni, Karpukhin, Goyal, K{\"u}ttler, Lewis, Yih, Rockt{\"a}schel, et~al\mbox{.}}{Lewis et~al\mbox{.}}{2020}]%
        {lewis2020retrieval}
\bibfield{author}{\bibinfo{person}{Patrick Lewis}, \bibinfo{person}{Ethan Perez}, \bibinfo{person}{Aleksandra Piktus}, \bibinfo{person}{Fabio Petroni}, \bibinfo{person}{Vladimir Karpukhin}, \bibinfo{person}{Naman Goyal}, \bibinfo{person}{Heinrich K{\"u}ttler}, \bibinfo{person}{Mike Lewis}, \bibinfo{person}{Wen-tau Yih}, \bibinfo{person}{Tim Rockt{\"a}schel}, {et~al\mbox{.}}} \bibinfo{year}{2020}\natexlab{}.
\newblock \showarticletitle{Retrieval-augmented generation for knowledge-intensive nlp tasks}.
\newblock \bibinfo{journal}{\emph{Advances in Neural Information Processing Systems}}  \bibinfo{volume}{33} (\bibinfo{year}{2020}), \bibinfo{pages}{9459--9474}.
\newblock


\bibitem[\protect\citeauthoryear{Li, Zhang, Sun, Wang, Li, Zhang, and Lin}{Li et~al\mbox{.}}{2019}]%
        {li2019approximate}
\bibfield{author}{\bibinfo{person}{Wen Li}, \bibinfo{person}{Ying Zhang}, \bibinfo{person}{Yifang Sun}, \bibinfo{person}{Wei Wang}, \bibinfo{person}{Mingjie Li}, \bibinfo{person}{Wenjie Zhang}, {and} \bibinfo{person}{Xuemin Lin}.} \bibinfo{year}{2019}\natexlab{}.
\newblock \showarticletitle{Approximate nearest neighbor search on high dimensional data—experiments, analyses, and improvement}.
\newblock \bibinfo{journal}{\emph{IEEE Transactions on Knowledge and Data Engineering}} \bibinfo{volume}{32}, \bibinfo{number}{8} (\bibinfo{year}{2019}), \bibinfo{pages}{1475--1488}.
\newblock


\bibitem[\protect\citeauthoryear{Liang, Wang, Lu, Yang, Li, and Li}{Liang et~al\mbox{.}}{2019}]%
        {liang2019cognitive}
\bibfield{author}{\bibinfo{person}{Shengwen Liang}, \bibinfo{person}{Ying Wang}, \bibinfo{person}{Youyou Lu}, \bibinfo{person}{Zhe Yang}, \bibinfo{person}{Huawei Li}, {and} \bibinfo{person}{Xiaowei Li}.} \bibinfo{year}{2019}\natexlab{}.
\newblock \showarticletitle{Cognitive $\{$SSD$\}$: A deep learning engine for $\{$In-Storage$\}$ data retrieval}. In \bibinfo{booktitle}{\emph{2019 USENIX Annual Technical Conference (USENIX ATC 19)}}. \bibinfo{pages}{395--410}.
\newblock


\bibitem[\protect\citeauthoryear{Liang, Wang, Yuan, Liu, Li, and Li}{Liang et~al\mbox{.}}{2022}]%
        {liang2022vstore}
\bibfield{author}{\bibinfo{person}{Shengwen Liang}, \bibinfo{person}{Ying Wang}, \bibinfo{person}{Ziming Yuan}, \bibinfo{person}{Cheng Liu}, \bibinfo{person}{Huawei Li}, {and} \bibinfo{person}{Xiaowei Li}.} \bibinfo{year}{2022}\natexlab{}.
\newblock \showarticletitle{VStore: in-storage graph based vector search accelerator}. In \bibinfo{booktitle}{\emph{Proceedings of the 59th ACM/IEEE Design Automation Conference}}. \bibinfo{pages}{997--1002}.
\newblock


\bibitem[\protect\citeauthoryear{Lin and Zhao}{Lin and Zhao}{2019}]%
        {lin2019graph}
\bibfield{author}{\bibinfo{person}{Peng-Cheng Lin} {and} \bibinfo{person}{Wan-Lei Zhao}.} \bibinfo{year}{2019}\natexlab{}.
\newblock \showarticletitle{Graph based nearest neighbor search: Promises and failures}.
\newblock \bibinfo{journal}{\emph{arXiv preprint arXiv:1904.02077}} (\bibinfo{year}{2019}).
\newblock


\bibitem[\protect\citeauthoryear{Liu, Zeng, Chen, Ainihaer, Ramasami, Chen, Xu, Wu, and Wang}{Liu et~al\mbox{.}}{2025}]%
        {liu2025tigervector}
\bibfield{author}{\bibinfo{person}{Shige Liu}, \bibinfo{person}{Zhifang Zeng}, \bibinfo{person}{Li Chen}, \bibinfo{person}{Adil Ainihaer}, \bibinfo{person}{Arun Ramasami}, \bibinfo{person}{Songting Chen}, \bibinfo{person}{Yu Xu}, \bibinfo{person}{Mingxi Wu}, {and} \bibinfo{person}{Jianguo Wang}.} \bibinfo{year}{2025}\natexlab{}.
\newblock \showarticletitle{TigerVector: Supporting Vector Search in Graph Databases for Advanced RAGs}.
\newblock \bibinfo{journal}{\emph{arXiv preprint arXiv:2501.11216}} (\bibinfo{year}{2025}).
\newblock


\bibitem[\protect\citeauthoryear{Liu, Ni, Leng, Feng, Guo, Chen, Li, Guo, and Zhu}{Liu et~al\mbox{.}}{2023}]%
        {liu2023juno}
\bibfield{author}{\bibinfo{person}{Zihan Liu}, \bibinfo{person}{Wentao Ni}, \bibinfo{person}{Jingwen Leng}, \bibinfo{person}{Yu Feng}, \bibinfo{person}{Cong Guo}, \bibinfo{person}{Quan Chen}, \bibinfo{person}{Chao Li}, \bibinfo{person}{Minyi Guo}, {and} \bibinfo{person}{Yuhao Zhu}.} \bibinfo{year}{2023}\natexlab{}.
\newblock \showarticletitle{JUNO: Optimizing High-Dimensional Approximate Nearest Neighbour Search with Sparsity-Aware Algorithm and Ray-Tracing Core Mapping}.
\newblock \bibinfo{journal}{\emph{arXiv preprint arXiv:2312.01712}} (\bibinfo{year}{2023}).
\newblock


\bibitem[\protect\citeauthoryear{Lu, Fang, Farahpour, and Shannon}{Lu et~al\mbox{.}}{2020}]%
        {lu2020chip}
\bibfield{author}{\bibinfo{person}{Alec Lu}, \bibinfo{person}{Zhenman Fang}, \bibinfo{person}{Nazanin Farahpour}, {and} \bibinfo{person}{Lesley Shannon}.} \bibinfo{year}{2020}\natexlab{}.
\newblock \showarticletitle{CHIP-KNN: A configurable and high-performance k-nearest neighbors accelerator on cloud FPGAs}. In \bibinfo{booktitle}{\emph{2020 International Conference on Field-Programmable Technology (ICFPT)}}. IEEE, \bibinfo{pages}{139--147}.
\newblock


\bibitem[\protect\citeauthoryear{Lu, Kudo, Xiao, and Ishikawa}{Lu et~al\mbox{.}}{2021}]%
        {lu2021hvs}
\bibfield{author}{\bibinfo{person}{Kejing Lu}, \bibinfo{person}{Mineichi Kudo}, \bibinfo{person}{Chuan Xiao}, {and} \bibinfo{person}{Yoshiharu Ishikawa}.} \bibinfo{year}{2021}\natexlab{}.
\newblock \showarticletitle{HVS: hierarchical graph structure based on voronoi diagrams for solving approximate nearest neighbor search}.
\newblock \bibinfo{journal}{\emph{Proceedings of the VLDB Endowment}} \bibinfo{volume}{15}, \bibinfo{number}{2} (\bibinfo{year}{2021}), \bibinfo{pages}{246--258}.
\newblock


\bibitem[\protect\citeauthoryear{Lu, Shen, Chen, and Ooi}{Lu et~al\mbox{.}}{2012}]%
        {lu2012efficient}
\bibfield{author}{\bibinfo{person}{Wei Lu}, \bibinfo{person}{Yanyan Shen}, \bibinfo{person}{Su Chen}, {and} \bibinfo{person}{Beng~Chin Ooi}.} \bibinfo{year}{2012}\natexlab{}.
\newblock \showarticletitle{Efficient processing of k nearest neighbor joins using mapreduce}.
\newblock \bibinfo{journal}{\emph{arXiv preprint arXiv:1207.0141}} (\bibinfo{year}{2012}).
\newblock


\bibitem[\protect\citeauthoryear{Malkov, Ponomarenko, Logvinov, and Krylov}{Malkov et~al\mbox{.}}{2014}]%
        {malkov2014approximate}
\bibfield{author}{\bibinfo{person}{Yury Malkov}, \bibinfo{person}{Alexander Ponomarenko}, \bibinfo{person}{Andrey Logvinov}, {and} \bibinfo{person}{Vladimir Krylov}.} \bibinfo{year}{2014}\natexlab{}.
\newblock \showarticletitle{Approximate nearest neighbor algorithm based on navigable small world graphs}.
\newblock \bibinfo{journal}{\emph{Information Systems}}  \bibinfo{volume}{45} (\bibinfo{year}{2014}), \bibinfo{pages}{61--68}.
\newblock


\bibitem[\protect\citeauthoryear{Malkov and Yashunin}{Malkov and Yashunin}{2018}]%
        {malkov2018efficient}
\bibfield{author}{\bibinfo{person}{Yu~A Malkov} {and} \bibinfo{person}{Dmitry~A Yashunin}.} \bibinfo{year}{2018}\natexlab{}.
\newblock \showarticletitle{Efficient and robust approximate nearest neighbor search using hierarchical navigable small world graphs}.
\newblock \bibinfo{journal}{\emph{IEEE transactions on pattern analysis and machine intelligence}} \bibinfo{volume}{42}, \bibinfo{number}{4} (\bibinfo{year}{2018}), \bibinfo{pages}{824--836}.
\newblock


\bibitem[\protect\citeauthoryear{Mehta and Sahni}{Mehta and Sahni}{2004}]%
        {mehta2004handbook}
\bibfield{author}{\bibinfo{person}{Dinesh~P Mehta} {and} \bibinfo{person}{Sartaj Sahni}.} \bibinfo{year}{2004}\natexlab{}.
\newblock \bibinfo{booktitle}{\emph{Handbook of data structures and applications}}.
\newblock \bibinfo{publisher}{Chapman and Hall/CRC}.
\newblock


\bibitem[\protect\citeauthoryear{Meyer and Sanders}{Meyer and Sanders}{2003}]%
        {meyer2003delta}
\bibfield{author}{\bibinfo{person}{Ulrich Meyer} {and} \bibinfo{person}{Peter Sanders}.} \bibinfo{year}{2003}\natexlab{}.
\newblock \showarticletitle{$\Delta$-stepping: a parallelizable shortest path algorithm}.
\newblock \bibinfo{journal}{\emph{Journal of Algorithms}} \bibinfo{volume}{49}, \bibinfo{number}{1} (\bibinfo{year}{2003}), \bibinfo{pages}{114--152}.
\newblock


\bibitem[\protect\citeauthoryear{Mohoney, Pacaci, Chowdhury, Mousavi, Ilyas, Minhas, Pound, and Rekatsinas}{Mohoney et~al\mbox{.}}{2023}]%
        {mohoney2023high}
\bibfield{author}{\bibinfo{person}{Jason Mohoney}, \bibinfo{person}{Anil Pacaci}, \bibinfo{person}{Shihabur~Rahman Chowdhury}, \bibinfo{person}{Ali Mousavi}, \bibinfo{person}{Ihab~F Ilyas}, \bibinfo{person}{Umar~Farooq Minhas}, \bibinfo{person}{Jeffrey Pound}, {and} \bibinfo{person}{Theodoros Rekatsinas}.} \bibinfo{year}{2023}\natexlab{}.
\newblock \showarticletitle{High-Throughput Vector Similarity Search in Knowledge Graphs}.
\newblock \bibinfo{journal}{\emph{Proceedings of the ACM on Management of Data}} \bibinfo{volume}{1}, \bibinfo{number}{2} (\bibinfo{year}{2023}), \bibinfo{pages}{1--25}.
\newblock


\bibitem[\protect\citeauthoryear{Ouyang, Wen, Qin, Chang, Zhang, and Lin}{Ouyang et~al\mbox{.}}{2020}]%
        {ouyang2020progressive}
\bibfield{author}{\bibinfo{person}{Dian Ouyang}, \bibinfo{person}{Dong Wen}, \bibinfo{person}{Lu Qin}, \bibinfo{person}{Lijun Chang}, \bibinfo{person}{Ying Zhang}, {and} \bibinfo{person}{Xuemin Lin}.} \bibinfo{year}{2020}\natexlab{}.
\newblock \showarticletitle{Progressive top-k nearest neighbors search in large road networks}. In \bibinfo{booktitle}{\emph{Proceedings of the 2020 ACM SIGMOD International Conference on Management of Data}}. \bibinfo{pages}{1781--1795}.
\newblock


\bibitem[\protect\citeauthoryear{Pan, Wang, and Li}{Pan et~al\mbox{.}}{2023}]%
        {pan2023survey}
\bibfield{author}{\bibinfo{person}{James~Jie Pan}, \bibinfo{person}{Jianguo Wang}, {and} \bibinfo{person}{Guoliang Li}.} \bibinfo{year}{2023}\natexlab{}.
\newblock \showarticletitle{Survey of vector database management systems}.
\newblock \bibinfo{journal}{\emph{arXiv preprint arXiv:2310.14021}} (\bibinfo{year}{2023}).
\newblock


\bibitem[\protect\citeauthoryear{Patel, Choukse, Zhang, Goiri, Shah, Maleki, and Bianchini}{Patel et~al\mbox{.}}{2023}]%
        {patel2023splitwise}
\bibfield{author}{\bibinfo{person}{Pratyush Patel}, \bibinfo{person}{Esha Choukse}, \bibinfo{person}{Chaojie Zhang}, \bibinfo{person}{{\'I}{\~n}igo Goiri}, \bibinfo{person}{Aashaka Shah}, \bibinfo{person}{Saeed Maleki}, {and} \bibinfo{person}{Ricardo Bianchini}.} \bibinfo{year}{2023}\natexlab{}.
\newblock \showarticletitle{Splitwise: Efficient generative llm inference using phase splitting}.
\newblock \bibinfo{journal}{\emph{arXiv preprint arXiv:2311.18677}} (\bibinfo{year}{2023}).
\newblock


\bibitem[\protect\citeauthoryear{Peng, Chen, Wang, Yang, Weitze, Geng, Li, Bi, Song, Jiang, et~al\mbox{.}}{Peng et~al\mbox{.}}{2021}]%
        {peng2021optimizing}
\bibfield{author}{\bibinfo{person}{Hongwu Peng}, \bibinfo{person}{Shiyang Chen}, \bibinfo{person}{Zhepeng Wang}, \bibinfo{person}{Junhuan Yang}, \bibinfo{person}{Scott~A Weitze}, \bibinfo{person}{Tong Geng}, \bibinfo{person}{Ang Li}, \bibinfo{person}{Jinbo Bi}, \bibinfo{person}{Minghu Song}, \bibinfo{person}{Weiwen Jiang}, {et~al\mbox{.}}} \bibinfo{year}{2021}\natexlab{}.
\newblock \showarticletitle{Optimizing fpga-based accelerator design for large-scale molecular similarity search (special session paper)}. In \bibinfo{booktitle}{\emph{2021 IEEE/ACM International Conference On Computer Aided Design (ICCAD)}}. IEEE, \bibinfo{pages}{1--7}.
\newblock


\bibitem[\protect\citeauthoryear{Peng, Choi, Chan, Yang, and Xu}{Peng et~al\mbox{.}}{2023a}]%
        {peng2023efficient}
\bibfield{author}{\bibinfo{person}{Yun Peng}, \bibinfo{person}{Byron Choi}, \bibinfo{person}{Tsz~Nam Chan}, \bibinfo{person}{Jianye Yang}, {and} \bibinfo{person}{Jianliang Xu}.} \bibinfo{year}{2023}\natexlab{a}.
\newblock \showarticletitle{Efficient approximate nearest neighbor search in multi-dimensional databases}.
\newblock \bibinfo{journal}{\emph{Proceedings of the ACM on Management of Data}} \bibinfo{volume}{1}, \bibinfo{number}{1} (\bibinfo{year}{2023}), \bibinfo{pages}{1--27}.
\newblock


\bibitem[\protect\citeauthoryear{Peng, Zhang, Li, Jin, and Ren}{Peng et~al\mbox{.}}{2023b}]%
        {peng2023iqan}
\bibfield{author}{\bibinfo{person}{Zhen Peng}, \bibinfo{person}{Minjia Zhang}, \bibinfo{person}{Kai Li}, \bibinfo{person}{Ruoming Jin}, {and} \bibinfo{person}{Bin Ren}.} \bibinfo{year}{2023}\natexlab{b}.
\newblock \showarticletitle{iqan: Fast and accurate vector search with efficient intra-query parallelism on multi-core architectures}. In \bibinfo{booktitle}{\emph{Proceedings of the 28th ACM SIGPLAN Annual Symposium on Principles and Practice of Parallel Programming}}. \bibinfo{pages}{313--328}.
\newblock


\bibitem[\protect\citeauthoryear{Putnam, Caulfield, Chung, Chiou, Constantinides, Demme, Esmaeilzadeh, Fowers, Gopal, Gray, Haselman, Hauck, Heil, Hormati, Kim, Lanka, Larus, Peterson, Pope, Smith, Thong, Xiao, and Burger}{Putnam et~al\mbox{.}}{2014}]%
        {putnam2014reconfigurable}
\bibfield{author}{\bibinfo{person}{Andrew Putnam}, \bibinfo{person}{Adrian~M Caulfield}, \bibinfo{person}{Eric~S Chung}, \bibinfo{person}{Derek Chiou}, \bibinfo{person}{Kypros Constantinides}, \bibinfo{person}{John Demme}, \bibinfo{person}{Hadi Esmaeilzadeh}, \bibinfo{person}{Jeremy Fowers}, \bibinfo{person}{Gopi~Prashanth Gopal}, \bibinfo{person}{Jan Gray}, \bibinfo{person}{Michael Haselman}, \bibinfo{person}{Scott Hauck}, \bibinfo{person}{Stephen Heil}, \bibinfo{person}{Amir Hormati}, \bibinfo{person}{Joo-Young Kim}, \bibinfo{person}{Sitaram Lanka}, \bibinfo{person}{James Larus}, \bibinfo{person}{Eric Peterson}, \bibinfo{person}{Simon Pope}, \bibinfo{person}{Aaron Smith}, \bibinfo{person}{Jason Thong}, \bibinfo{person}{Phillip~Yi Xiao}, {and} \bibinfo{person}{Doug Burger}.} \bibinfo{year}{2014}\natexlab{}.
\newblock \showarticletitle{A reconfigurable fabric for accelerating large-scale datacenter services}.
\newblock \bibinfo{journal}{\emph{ACM SIGARCH Computer Architecture News}} \bibinfo{volume}{42}, \bibinfo{number}{3} (\bibinfo{year}{2014}), \bibinfo{pages}{13--24}.
\newblock


\bibitem[\protect\citeauthoryear{Ren, Zhang, and Li}{Ren et~al\mbox{.}}{2020}]%
        {ren2020hm}
\bibfield{author}{\bibinfo{person}{Jie Ren}, \bibinfo{person}{Minjia Zhang}, {and} \bibinfo{person}{Dong Li}.} \bibinfo{year}{2020}\natexlab{}.
\newblock \showarticletitle{Hm-ann: Efficient billion-point nearest neighbor search on heterogeneous memory}.
\newblock \bibinfo{journal}{\emph{Advances in Neural Information Processing Systems}}  \bibinfo{volume}{33} (\bibinfo{year}{2020}), \bibinfo{pages}{10672--10684}.
\newblock


\bibitem[\protect\citeauthoryear{Roy, Bindschaedler, Malicevic, and Zwaenepoel}{Roy et~al\mbox{.}}{2015}]%
        {roy2015chaos}
\bibfield{author}{\bibinfo{person}{Amitabha Roy}, \bibinfo{person}{Laurent Bindschaedler}, \bibinfo{person}{Jasmina Malicevic}, {and} \bibinfo{person}{Willy Zwaenepoel}.} \bibinfo{year}{2015}\natexlab{}.
\newblock \showarticletitle{Chaos: Scale-out graph processing from secondary storage}. In \bibinfo{booktitle}{\emph{Proceedings of the 25th Symposium on Operating Systems Principles}}. \bibinfo{pages}{410--424}.
\newblock


\bibitem[\protect\citeauthoryear{Sivic and Zisserman}{Sivic and Zisserman}{2003}]%
        {IVF}
\bibfield{author}{\bibinfo{person}{Josef Sivic} {and} \bibinfo{person}{Andrew Zisserman}.} \bibinfo{year}{2003}\natexlab{}.
\newblock \showarticletitle{Video Google: A text retrieval approach to object matching in videos}. In \bibinfo{booktitle}{\emph{Computer Vision, IEEE International Conference on}}, Vol.~\bibinfo{volume}{3}. IEEE Computer Society, \bibinfo{pages}{1470--1470}.
\newblock


\bibitem[\protect\citeauthoryear{Suchal and N{\'a}vrat}{Suchal and N{\'a}vrat}{2010}]%
        {suchal2010full}
\bibfield{author}{\bibinfo{person}{J{\'a}n Suchal} {and} \bibinfo{person}{Pavol N{\'a}vrat}.} \bibinfo{year}{2010}\natexlab{}.
\newblock \showarticletitle{Full text search engine as scalable k-nearest neighbor recommendation system}. In \bibinfo{booktitle}{\emph{IFIP International Conference on Artificial Intelligence in Theory and Practice}}. Springer, \bibinfo{pages}{165--173}.
\newblock


\bibitem[\protect\citeauthoryear{Sun, Wang, Qin, Zhang, and Lin}{Sun et~al\mbox{.}}{2014}]%
        {sun2014srs}
\bibfield{author}{\bibinfo{person}{Yifang Sun}, \bibinfo{person}{Wei Wang}, \bibinfo{person}{Jianbin Qin}, \bibinfo{person}{Ying Zhang}, {and} \bibinfo{person}{Xuemin Lin}.} \bibinfo{year}{2014}\natexlab{}.
\newblock \showarticletitle{SRS: solving c-approximate nearest neighbor queries in high dimensional euclidean space with a tiny index}.
\newblock \bibinfo{journal}{\emph{Proceedings of the VLDB Endowment}} (\bibinfo{year}{2014}).
\newblock


\bibitem[\protect\citeauthoryear{Vora, Xu, and Gupta}{Vora et~al\mbox{.}}{2016}]%
        {vora2016load}
\bibfield{author}{\bibinfo{person}{Keval Vora}, \bibinfo{person}{Guoqing Xu}, {and} \bibinfo{person}{Rajiv Gupta}.} \bibinfo{year}{2016}\natexlab{}.
\newblock \showarticletitle{Load the Edges You Need: A Generic $\{$I/O$\}$ Optimization for Disk-based Graph Processing}. In \bibinfo{booktitle}{\emph{2016 USENIX Annual Technical Conference (USENIX ATC 16)}}. \bibinfo{pages}{507--522}.
\newblock


\bibitem[\protect\citeauthoryear{Wang, Yi, Guo, Jin, Xu, Li, Wang, Guo, Li, Xu, et~al\mbox{.}}{Wang et~al\mbox{.}}{2021b}]%
        {wang2021milvus}
\bibfield{author}{\bibinfo{person}{Jianguo Wang}, \bibinfo{person}{Xiaomeng Yi}, \bibinfo{person}{Rentong Guo}, \bibinfo{person}{Hai Jin}, \bibinfo{person}{Peng Xu}, \bibinfo{person}{Shengjun Li}, \bibinfo{person}{Xiangyu Wang}, \bibinfo{person}{Xiangzhou Guo}, \bibinfo{person}{Chengming Li}, \bibinfo{person}{Xiaohai Xu}, {et~al\mbox{.}}} \bibinfo{year}{2021}\natexlab{b}.
\newblock \showarticletitle{Milvus: A purpose-built vector data management system}. In \bibinfo{booktitle}{\emph{Proceedings of the 2021 International Conference on Management of Data}}. \bibinfo{pages}{2614--2627}.
\newblock


\bibitem[\protect\citeauthoryear{Wang, Xu, Yi, Wu, Peng, Ke, Gao, Xu, Guo, and Xie}{Wang et~al\mbox{.}}{2024}]%
        {wang2024starling}
\bibfield{author}{\bibinfo{person}{Mengzhao Wang}, \bibinfo{person}{Weizhi Xu}, \bibinfo{person}{Xiaomeng Yi}, \bibinfo{person}{Songlin Wu}, \bibinfo{person}{Zhangyang Peng}, \bibinfo{person}{Xiangyu Ke}, \bibinfo{person}{Yunjun Gao}, \bibinfo{person}{Xiaoliang Xu}, \bibinfo{person}{Rentong Guo}, {and} \bibinfo{person}{Charles Xie}.} \bibinfo{year}{2024}\natexlab{}.
\newblock \showarticletitle{Starling: An i/o-efficient disk-resident graph index framework for high-dimensional vector similarity search on data segment}.
\newblock \bibinfo{journal}{\emph{Proceedings of the ACM on Management of Data}} \bibinfo{volume}{2}, \bibinfo{number}{1} (\bibinfo{year}{2024}), \bibinfo{pages}{1--27}.
\newblock


\bibitem[\protect\citeauthoryear{Wang, Xu, Yue, and Wang}{Wang et~al\mbox{.}}{2021a}]%
        {wang2021comprehensive}
\bibfield{author}{\bibinfo{person}{Mengzhao Wang}, \bibinfo{person}{Xiaoliang Xu}, \bibinfo{person}{Qiang Yue}, {and} \bibinfo{person}{Yuxiang Wang}.} \bibinfo{year}{2021}\natexlab{a}.
\newblock \showarticletitle{A comprehensive survey and experimental comparison of graph-based approximate nearest neighbor search}.
\newblock \bibinfo{journal}{\emph{Proceedings of the VLDB Endowment}} (\bibinfo{year}{2021}).
\newblock


\bibitem[\protect\citeauthoryear{Wang, Zhang, Zhang, Lin, and Cheema}{Wang et~al\mbox{.}}{2015}]%
        {wang2015optimal}
\bibfield{author}{\bibinfo{person}{Xiaoyang Wang}, \bibinfo{person}{Ying Zhang}, \bibinfo{person}{Wenjie Zhang}, \bibinfo{person}{Xuemin Lin}, {and} \bibinfo{person}{Muhammad~Aamir Cheema}.} \bibinfo{year}{2015}\natexlab{}.
\newblock \showarticletitle{Optimal spatial dominance: an effective search of nearest neighbor candidates}. In \bibinfo{booktitle}{\emph{Proceedings of the 2015 ACM SIGMOD International Conference on Management of Data}}. \bibinfo{pages}{923--938}.
\newblock


\bibitem[\protect\citeauthoryear{Wang, Li, Zheng, Song, Li, Chang, Li, Chen, et~al\mbox{.}}{Wang et~al\mbox{.}}{2023}]%
        {wang2023storage}
\bibfield{author}{\bibinfo{person}{Yitu Wang}, \bibinfo{person}{Shiyu Li}, \bibinfo{person}{Qilin Zheng}, \bibinfo{person}{Linghao Song}, \bibinfo{person}{Zongwang Li}, \bibinfo{person}{Andrew Chang}, \bibinfo{person}{Hai Li}, \bibinfo{person}{Yiran Chen}, {et~al\mbox{.}}} \bibinfo{year}{2023}\natexlab{}.
\newblock \showarticletitle{In-Storage Acceleration of Graph-Traversal-Based Approximate Nearest Neighbor Search}.
\newblock \bibinfo{journal}{\emph{arXiv preprint arXiv:2312.03141}} (\bibinfo{year}{2023}).
\newblock


\bibitem[\protect\citeauthoryear{Wei, Wu, Wang, Lou, Zhan, Li, and Cai}{Wei et~al\mbox{.}}{2020}]%
        {adb-v}
\bibfield{author}{\bibinfo{person}{Chuangxian Wei}, \bibinfo{person}{Bin Wu}, \bibinfo{person}{Sheng Wang}, \bibinfo{person}{Renjie Lou}, \bibinfo{person}{Chaoqun Zhan}, \bibinfo{person}{Feifei Li}, {and} \bibinfo{person}{Yuanzhe Cai}.} \bibinfo{year}{2020}\natexlab{}.
\newblock \showarticletitle{AnalyticDB-V: a hybrid analytical engine towards query fusion for structured and unstructured data}.
\newblock \bibinfo{journal}{\emph{Proceedings of the VLDB Endowment}} \bibinfo{volume}{13}, \bibinfo{number}{12} (\bibinfo{year}{2020}), \bibinfo{pages}{3152--3165}.
\newblock


\bibitem[\protect\citeauthoryear{Wieschollek, Wang, Sorkine-Hornung, and Lensch}{Wieschollek et~al\mbox{.}}{2016}]%
        {wieschollek2016efficient}
\bibfield{author}{\bibinfo{person}{Patrick Wieschollek}, \bibinfo{person}{Oliver Wang}, \bibinfo{person}{Alexander Sorkine-Hornung}, {and} \bibinfo{person}{Hendrik Lensch}.} \bibinfo{year}{2016}\natexlab{}.
\newblock \showarticletitle{Efficient large-scale approximate nearest neighbor search on the gpu}. In \bibinfo{booktitle}{\emph{Proceedings of the IEEE Conference on Computer Vision and Pattern Recognition}}. \bibinfo{pages}{2027--2035}.
\newblock


\bibitem[\protect\citeauthoryear{Wu, Jin, and Zhang}{Wu et~al\mbox{.}}{2014}]%
        {wu2014fast}
\bibfield{author}{\bibinfo{person}{Yubao Wu}, \bibinfo{person}{Ruoming Jin}, {and} \bibinfo{person}{Xiang Zhang}.} \bibinfo{year}{2014}\natexlab{}.
\newblock \showarticletitle{Fast and unified local search for random walk based k-nearest-neighbor query in large graphs}. In \bibinfo{booktitle}{\emph{Proceedings of the 2014 ACM SIGMOD international conference on Management of Data}}. \bibinfo{pages}{1139--1150}.
\newblock


\bibitem[\protect\citeauthoryear{Xiong, Xiong, Li, Tang, Liu, Bennett, Ahmed, and Overwijk}{Xiong et~al\mbox{.}}{2020}]%
        {xiong2020approximate}
\bibfield{author}{\bibinfo{person}{Lee Xiong}, \bibinfo{person}{Chenyan Xiong}, \bibinfo{person}{Ye Li}, \bibinfo{person}{Kwok-Fung Tang}, \bibinfo{person}{Jialin Liu}, \bibinfo{person}{Paul Bennett}, \bibinfo{person}{Junaid Ahmed}, {and} \bibinfo{person}{Arnold Overwijk}.} \bibinfo{year}{2020}\natexlab{}.
\newblock \showarticletitle{Approximate nearest neighbor negative contrastive learning for dense text retrieval}.
\newblock \bibinfo{journal}{\emph{arXiv preprint arXiv:2007.00808}} (\bibinfo{year}{2020}).
\newblock


\bibitem[\protect\citeauthoryear{Xu, Chen, Hsu, Kang, Zhou, Pinge, Yu, and Rosing}{Xu et~al\mbox{.}}{2023}]%
        {xu2023proxima}
\bibfield{author}{\bibinfo{person}{Weihong Xu}, \bibinfo{person}{Junwei Chen}, \bibinfo{person}{Po-Kai Hsu}, \bibinfo{person}{Jaeyoung Kang}, \bibinfo{person}{Minxuan Zhou}, \bibinfo{person}{Sumukh Pinge}, \bibinfo{person}{Shimeng Yu}, {and} \bibinfo{person}{Tajana Rosing}.} \bibinfo{year}{2023}\natexlab{}.
\newblock \showarticletitle{Proxima: Near-storage Acceleration for Graph-based Approximate Nearest Neighbor Search in 3D NAND}.
\newblock \bibinfo{journal}{\emph{arXiv preprint arXiv:2312.04257}} (\bibinfo{year}{2023}).
\newblock


\bibitem[\protect\citeauthoryear{Yang, Cheema, Lin, and Wang}{Yang et~al\mbox{.}}{2015}]%
        {yang2015reverse}
\bibfield{author}{\bibinfo{person}{Shiyu Yang}, \bibinfo{person}{Muhammad~Aamir Cheema}, \bibinfo{person}{Xuemin Lin}, {and} \bibinfo{person}{Wei Wang}.} \bibinfo{year}{2015}\natexlab{}.
\newblock \showarticletitle{Reverse k nearest neighbors query processing: experiments and analysis}.
\newblock \bibinfo{journal}{\emph{Proceedings of the VLDB Endowment}} \bibinfo{volume}{8}, \bibinfo{number}{5} (\bibinfo{year}{2015}), \bibinfo{pages}{605--616}.
\newblock


\bibitem[\protect\citeauthoryear{Yang, Li, Fang, and Wei}{Yang et~al\mbox{.}}{2020}]%
        {yang2020pase}
\bibfield{author}{\bibinfo{person}{Wen Yang}, \bibinfo{person}{Tao Li}, \bibinfo{person}{Gai Fang}, {and} \bibinfo{person}{Hong Wei}.} \bibinfo{year}{2020}\natexlab{}.
\newblock \showarticletitle{Pase: Postgresql ultra-high-dimensional approximate nearest neighbor search extension}. In \bibinfo{booktitle}{\emph{Proceedings of the 2020 ACM SIGMOD international conference on management of data}}. \bibinfo{pages}{2241--2253}.
\newblock


\bibitem[\protect\citeauthoryear{Zeng, Luo, Ning, Han, Jiang, Tang, Wang, Chen, and Guo}{Zeng et~al\mbox{.}}{2022}]%
        {zeng2022faery}
\bibfield{author}{\bibinfo{person}{Chaoliang Zeng}, \bibinfo{person}{Layong Luo}, \bibinfo{person}{Qingsong Ning}, \bibinfo{person}{Yaodong Han}, \bibinfo{person}{Yuhang Jiang}, \bibinfo{person}{Ding Tang}, \bibinfo{person}{Zilong Wang}, \bibinfo{person}{Kai Chen}, {and} \bibinfo{person}{Chuanxiong Guo}.} \bibinfo{year}{2022}\natexlab{}.
\newblock \showarticletitle{$\{$FAERY$\}$: An $\{$FPGA-accelerated$\}$ Embedding-based Retrieval System}. In \bibinfo{booktitle}{\emph{16th USENIX Symposium on Operating Systems Design and Implementation (OSDI 22)}}. \bibinfo{pages}{841--856}.
\newblock


\bibitem[\protect\citeauthoryear{Zeng, Zhu, Liu, Zhang, Dai, Zhou, Li, Ning, Xie, Yang, et~al\mbox{.}}{Zeng et~al\mbox{.}}{2023}]%
        {zeng2023df}
\bibfield{author}{\bibinfo{person}{Shulin Zeng}, \bibinfo{person}{Zhenhua Zhu}, \bibinfo{person}{Jun Liu}, \bibinfo{person}{Haoyu Zhang}, \bibinfo{person}{Guohao Dai}, \bibinfo{person}{Zixuan Zhou}, \bibinfo{person}{Shuangchen Li}, \bibinfo{person}{Xuefei Ning}, \bibinfo{person}{Yuan Xie}, \bibinfo{person}{Huazhong Yang}, {et~al\mbox{.}}} \bibinfo{year}{2023}\natexlab{}.
\newblock \showarticletitle{DF-GAS: a Distributed FPGA-as-a-Service Architecture towards Billion-Scale Graph-based Approximate Nearest Neighbor Search}.
\newblock  (\bibinfo{year}{2023}).
\newblock


\bibitem[\protect\citeauthoryear{Zhang, Khoram, and Li}{Zhang et~al\mbox{.}}{2018}]%
        {zhang2018efficient}
\bibfield{author}{\bibinfo{person}{Jialiang Zhang}, \bibinfo{person}{Soroosh Khoram}, {and} \bibinfo{person}{Jing Li}.} \bibinfo{year}{2018}\natexlab{}.
\newblock \showarticletitle{Efficient large-scale approximate nearest neighbor search on OpenCL FPGA}. In \bibinfo{booktitle}{\emph{Proceedings of the IEEE Conference on Computer Vision and Pattern Recognition}}. \bibinfo{pages}{4924--4932}.
\newblock


\bibitem[\protect\citeauthoryear{Zhang, Wahib, Zhang, and Matsuoka}{Zhang et~al\mbox{.}}{2020}]%
        {zhang2020study}
\bibfield{author}{\bibinfo{person}{Lingqi Zhang}, \bibinfo{person}{Mohamed Wahib}, \bibinfo{person}{Haoyu Zhang}, {and} \bibinfo{person}{Satoshi Matsuoka}.} \bibinfo{year}{2020}\natexlab{}.
\newblock \showarticletitle{A study of single and multi-device synchronization methods in Nvidia GPUs}. In \bibinfo{booktitle}{\emph{2020 IEEE International Parallel and Distributed Processing Symposium (IPDPS)}}. IEEE, \bibinfo{pages}{483--493}.
\newblock


\bibitem[\protect\citeauthoryear{Zhao, Tan, and Li}{Zhao et~al\mbox{.}}{2020}]%
        {zhao2020song}
\bibfield{author}{\bibinfo{person}{Weijie Zhao}, \bibinfo{person}{Shulong Tan}, {and} \bibinfo{person}{Ping Li}.} \bibinfo{year}{2020}\natexlab{}.
\newblock \showarticletitle{Song: Approximate nearest neighbor search on gpu}. In \bibinfo{booktitle}{\emph{2020 IEEE 36th International Conference on Data Engineering (ICDE)}}. IEEE, \bibinfo{pages}{1033--1044}.
\newblock


\bibitem[\protect\citeauthoryear{Zhao, Tian, Huang, Zheng, and Zhou}{Zhao et~al\mbox{.}}{2023}]%
        {zhao2023towards}
\bibfield{author}{\bibinfo{person}{Xi Zhao}, \bibinfo{person}{Yao Tian}, \bibinfo{person}{Kai Huang}, \bibinfo{person}{Bolong Zheng}, {and} \bibinfo{person}{Xiaofang Zhou}.} \bibinfo{year}{2023}\natexlab{}.
\newblock \showarticletitle{Towards efficient index construction and approximate nearest neighbor search in high-dimensional spaces}.
\newblock \bibinfo{journal}{\emph{Proceedings of the VLDB Endowment}} \bibinfo{volume}{16}, \bibinfo{number}{8} (\bibinfo{year}{2023}), \bibinfo{pages}{1979--1991}.
\newblock


\bibitem[\protect\citeauthoryear{Zheng, Guo, Tung, and Wu}{Zheng et~al\mbox{.}}{2016}]%
        {zheng2016lazylsh}
\bibfield{author}{\bibinfo{person}{Yuxin Zheng}, \bibinfo{person}{Qi Guo}, \bibinfo{person}{Anthony~KH Tung}, {and} \bibinfo{person}{Sai Wu}.} \bibinfo{year}{2016}\natexlab{}.
\newblock \showarticletitle{Lazylsh: Approximate nearest neighbor search for multiple distance functions with a single index}. In \bibinfo{booktitle}{\emph{Proceedings of the 2016 International Conference on Management of Data}}. \bibinfo{pages}{2023--2037}.
\newblock


\bibitem[\protect\citeauthoryear{Zhong, Liu, Chen, Hu, Zhu, Liu, Jin, and Zhang}{Zhong et~al\mbox{.}}{2024}]%
        {zhong2024distserve}
\bibfield{author}{\bibinfo{person}{Yinmin Zhong}, \bibinfo{person}{Shengyu Liu}, \bibinfo{person}{Junda Chen}, \bibinfo{person}{Jianbo Hu}, \bibinfo{person}{Yibo Zhu}, \bibinfo{person}{Xuanzhe Liu}, \bibinfo{person}{Xin Jin}, {and} \bibinfo{person}{Hao Zhang}.} \bibinfo{year}{2024}\natexlab{}.
\newblock \showarticletitle{Distserve: Disaggregating prefill and decoding for goodput-optimized large language model serving}.
\newblock \bibinfo{journal}{\emph{arXiv preprint arXiv:2401.09670}} (\bibinfo{year}{2024}).
\newblock


\bibitem[\protect\citeauthoryear{Zhu, Yang, Wang, and Lee}{Zhu et~al\mbox{.}}{2016}]%
        {zhu2016range}
\bibfield{author}{\bibinfo{person}{Huaijie Zhu}, \bibinfo{person}{Xiaochun Yang}, \bibinfo{person}{Bin Wang}, {and} \bibinfo{person}{Wang-Chien Lee}.} \bibinfo{year}{2016}\natexlab{}.
\newblock \showarticletitle{Range-based obstructed nearest neighbor queries}. In \bibinfo{booktitle}{\emph{Proceedings of the 2016 International Conference on Management of Data}}. \bibinfo{pages}{2053--2068}.
\newblock


\bibitem[\protect\citeauthoryear{Zuo and Deng}{Zuo and Deng}{2023}]%
        {zuo2023arkgraph}
\bibfield{author}{\bibinfo{person}{Chaoji Zuo} {and} \bibinfo{person}{Dong Deng}.} \bibinfo{year}{2023}\natexlab{}.
\newblock \showarticletitle{ARKGraph: All-Range Approximate K-Nearest-Neighbor Graph}.
\newblock \bibinfo{journal}{\emph{Proceedings of the VLDB Endowment}} \bibinfo{volume}{16}, \bibinfo{number}{10} (\bibinfo{year}{2023}), \bibinfo{pages}{2645--2658}.
\newblock


\end{thebibliography}

\end{document}